\documentclass[apj]{emulateapj}

\usepackage{xspace}
\usepackage{graphicx}
\usepackage{color}
\usepackage{amsmath}

\usepackage[T1]{fontenc}
\usepackage{aecompl}

\usepackage{hyperref}

\newcommand{\Ha}{{{\rm H}\ensuremath{\,\alpha}}}

\newcommand{\kpc}{\,{\rm kpc}}
\newcommand{\sqkpc}{\,{\rm kpc}^{-2}}

\newcommand{\Msun}{\,{\rm M}_{\odot}}
\newcommand{\peryr}{\,{\rm yr}^{-1}}
\newcommand{\Msunyr}{\ensuremath{\,\Msun\, {\rm yr}^{-1}}}
\newcommand{\ang}{\ensuremath{\textrm{\AA}}\xspace}
\newcommand{\Mstellar}{\ensuremath{\mathcal{M}_{*}}}
\newcommand{\Mgas}{\ensuremath{\mathcal{M}_{\rm gas}}}

\newcommand{\fluxys}{\ensuremath{10^{-17}\, {\rm erg}\, {\rm s}^{-1}\,
    {\rm cm}^{-2}}}
\newcommand{\flux}{\, {\rm erg}\, {\rm s}^{-1}\,
    {\rm cm}^{-2}}
\newcommand{\ergsec}{\ensuremath{\,{\rm erg}\,\, {\rm s}^{-1}}}
\newcommand{\kms}{\ensuremath{\, {\rm km\, s^{-1}}}}
\newcommand{\kelvin}{\mathrm{{}^{\circ}K}}

\newcommand{\Vshear}{V_{\rm shear}}
\newcommand{\Vcirc}{V_{2.2R_r}}
\newcommand{\sigmean}{\texorpdfstring{\ensuremath{\sigma_{\rm m}}}{sigma\_m}}
\newcommand{\sigmeancorr}{\texorpdfstring{\ensuremath{\sigma_{\rm m,corr}}}{sigma\_m,corr}}

\newcommand{\sigpix}{\sigma_{\rm pix}}

\newcommand{\erf}{\mathop{\mathrm{erf}}}

\usepackage{boolexpr}

\newcommand{\samplesize}{67}

\newcommand{\n}[1]{%
  \switch[\pdfstrcmp{#1}]%
  \case{{sample}}67%
  \case{{disklike}}56
  \case{{compact}}25
  \case{{rd}}25%
  \case{{pr}}6%
  \case{{mingasfrac}}0.06%
  \case{{maxgasfrac}}0.77%
  \case{{minstellarmass}}1.09
  \case{{maxstellarmass}}65.0%
  \case{{sfrcut}}35%
  \otherwise\textbf{?}%
  \endswitch%
}

\shorttitle{DYNAMO I: Resolved H$\alpha$ Kinematics}

\shortauthors{Green et al.}

\journalinfo{Accepted for publication in MNRAS 2013 October 02}
\submitted{}

\begin{document}

\title{DYNAMO I: A Sample of \Ha-Luminous Galaxies with Resolved Kinematics}

\author{
  Andrew W. Green$^{1}$, %
  Karl Glazebrook$^2$, %
  Peter J. McGregor$^3$, %
  Ivana Damjanov$^4$, %
  Emily Wisnioski$^5$, %
  Roberto G. Abraham$^6$, %
  Matthew Colless$^3$, %
  Robert G. Sharp$^3$ and %
  Robert A. Crain$^7$, %
  Gregory B. Poole$^8$, %
  Patrick J. McCarthy$^{9,10}$ %
}

\affil{$^1$Australian Astronomical Observatory, PO Box 970, North Ryde, NSW
  1670, Australia}
  
\affil{$^2$Centre for Astrophysics and Supercomputing, Swinburne
  University, Hawthorn VIC 3122, Australia}
  
\affil{$^3$Research School of Astronomy and Astrophysics, Australian
  National University, Cotter Rd, Weston, ACT 2611, Australia}

\affil{$^4$Harvard-Smithsonian CfA, 60 Garden St., MS-20, Cambridge, MA 02138, USA}

\affil{$^{5}$Max-Plank-Institut f\"{u}r extraterrestrische Physik, Postfach
  1312, Giessenbachstr., D-85741 Garching, Germany}

\affil{$^6$Department of Astronomy and Astrophysics, University of
    Toronto, 50 St George St, Toronto, ON M5S3H4, Canada}

\affil{$^7$Leiden Observatory, PO Box 9513, NL-2300 RA, Leiden, The Netherlands}

\affil{$^8$School of Physics, The University of Melbourne, Parkville VIC 3010, Australia}

 \affil{$^9$Observatories of the Carnegie Institution of Washington, 813
    Santa Barbara Street, Pasadena, CA 91101, USA}

  \affil{$^{10}$Giant Magellan Telescope Organisation, 251 S. Lake
    Avenue, Suite 300, Pasadena, CA 91101, USA}

\email{andrew.green@aao.gov.au}

\begin{abstract}

  {\small DYNAMO} is a multi-wavelength, spatially-resolved survey of
  local ($z \sim 0.1$) star-forming galaxies designed to study
  evolution through comparison with samples at $z \simeq 2$. Half of
  the sample has integrated \Ha{} luminosities of $>10^{42}\ergsec$,
  the typical lower limit for resolved spectroscopy at $z \simeq
  2$. The sample covers a range in stellar mass
  ($10^9$--$10^{11}\Msun$) and star-formation rate
  (0.2--$100\Msunyr$).  In this first paper of a series, we present
  integral-field spectroscopy of \Ha{} emission for the sample of
  \samplesize\ galaxies. We infer gas fractions in our sample as high
  as $\simeq 0.8$, higher than typical for local galaxies. Gas
  fraction correlates with stellar mass in galaxies with
  star-formation rates below $10\Msunyr$, as found by {\small
    COLDGASS}, but galaxies with higher star-formation rates have
  higher than expected gas fractions. There is only a weak
  correlation, if any, between gas fraction and gas velocity
  dispersion. Galaxies in the sample visually classified as disc-like
  are offset from the local stellar-mass Tully-Fisher relation to
  higher circular velocities, but this offset vanishes when both gas
  and stars are included in the baryonic Tully-Fisher relation. The
  mean gas velocity dispersion of the sample is $\simeq 50\kms$, and
  $V/\sigma$ ranges from 2 to 10 for most of the discs, similar to
  `turbulent' galaxies at high redshift. Half of our sample show
  disc-like rotation, while $\sim$20 per~cent show no signs of
  rotation. The division between rotating and non-rotating is
  approximately equal for the sub-samples with either star-formation
  rates $>10 \Msunyr$, or specific-star-formation rates typical of the
  star-formation `main sequence' at $z \simeq 2$. Across our whole
  sample, we find good correlation between the dominance of
  `turbulence' in galaxy discs (as expressed by $V/\sigma$) and gas
  fraction as has been predicted for marginally stable Toomre
  discs. Comparing our sample with many others at low- and
  high-redshift reveals a correlation between gas velocity dispersion
  and star formation rate. These findings suggest the {\small DYNAMO}
  discs are excellent candidates for local galaxies similar to
  turbulent $z\simeq 2$ disc galaxies.

\end{abstract}


\section{Introduction}\label{sec:introduction}

At high redshift, galaxies have much higher star-formation rates than
galaxies today \citep{1996MNRAS.283.1388M, 2006ApJ...651..142H}.
Massive galaxies in particular are very strongly star forming, unlike
their modern counterparts \citep{2005ApJ...625...23B,
  2005ApJ...619L.135J}. It has long been known that star-forming
galaxies at high redshift also exhibit very different physical
morphologies from the local Hubble sequence, and a large number have a
clumpy and irregular morphology \citep{1996ApJS..107....1A,
  1996MNRAS.279L..47A, 2003ApJS..147....1C, 1996AJ....112..359V}. Such
structures could be associated with merging objects and early
modelling suggested that such merging would be the primary mechanism
of mass growth of massive galaxies \citep{1996MNRAS.282L..27B,
  2000MNRAS.319..168C}. More recently, clumpy morphologies have
been viewed as clumpy star-formation patterns in disc galaxies
\citep{1998Natur.392..253N, 2006ApJ...650..644E, 2004ApJ...604L..21E,
  2005ApJ...627..632E,2006Natur.442..786G, 2009ApJ...701..306E,
  2009ApJ...707L...1B, 2011ApJ...733..101G} with the growth of mass
dominated by \emph{in situ} star formation from cosmological accretion
\citep{2009ApJ...703..785D, 2009Natur.457..451D, 2010MNRAS.404.2151C}. Kinematic studies offer the possibility to distinguish clumpy discs from mergers.

Scenarios that explain the differences seen in high-redshift galaxies
have been tested with spatially resolved kinematics, primarily probing
the strong \Ha\ emission line associated with star formation
\citep[see][for a review]{2013arXiv1305.2469G}. In the clumpy disc
picture, galaxies may be photometrically irregular but kinematically
regular, and numerous examples have now been observed
\citep{2006ApJ...647..128E, 2006Natur.442..786G, 2007ApJ...658...78W,
  2009ApJ...706.1364F, 2011MNRAS.417.2601W, 2011ApJ...733..101G}. A
fraction of 30--50 per~cent of massive galaxies at $z>1$ may be discs, with
the fraction likely increasing with stellar mass
\citep{2009ApJ...697.2057L, 2009ApJ...706.1364F}. Despite the
demonstrated existence of disc galaxies, the observed merger rate is
also high, with typically 30--50 per~cent of galaxies in integral-field
spectroscopic surveys demonstrating merger-like kinematics
\citep{2008A&A...477..789Y, 2009ApJ...706.1364F, 2012A&A...546A.118V}.

The rotation velocities, $V$, of disc galaxies at high redshift are similar to those of local galaxies, with little or no evolution in the Tully-Fisher relation at fixed stellar mass \citep{2007ApJ...660L..35K, 2009ApJ...697..115C, 2010A&A...510A..68P, 2011A&A...528A..88G, 2011ApJ...741..115M, 2012A&A...546A.118V}. An unexpected finding was the high velocity dispersion, $\sigma$, of many high-redshift discs. Values of 50--100$\kms$ \citep{2006Natur.442..786G, 2007ApJ...656....1L, 2009ApJ...697.2057L, 2009ApJ...706.1364F, 2011A&A...528A..88G, 2012A&A...546A.118V} are regularly observed. With typical ratios $V/\sigma \sim 1$--5, such discs are dynamically much `hotter' than local spiral discs where $V/\sigma \sim$ 10--20 \citep{2010MNRAS.401.2113E,2006ApJS..166..505A}. This difference has been attributed to high fractions (by mass) of  gas \citep{2011ApJ...733..101G}.  Such high fractions of gas have been observed \citep{2010Natur.463..781T,2013ApJ...768...74T,2010ApJ...713..686D,2013arXiv1301.0371C} and in some cases resolved molecular gas observations have confirmed the high-dispersion discs seen in \Ha\ \citep{2011ApJ...742...11S,2012ApJ...760...11H}.  

Also at high redshift, a significant population of
`dispersion-dominated' galaxies with $\sigma \gtrsim V$ and stellar
masses $>10^{10}\Msun$ has been identified
\citep{2007ApJ...656....1L,2009ApJ...697.2057L}. Dispersion dominated
galaxies of this mass only appear above $z\gtrsim 1$
\citep{2012ApJ...758..106K}. These are generally quite compact
(half-light radii $<2\kpc$) and may be unresolved small discs
\citep{2013ApJ...767..104N}.

High velocity dispersion seems to be a universal feature of galaxies
at high redshift. The velocity dispersions are supersonic and likely
represent a highly turbulent interstellar medium. A large turbulent
velocity dispersion and pressure support gives rise to a large Jeans
mass \citep[$10^8$--$10^9 \Msun$; see][]{2009ApJ...701..306E} for
gravitational collapse and hence imply that galactic star formation
will be dominated by a handful of giant clumps, consistent with the
irregular morphologies. The physical mechanism sustaining the
turbulence, which would otherwise decay quickly, is not yet determined
but may be due to star-formation feedback \citep{2009ApJ...699.1660L,
  2013arXiv1304.7734L, 2010Natur.467..684G, 2011A&A...534L...4L},
initial gravitational collapse \citep{2010ApJ...712..294E}, ongoing
cosmic accretion \citep{2010ApJ...719.1230A}, gravitational
instabilities \citep{2004ApJ...611...20I, 2010MNRAS.409.1088B,
  2010MNRAS.404.2151C, 2010ApJ...719.1230A} or some combination
thereof.

It is important to note that integral-field spectroscopy at high
redshift is intrinsically difficult and is subject to selection
effects, resolution and surface-brightness limitations. The technique
of Adaptive Optics \citep[AO; reviewed by][]{2012ARA&A..50..305D} has
allowed observations at $\sim 0.1$ arcsec  resolution ($0.8\kpc$ for
$1<z<3$). However, these observations are possible only in a limited
number of cases due to guide star and other constraints; the largest
AO samples number 10--35 objects \citep{2009ApJ...697.2057L,
  2011MNRAS.417.2601W, 2013ApJ...767..104N}. Techniques using adaptive
optics only correct a fraction of the light into a compact PSF
and necessitate finer detector sampling; hence the observations are
only sensitive to features with the highest surface
brightness. Nonetheless, they are essential for resolved spectroscopic
studies of galaxy sub-structures such as clumps.  Techniques using
adaptive optics have also been enhanced in many cases by the
gravitational lensing of giant clusters of galaxies, delivering
200--300 pc resolution and probing sub-luminous galaxies
\citep{2008Natur.455..775S, 2010MNRAS.404.1247J, 2011ApJ...732L..14Y,
  2012ApJ...759...66Y, 2007MNRAS.376..479S, 2009MNRAS.400.1121S,
  2006ApJ...650..661N, 2007ApJ...657..725N}, albeit with highly
anamorphic magnification.  

The majority of integral-field spectroscopy at high redshift has been
performed \emph{without} the resolution improvements provided by
adaptive optics. Whilst `natural seeing' observations have fewer
complications, they only offer 5--8 kpc resolution. Since the point-spread function
is comparable to the size of the galaxies studied, astronomers must
rely to a much greater extent on model fitting to describe each
galaxy's physical parameters \citep{2009ApJ...697..115C,
  2012A&A...539A..92E, 2010MNRAS.402.2291L}. The models used include
many assumptions about the real physical structure of the galaxies to
which they are applied.  With or without adaptive optics, instrument
sensitivity limits most integral-field spectroscopic surveys to
high-redshift galaxies with \Ha\ luminosities of $>10^{42}\ergsec$
\citep[e.g.,][]{2009ApJ...706.1364F, 2009ApJ...697.2057L,
  2009ApJ...699..421W}.  Although only a few percent of local galaxies
reach this luminosity, it does correspond to `normal' galaxies at
$z\sim 2$ in that they follow the star-formation rate--stellar mass
`main sequence' \citep{2007ApJ...660L..43N, 2007ApJ...670..156D}.

Integral-field spectroscopic data of local galaxies comparable to
existing high-redshift samples can clarify the interpretation of
high-redshift-galaxy data. A sample of low-redshift galaxies would
allow the methods applied to high-redshift galaxies to be tested on
data with higher signal-to-noise-ratio and higher resolution (e.g.,
fitting 2D disc models or testing for mergers with kinemetry as in
\citealt{2008ApJ...682..231S}), and on artificially redshifted data
(where the appearance of a galaxy at high-redshift is simulated with
all the noise, PSF, resolution and sensitivity effects
included). Until recently, integral-field spectroscopy of local
galaxies has not been widely available, as historical work has relied
on long-slit spectroscopy. However, surveys such as ATLAS$^\mathrm{3D}$
\citep{2011MNRAS.413..813C}, {\small GHASP}
\citep{2010MNRAS.401.2113E}, {\small CALIFA}
\citep{2012A&A...538A...8S} and {\small DISKMASS}
\citep{2010ApJ...716..198B} have now provided integral-field
spectroscopy of well-resolved, large, local galaxies (within $\sim
100$ Mpc and as large as $\sim$ 1 arcmin on the sky).  None of these
samples includes galaxies with star-formation rates comparable to $z
\sim 2$ objects. It is desirable to construct such a sample to allow
the effects of star-formation rate and temporal evolution to be
considered separately, for example.

There are several samples of local luminous and ultra-luminous infrared galaxies (LIRGs/ULIRGs) observed with integral-field
spectroscopy that can be used for comparison with high-redshift samples. \cite{2005ApJ...621..725C} compared the stellar kinematics with the ionised- and molecular-gas kinematics of 11 ultra-luminous infrared galaxies (ULIRGs) in order to gain insight into similar measurements of $z \simeq 2$ galaxies.  \cite{2008A&A...479..687A} have observed 42 (U)LIRGs selected by their
far-infrared emission. More recently, \cite{2012A&A...541A..20A} and \cite{2012MNRAS.424..416W} have furthered the comparisons with larger and more detailed samples, and applied many of the analysis popularly employed for high-redshift samples to the local ULIRGs, respectively. These samples of ULIRGs have greatly expanded our understanding of highly star-forming galaxies in the nearby universe.

In addition to selections based on infrared luminosity, there is also
integral-field spectroscopy of galaxy samples selected on ultraviolet
luminosity. \cite{2009ApJ...699L.118B} and \cite{2010ApJ...724.1373G}
obtained adaptive-optics-corrected, integral-field spectroscopic data
of 19 galaxies at $z\sim 0.2$. These galaxies were selected via
near-ultraviolet emission to be analogous to $z\sim 3$ `Lyman-Break
Galaxy' dropout selections \citep{1996ApJ...462L..17S} and selected to
have compact morphologies with high surface
brightnesses. Interestingly infrared and ultraviolet selected samples
include some galaxies that are clearly mergers, but could be
mis-classified as discs in simulated $z\sim 2$ observations
\citep{2012A&A...542A..54B, 2010ApJ...724.1373G}. The `Lyman-Break
Analogs' show only weak velocity shears ($V/\sigma \sim 1$) and could
be similar to high-redshift, dispersion-dominated galaxies. Neither
sample probes large, rotationally dominated galaxies with high stellar
mass.

In this paper, we present integral-field spectroscopy of a new sample
of nearby galaxies selected to have high star-formation rates.  We
select our sample primarily by \Ha\ luminosity in order to
maximise the physical overlap with high-redshift
galaxies. The spatial resolution of our data (1--$3\kpc$) is
comparable to high-redshift samples observed with adaptive
optics. This paper provides the core description of the survey and
some initial scientific results. We also expand on the early results
on the relation of star-formation and velocity dispersion presented
in \cite{2010Natur.467..684G} and make a case that a number of these
galaxies are rotationally supported turbulent discs.  Later papers in
the series will delve more deeply into other physical comparisons.

The plan of this paper is as follows. We begin in
\S~\ref{sec:dynamo-sample} by describing the sample in detail,
including the selection criteria,
properties of the sample, and some of the sample selection biases. Section
\ref{sec:observations} presents the optical integral-field
spectroscopy of these galaxies, and describes the removal of the instrumental
signature from the data. The data analysis methods are described in
\S~\ref{sec:methods}. Results from these data are presented in
\S~\ref{sec:results}, including star-formation rates
(\S~\ref{sec:star-formation-rates}), estimates of gas content
(\S~\ref{sec:gas-mass}) and the Tully-Fisher relation
(\S~\ref{sec:tf-relation}). The relationship between galaxy
star-formation rate and gas turbulence which was highlighted in
\cite{2010Natur.467..684G} is expanded in \S~\ref{sec:lsig-theory}. We discuss
our results and outline the plan for this series of papers in
\S~\ref{sec:discussion}. Section~\ref{sec:conclusions} summarizes our
conclusions.

Throughout this paper, we will use the cosmology given by ($H_0 = 71
{\rm \, km\, s^{-1}\, Mpc^{-1}}$, $\Omega_M=0.27$, $\Omega_\lambda =
0.73$). We use a \cite{2003PASP..115..763C} initial-mass function (IMF). For conversions
from other IMFs, we adopt the following mass ratios: Salpeter -- 1/1.8
\citep{2006ApJ...646..107E}, Kroupa 2001 -- 1/0.88
\citep{Lemoine-Busserolle2010}, diet Salpeter -- 1/1.19
\citep{2009ApJ...697..115C}.

\section{The DYNAMO sample}
\label{sec:dynamo-sample}

\subsection{Sample selection}
\label{sec:target-selection}

We have selected a representative sample of \n{sample} galaxies
classified as star forming in the MPA-JHU Value Added
Catalog\footnote{\url{http://www.mpa-garching.mpg.de/SDSS/DR4/}} of
the Sloan Digital Sky Survey
\citep[SDSS,][]{2000AJ....120.1579Y}. This value-added catalog
provides stellar masses, metallicities, and star-formation rates,
which are based on improved fits to the original SDSS spectra
\citep{2004MNRAS.351.1151B, 2004ApJ...613..898T, Kauffmann2003a}. Our
sample has been designed to include the most highly \Ha-luminous,
star-forming galaxies, as well as lower \Ha-luminosity galaxies more
common in the local universe.

\begin{figure}
  \centering
  \includegraphics[width=0.9\linewidth]{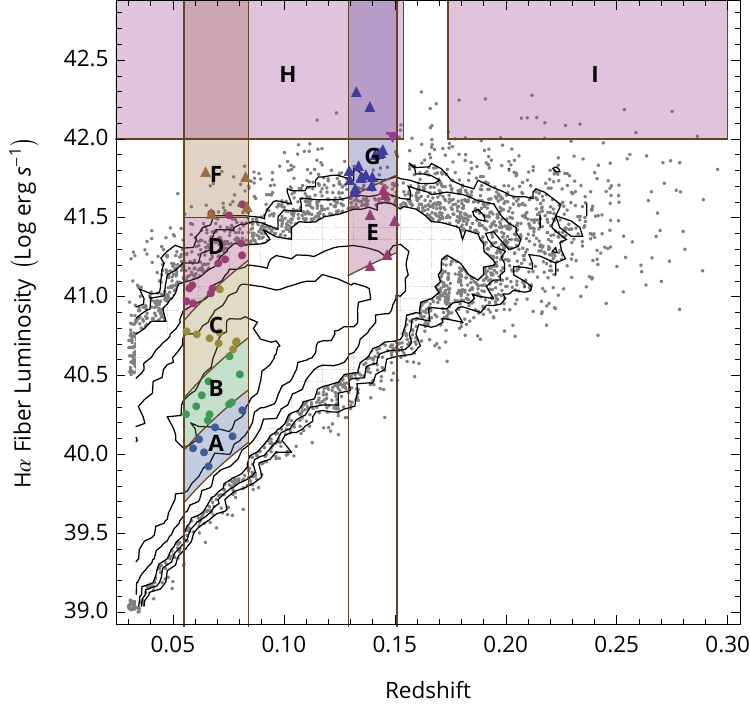}
  \caption{%
    \label{fig:selection}%
    Target selection windows in \Ha{} luminosity (measured within the
    3 arcsec SDSS fibre aperture) and redshift. The population of all
    star-forming (classified as `{\bf SF}' by
    \citealt{2004MNRAS.351.1151B}) galaxies in SDSS is shown by the
    small grey points and the logarithmic density contours. Galaxies
    making up the sample here were randomly selected from this
    population inside each shaded window, and are shown with coloured
    points and shapes corresponding to the selection window. Note that
    the selection windows for regions F and G overlap with that for
    H. Regions A, B, C, D, E, and G have curved limits because they
    are flux-limited regions instead of luminosity-limited regions. %
  }
\end{figure}

The sample galaxies were selected by redshift and \Ha{} emission flux
or luminosity {\it as measured in the SDSS 3.0-arcsec-diameter fibre
  aperture}.  Galaxies have been selected principally in two redshift
($z$) ranges that avoid placing their \Ha{} emission on significant
night-sky emission and absorption features; $0.055 < z < 0.084$ and
$0.129 < z < 0.151$. Within these redshift constraints, a series of
selection windows in flux or luminosity of the \Ha{} emission line
were defined. Five to ten galaxies were chosen within each window. To
include rarer galaxies with high H$\alpha$ luminosities, the redshift
constraint was relaxed to $z < 0.154$ or $0.174 < z < 0.3$, which
avoids only the most significant telluric absorption feature. Thus
equal numbers of both rare and more common galaxies were selected,
despite significant change in the underlying number density with
luminosity.  Each window is identified by a letter (A through I) in
Figure~\ref{fig:selection}. Although galaxies were selected in window
I, none have been observed for this paper, but will ultimately be
included in the {\small DYNAMO} sample. The criteria for the selection
windows and the number of galaxies in each window are shown in Table
\ref{tab:target-categories}. Figure~\ref{fig:selection} shows the
selection windows and locations of the selected galaxies, as well as
the distribution of SDSS star-forming galaxies from
\cite{2004MNRAS.351.1151B}.

\begin{deluxetable}{lccc}[t]
\tablecaption{Target Categories}
\tablewidth{0pt}
\tablecolumns{4}
\tablehead{
  \colhead{} &
  \multicolumn{2}{c}{Selection Criteria} &
  \colhead{} \\
  \colhead{Cat.} &
  \colhead{$z$}  &
  \colhead{\Ha{}\tablenotemark{a}} &
  \colhead{$n_{\text{selected}}$}
}

\startdata
A  &  $0.055<z<0.084$  &   $65 < f < 140$  & 7 \\
B  &  $0.055<z<0.084$  &  $140 < f < 300$  & 10 \\
C  &  $0.055<z<0.084$  &  $300 < f < 930$  & 9 \\
D  &  $0.055<z<0.084$  &  $930 < f$        & 13 \\
E  &  $0.129<z<0.151$  &  $300 < f < 930$  & 6  \\
F  &  $0.055<z<0.084$  &  $ 41.5 < L$      & 7  \\
G  &  $0.129<z<0.151$  &  $930 < f $       & 17 \\
H  &        $z<0.154$  &  $ 42.0 < L$      & 1 \\
I  &  $0.174<z$        &  $ 42.0 < L$      & 0\tablenotemark{b} \\[6pt]
{\bf Total} &  &   & 67
\enddata

\label{tab:target-categories}%
\tablenotetext{a}{For a galaxy to be selected, its \Ha\ emission was
  required to meet either a flux range or a luminosity range in the
  MPA-JHU Value Added Catalogue measurements of SDSS spectra. Flux
  requirements are denoted by $f$ and have units of \fluxys, while
  luminosity requirements are denoted by $L$ and have units of $\log
  \ergsec$}%
\tablenotetext{b}{Although galaxies were selected in window I, none of
  them has been observed at the time of writing. Window I is included for
  completeness in subsequent papers of this series.}

\end{deluxetable}

We target the \Ha{} emission line specifically because it is both
easily detected, and makes a good probe of star formation and gas
kinematics. One of our primary goals is to compare our sample with
galaxy samples at high redshift.  Where possible, most high-redshift
integral-field spectroscopy uses the same \Ha{} emission line for the
same reasons. For $z \simeq 2$ galaxies, the typical detection limit
in one hour using integral-field spectroscopy corresponds to a
star-formation rate of $1\Msunyr\sqkpc$, even on 8--$10\, {\rm m}$
telescopes \citep{Law2007}. Consequently, high-redshift samples are
often effectively limited to H$\alpha$ luminosities of at least
$10^{42}\ergsec$ by the observed flux of the emission line,
independent of other criteria\footnote{For example, see figure~1 of
  \cite{2011MNRAS.417.2601W} for a graphical illustration of this.}. At lower
redshift, this limitation is alleviated, so lower luminosity galaxies
can be detected with integral-field spectroscopy. The SDSS fibre typically does
not sample all of the \Ha{} emission from a galaxy. We will
show in \S~\ref{sec:aperture-effects} that galaxies with \Ha{}
luminosities within the SDSS fibre of $10^{41.27}\ergsec$ have total
\Ha{} luminosities representing the typical high-redshift limit, so
our sample overlaps samples at high redshift.

The galaxies of our sample are listed in Table~\ref{tab:observations},
including the window from which they were selected and their SDSS
designation. Table~\ref{tab:basic-sfr} lists the redshift of the
galaxy from the SDSS DR4 database \citep{2006ApJS..162...38A}. The
stellar masses, $\mathcal{M}_*$, of our galaxies range from
\n{minstellarmass} to $\n{maxstellarmass}\times 10^{9} \Msun$ after
IMF correction \citep{Kauffmann2003a}. The stellar masses, as well as
\Ha\ luminosities measured within the fibre ($L_{\rm fiber}(\Ha)$,
from \citealt{2004ApJ...613..898T}), and aperture-corrected
star-formation rates (${\rm SFR}_{\rm B04}$, from
\citealt{2004MNRAS.351.1151B}) are also listed in
Table~\ref{tab:basic-sfr}.

\begin{figure*}
  \centering
  \includegraphics[width=0.8\linewidth]{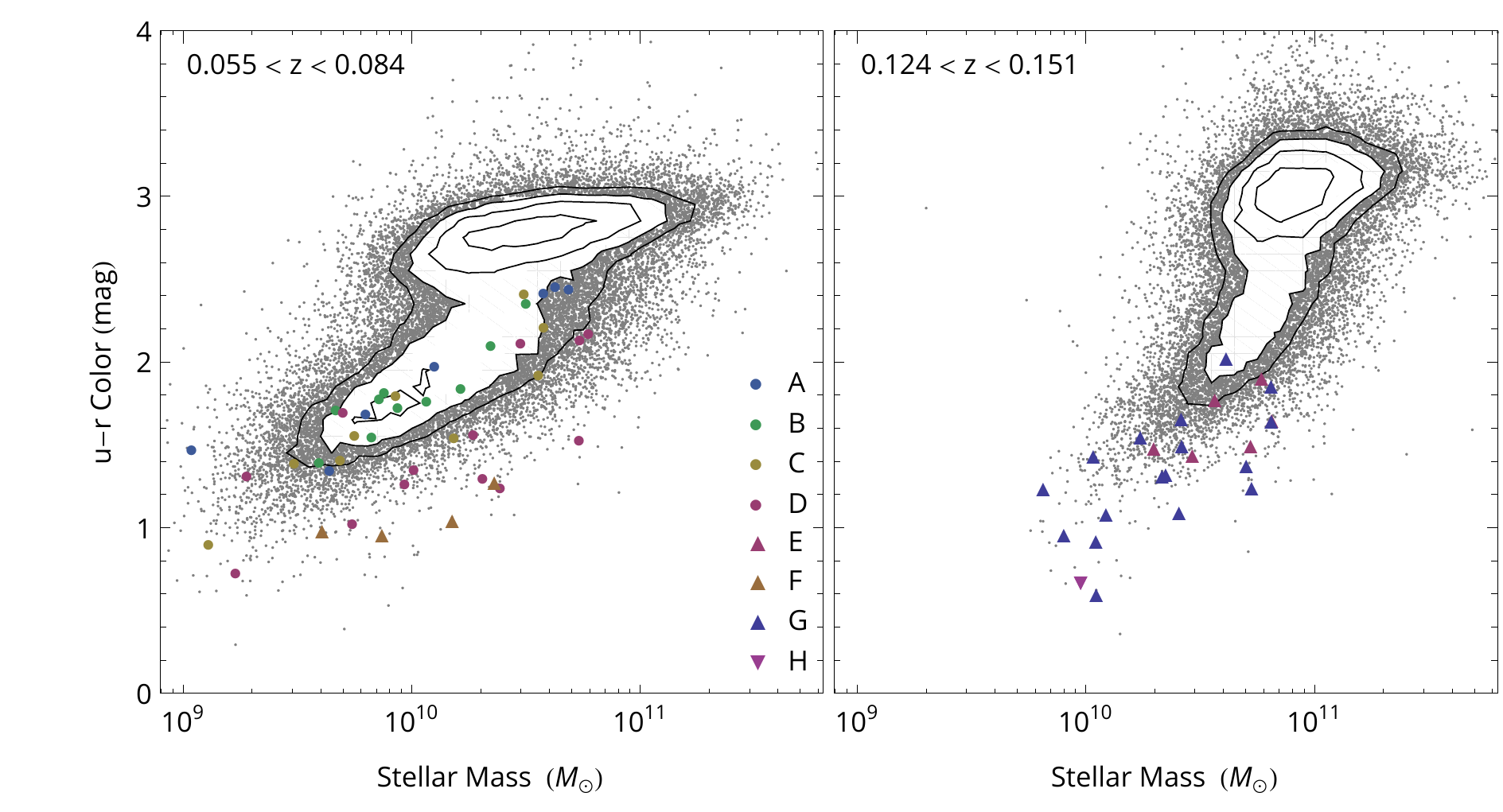}

  \caption{%
    \label{fig:cmd}%
    The $u-r$ colours and masses for the SDSS spectroscopic sample are
    shown by the grey points and contours for the two redshift ranges
    of our selection criteria. Galaxies making up the {\small DYNAMO} selection
    within each redshift range are highlighted with coloured
    symbols. The symbol shape and colour correspond to the selection
    window as shown in the key, and match the symbol coding of
    Figure~\ref{fig:selection}. The impact of the limit of SDSS
    spectroscopy to apparent $r$-band magnitude less than 17.7 is
    reflected in the range of stellar masses included in each redshift
    range: $\gtrsim 10^9$ for the lower-redshift range, and $\gtrsim
    10^{10}$ for the higher-redshift range.%
  }

\end{figure*}

Figure~\ref{fig:cmd} shows the colour-mass diagram for the two main
redshift bands of the sample. SDSS $u-r$ colour separates the red
sequence of galaxies from the blue cloud
\citep{2004ApJ...600..681B,2006MNRAS.373..469B}. Each panel shows the $u-r$ colour and
stellar masses for all SDSS galaxies (not just those that are star
forming) that meet the corresponding redshift requirement. The
galaxies of our sample have been highlighted within each redshift
range. For the lower-redshift range, the sample is representative of
the blue cloud, but in the higher-redshift range only fairly extreme
blue galaxies are included. The effect of the SDSS apparent magnitude
limit for spectroscopy is also noticeable, as only more massive
galaxies are included in the higher-redshift range. Overall, the
sample is fairly representative of the blue cloud at $z \sim 0.1$.

Our sample can be compared with samples of galaxies observed with
integral-field spectroscopy at high redshift in three different
ways. First, galaxies in our sample with \Ha{} luminosities above
$10^{42}\ergsec$ can be directly compared with similarly luminous
galaxies at high redshift. In this way, galaxies with similar
star-formation rates in different eras can be compared as a probe of
the physical processes regulating star formation. Second, a comparison
of the {\small DYNAMO} sample as a whole with representative samples at high
redshift could probe redshift evolution of galaxy properties. Even
though many high-redshift samples only probe galaxies with \Ha{}
luminosities of more than $10^{42}\ergsec$, this does not mean that
the $z \simeq 2$ samples are necessarily `tip of the iceberg'
samples covering only a highly biased fraction of the
population. Instead, the median star-formation rate for a fixed
stellar mass evolves upward \citep[by a factor of three from $z=0.36$
to $z=0.98$,][]{2007ApJ...660L..43N} and continues to increase to
$z\sim 2$ \citep{2007ApJ...670..156D}. Both our sample and samples at
high redshift can probe properties of the star-formation `main
sequence' -- the relationship between star-formation rate in galaxies
and their stellar mass \citep{2007ApJ...660L..43N}. Finally, we could
compare galaxies between eras based on their distance above the
star-formation main sequence, which could probe the origins of such
`star bursts.'  Each of these comparisons potentially test very
different kinds of `evolution' between $z \simeq 2$ and today. We
will explore these different comparisons below and in future papers of
this series.

\subsection{Sample selection biases}
\label{sec:sample-bias}

Our sample selection criteria impose a lower limit on the galaxy
continuum luminosity and stellar mass for included galaxies.  To be
included in the SDSS spectroscopic sample from which we have selected our sample,
the galaxy must have $r \lesssim 17.77$ mag
\citep{2002AJ....124.1810S}. This corresponds to an absolute magnitude of
$M_r < -19.4$ and $-21.5\, {\rm mag}$ in the two redshift windows
around 0.07 and 0.13, respectively. These limits on continuum
luminosity translate to implicit mass limits of roughly $10^{9}\Msun$
for the $0.055 < z < 0.084$ redshift range, and roughly $10^{10}\Msun$
for the $0.124 < z < 0.151$ redshift range (see
Figure~\ref{fig:cmd}). Particularly for the low redshift range, the
sample probes a broad range in stellar mass of $\sim 10^9$ to
$10^{11}\Msun$.

Galaxies hosting active nuclei have been excluded because such active
nuclei can significantly bias estimates of star-formation rate based
on the \Ha\ emission line. The star-forming sample of
\cite{2004MNRAS.351.1151B} already has active galactic nuclei
removed. To define galaxies as star forming,
\citeauthor{2004MNRAS.351.1151B} use the criteria of
\citet{Kauffmann2003c}, which are based on the
\citet*[BPT]{1981PASP...93....5B} diagram. To be included in the
sample, galaxies must have been detected at a level of $3\sigma$ in
each of four diagnostic lines (H$\alpha$ $\lambda$6563, [\ion{N}{2}]
$\lambda$6584, [\ion{O}{3}] $\lambda$5007, and H$\beta$
$\lambda$4861). The detection requirement in these four lines may
introduce some selection bias, as described in detail in
\cite{2004MNRAS.351.1151B}. The strongest bias is from the strength of
the [\ion{O}{3}] $\lambda$5007 line, which varies inversely with
galaxy stellar mass. Our star-forming requirement eliminates
approximately half of all galaxies with detectable \Ha{} emission from
star formation in SDSS.

No correction for dust extinction has been included in the
selection. This may bias the sample away from extremely dusty
galaxies, even if they have high rates of star formation. Also, our selection on \Ha\ luminosity is effectively a selection on \Ha\ central surface brightness because of the small aperture for the SDSS fibre spectroscopy. If they are sufficiently common, galaxies with significant star formation only in regions beyond the fiber coverage could still be included in our sample through our inclusion of galaxies with lower (central) luminosities. However, if such galaxies are rare, then they could be missed by our selection strategy.

\section{Observations}
\label{sec:observations}

Integral-field spectroscopic data were obtained using two different
telescopes; the 3.9~m Anglo-Australian Telescope and the ANU 2.3~m
Telescope, both situated at Siding Spring Observatory,
Australia. Table \ref{tab:obs-run} lists our observing runs and
instrument configurations. Table \ref{tab:observations} lists the
telescope used, observation date, and exposure time for each galaxy in
our sample. The observing procedures and data-reduction techniques are
described separately for each telescope.

\subsection{SPIRAL}

The SPIRAL Integral-Field Unit was used with the AAOmega spectrograph
\citep{2006SPIE.6269E..14S} on the Anglo-Australian Telescope. SPIRAL
is an array of $32\times16$ square, $0\farcs7$ lenslets situated at
the Cassegrain focus. This provides a contiguous integral field of
$22\farcs4 \times 11\farcs2$ on the sky. The long axis of the field
was set to be east--west for all observations. Light from the lenslets
is fed by fibre-optic cable into AAOmega, a two-arm spectrograph. The
incoming light is split by a dichroic mirror into the two spectrograph
arms. Each arm is fitted with interchangeable, volume-phase
holographic gratings. Both the grating angle and the camera angle are
set to achieve optimum throughput for the desired wavelength range.

We used the 570 nm dichroic and the 1700I grating in the red
spectrograph. These provided a nominal spectral resolving power of
$R\simeq 12,000$, and a wavelength coverage of 500\ang. The blaze
angle of the volume-phase holographic grating was set so that the
wavelength coverage included the \Ha{} line for the redshift range of
galaxies observed during a particular night. Data from the blue
spectrograph were not used.

Data were obtained over three observing runs on 13--16 July 2008, 5
June 2009, and 16--19 January 2010. Exposures of a quartz-halogen lamp
were obtained during each afternoon for spectral and spatial
flatfielding. Exposures of copper-argon and iron-argon arc lamps were
obtained for wavelength calibration. The typical instrumental
full-width at half maximum was 2.5 pixels (measured from arc lines) or
about $30\kms$. The wavelength calibration was confirmed to be stable
through the night using repeated arc-lamp observations and
measurements of sky lines in each exposure. Since the bias level on
the red AAOmega camera is stable and we observed no structure in the
bias frames, we chose not to record separate bias frames, but
subtracted an overscan bias in our data reduction.

We took three 1200~s exposures for targets with SDSS fibre \Ha\ fluxes
of greater than 
$3.0\times 10^{-15}\flux$
(selection windows C--I, Figure
\ref{fig:selection}). The telescope was offset by a single spatial
pixel between each dithered exposure to mitigate the effects of a few
dead fibres and detector defects. For fainter targets (selection
windows A and B, Figure~\ref{fig:selection}), we repeated this
procedure twice for a total on-source integration time of 7200~s.  In
two cases (selection IDs G~21-2 and E~0-2), the galaxy was much larger
than the SPIRAL field of view, and spatial mosaicing was
necessary. Mean seeing was 1.4 arcsec full width at half maximum,
which corresponds to a physical resolution of 1.5--$3.6\kpc$ depending
on redshift.

Wavelength- and spatially-calibrated data cubes were first extracted
from the raw data using the standard \textsc{2dfdr} data reduction
facility for this instrument \citep{2006SPIE.6269E..14S}. Briefly,
this uses an optimal-extraction routine to remove crosstalk between
fibres in the raw CCD image, and then rectifies the data using the
arc-lamp images. The basic reduction was then completed using
custom-written \textsc{idl}\footnote{\textsc{idl}: Interactive Data Language} scripts.

A sigma-clipped mean was used to simultaneously combine frames and
remove cosmic rays and other glitches. First, the frames were median
combined. This median provided the information to construct a noise
estimate for each pixel in the image (including read noise).  For each
frame, outliers greater than $8\sigma$ were masked.  The frames were
then combined by averaging the unmasked pixels across individual
frames.

A two-step, iterative process was used to subtract sky emission. First,
an approximate sky spectrum was generated by median combining spectra
from all spatial locations in the image. This approximate sky spectrum
was then subtracted from the whole cube, and the residual flux in each
spaxel summed. This residual flux image represents object photons. We
identified a residual surface-brightness limit in this image by eye, and deemed
spaxels with residual surface brightnesses below this limit to contain
only sky emission. The spectra for these spaxels (typically 20--40 per~cent
of the field of view) were then median combined to produce a final sky
spectrum. This final sky spectrum was then subtracted from the whole
cube.

The flux calibration for SPIRAL was achieved in two stages. First, a
spectro-photometric standard star was observed each night. With this
observation, we calibrated the relative throughput of the
telescope-instrument system as a function of wavelength. Second, the
flux in the central 3 arcsec of the galaxy was scaled to match the
flux measured by the SDSS pipeline. This scaling corrects for the
variation in fibre throughput due to flexing of the fibre bundle, and
corrects for varying transparency on many of the nights SPIRAL was
used.

\begin{deluxetable}{lrcl}

\tablecaption{\label{tab:obs-run}Observing Runs}
\tablewidth{0pt}
\tablecolumns{4}
\tabletypesize{\footnotesize}
\tablehead{
  \colhead{Program} &
  \colhead{Date} &
  \colhead{Moon\tablenotemark{a}} &
  \colhead{Weather}
}
\startdata
SPIRAL-08A & 13 Jul 2008 & 0.799 & Not photometric    \\
SPIRAL-08A & 14 Jul 2008 & 0.870 & Clear              \\
SPIRAL-08A & 15 Jul 2008 & 0.927 & Occasional Cloud   \\
SPIRAL-08A & 16 Jul 2008 & 0.970 & Clear              \\
SPIRAL-09A & 5 Jun 2009  & 0.952 & Occasional Cloud   \\
SPIRAL-DD\tablenotemark{b}& 16 Jan 2010  & 0.014 & Not photometric    \\
SPIRAL-DD         & 17 Jan 2010 & 0.044 & Windy, clear       \\
SPIRAL-DD         & 18 Jan 2010 & 0.090 & Clear, photometric \\
SPIRAL-DD         & 19 Jan 2010 & 0.150 & Not photometric    \\
WiFeS-09D  & 19 Jan 2010 & 0.150 & Photometric        \\
WiFeS-09D  & 20 Jan 2010 & 0.224 & Photometric        \\
WiFeS-09D  & 21 Jan 2010 & 0.309 & Photometric        \\
WiFeS-09D  & 22 Jan 2010 & 0.403 & Photometric        


\enddata

\tablenotetext{a}{Fraction of moon illuminated at midnight.} %
\tablenotetext{b}{DD -- Director's Discretionary time.} %

\end{deluxetable}

\subsection{WiFeS}

We used the Wide-Field Spectrograph
\citep[WiFeS;][]{2007Ap&SS.310..255D} on the ANU $2.3\,{\rm m}$
Telescope. WiFeS is an image-slicing integral-field
spectrograph. WiFeS provides a 25 by 38 arcsec field of view sampled
with either $1.0\times0.5$ or $1.0\times1.0$ arcsec spatial pixels
(the latter being achieved with 1$\times$2 CCD binning). The
spectrograph has two arms with interchangeable dichroics and fixed
gratings. We chose the 615 nm dichroic and the I7000 grating for the
red side, which provided a spectral resolving power of $R\simeq 7,000$
and 6832--9120\ang wavelength coverage. The data were taken on 16--24
January 2010. The mean seeing was 1.2 arcsec, corresponding to a
physical resolution of 1.3--$3.2\kpc$ for the redshift range of our
galaxies. Data from the blue spectrograph are not considered here.

Calibration frames were taken each afternoon, including bias frames,
quartz-iodide flat-field frames, and arc-lamp frames. The wavelength solutions
were checked against identical arc-lamp frames taken at the end of the
night, and were stable through the night. Except for bias subtraction, the
reductions took advantage of the WiFeS data-reduction pipeline
\citep{2010AP&SS.327..245D}, which includes flatfielding and both
spatial and spectral rectification.

For targets with SDSS fibre \Ha\ fluxes of greater than $3.0\times 10^{-15}
\flux$ (selection windows C--I, Figure~\ref{fig:selection}), we obtained
two 1800~s exposures with no on-detector binning. For fainter targets,
we recorded $4\times 1800\,{\rm s}$  exposures, with $1\times 2$ on-detector
pixel binning along the spatial axis. We dithered the telescope 2 arcsec in
different directions between exposures to ensure detector artefacts
could be averaged out. As the WiFeS field of view is larger than all
of our targets, we did not employ any mosaicing.

Flat fielding, sky subtraction, and rectification were all
accomplished using the standard WiFeS reduction pipeline
\citep{2010AP&SS.327..245D}, written in \textsc{iraf}\footnote{\textsc{iraf} is
  distributed by the National Optical Astronomy Observatories, which
  are operated by the Association of Universities for Research in
  Astronomy, Inc., under cooperative agreement with the National
  Science Foundation.} \citep{1993ASPC...52..173T}. For the sky
subtraction, a sky region was identified in each observation to
estimate the sky spectrum. Spectra in this region were combined to
create a master sky spectrum that was then subtracted from each
spectrum in the rectified data cube. 

The individual frames were median combined using {\em imcombine} in
\textsc{iraf} to form final data cubes, and to remove cosmic rays.

The data were flux calibrated using observations of a
spectro-photometric standard star taken on each night. The WiFeS
observations were obtained on photometric nights and WiFeS does not
suffer from the fibre-related throughput variations of SPIRAL so
accurate absolute flux calibrations were obtained. Repeat observations
of standards at the beginning and end of the night were used to
confirm the stability of the flux calibration. Fluxes measured from
the central 3 arcsec of each observed galaxy agree within a few
percent with those measured by SDSS.

\section{Data analysis}
\label{sec:methods}

\subsection{Emission-line fitting}\label{sec:emiss-line-fitt}

The wavelength range of our data generally includes five bright
emission lines that are all well-resolved spectrally; [\ion{N}{2}]
$\lambda6548$, \Ha\ $\lambda6563$, [\ion{N}{2}] $\lambda6584$,
[\ion{S}{2}] $\lambda6717$, and [\ion{S}{2}] $\lambda6731$. Continuum
emission is detected from a few objects. The continuum is estimated
using a 300-pixel moving median filter, and subtracted. This is the
same technique used by the SDSS pipeline
\citep{2006ApJS..162...38A}. The resulting spectrum contains only
emission lines.

Our custom \textsc{idl} code fits a synthetic spectrum consisting of five
Gaussian spectra (one for each line) to the observed spectrum. Line
widths in the synthetic spectrum are initially set to the instrumental
resolution, as determined by arc lines, and then the best fit
broadening is determined during the fitting process. Thus the
dispersion values we quote are intrinsically corrected for instrument
broadening. This approach is superior to the alternative of
subtracting the instrumental line width in quadrature, which is
problematic at low signal-to-noise ratios
\citep{2009ApJ...706.1364F}. The redshifts of the five lines are
assumed to be the same, but the flux and width of the lines are
allowed to vary independently. A Levenberg-Markwardt minimization
algorithm is used to fit the data \citep{2009ASPC..411..251M}.

Once the fitting is completed, a mask is created to exclude regions
having low signal-to-noise ratio from further analysis. First, we
compute the median absolute deviation of each reduced spectrum to give
an estimate of the typical noise at each spatial position in the data
cube. The median absolute deviation estimates the width of the
distribution of intensities in the continuum-subtracted spectrum. It
is robust against outliers, including emission lines in the spectrum,
bad pixels, and edge effects.  We then compute the significance of a
\Ha{} detection as the ratio of its total fit flux (integrated across
the line) to this width of the intensity distribution. Spectra with
signal-to-noise ratios less than three are masked automatically. This
mask was reviewed by eye interactively, and invalid fits were removed
(typically a few per galaxy). All measurements reported below were
performed within the unmasked region.

The results of our emission-line fitting are shown in the online data
for this paper as maps of continuum-subtracted \Ha{} emission,
velocity and velocity dispersion for each galaxy in our sample.

\begin{figure*}
  \centering
  \includegraphics[width=0.9\textwidth]{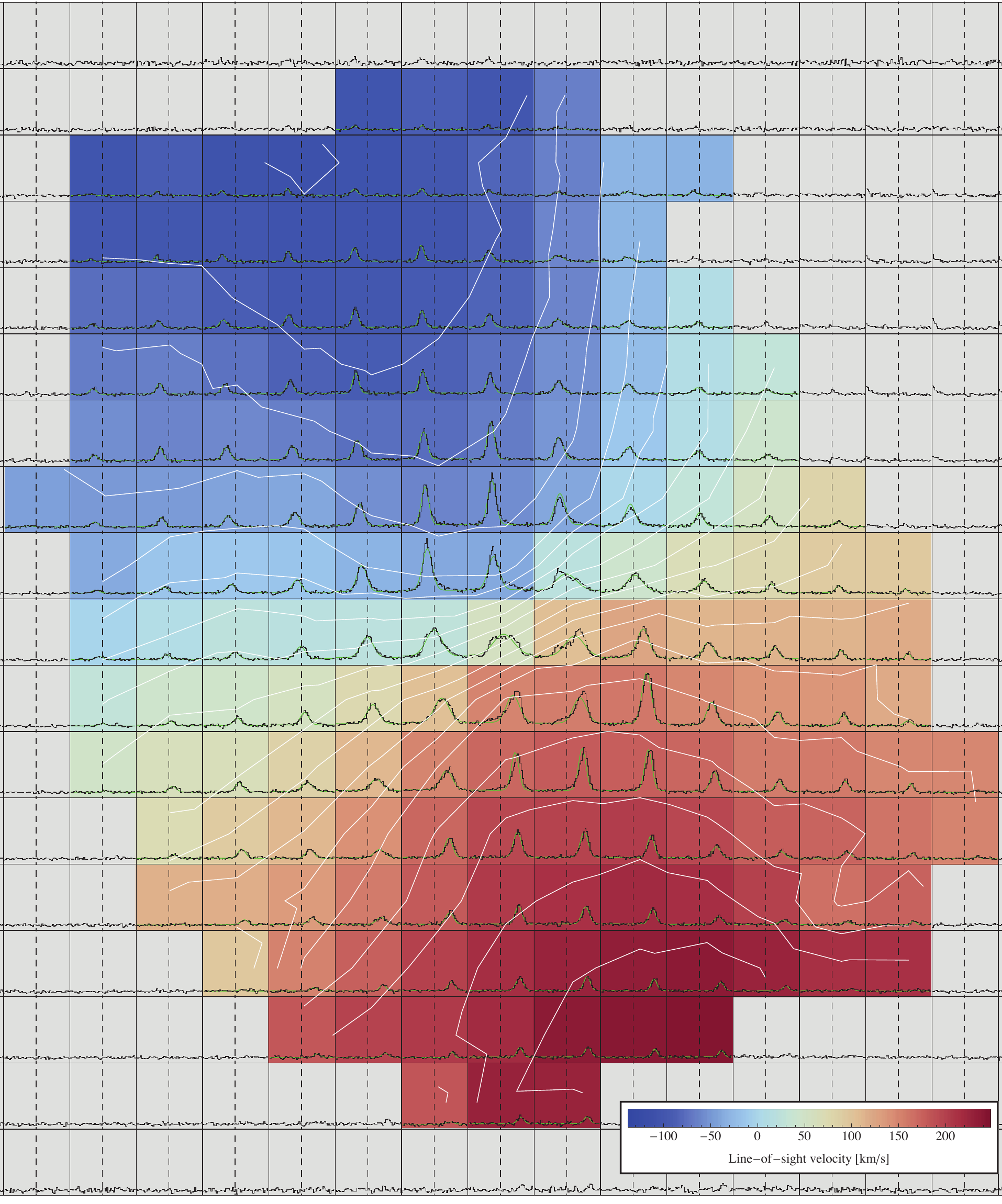}
  \caption{ %
    \label{fig:spectra-map} %
    The spatial map of spectra in the region around the \Ha{} emission
    line for D~15-3. This galaxy has an integrated star-formation rate
    of $13.7\Msunyr$, a stellar mass of $5.4\times10^{10}\Msun$, and a
    velocity dispersion of $\sigmean = 30.6\kms$. Each grid square
    shows the object spectrum (black) in the 14~\ang region around the
    \Ha{} emission line for the corresponding spatial location from
    our integral-field spectroscopic data. North is up, east is to the
    left. The flux scale (in arbitrary units) is the same in all
    squares. Under-plotted in green is the Gaussian fit
    (\S~\ref{sec:emiss-line-fitt}). The dashed vertical line in each
    square shows the systemic redshift of \Ha{} for this galaxy. The
    background colour of each square corresponds to the line-of-sight
    velocity of the \Ha\ emission in that spatial pixel relative to
    the systemic redshift as shown by the key in the lower left
    (i.e. the velocity map). Masked pixels where the fit to the \Ha\
    emission failed have a grey background. Overlaid in white are
    lines of constant velocity, separated by $20 \kms$. These reveal
    the `spider-diagram' shape to the velocity field, which helps
    identify this galaxy as a rotating disc. This galaxy shows the
    skewing and splitting of the emission line profile due to beam
    smearing near the centre.%
  }
\end{figure*}

Figure~\ref{fig:spectra-map} illustrates the quality of the data for
one of our galaxies, D~15-3. The spectrum in the region of the
\Ha{} emission line at each spatial location in the data is shown in
the corresponding box of the figure. Under-plotted in green is the best
fit to the \Ha{} emission line. Most of the spectra are well-fit by a
single Gaussian profile. Spectra in the top-left quadrant of
Figure~\ref{fig:spectra-map} are blue-shifted relative to the systemic
velocity (shown by the vertical dashed line in each box), and those in
the lower-right quadrant are predominantly redshifted. Some spectra
show more complex emission-line structure than a simple Gaussian. Many
of these are located near the rotation axis, and consist of blended
blue-shifted and redshifted emission. These spectra are fit by a single
Gaussian of larger width as a result of this beam smearing. Other subtle line
asymmetries are apparent in Figure~\ref{fig:spectra-map}, which we have ignored in this analysis.

\subsection{Aperture effects}
\label{sec:aperture-effects}

\begin{figure}
  \includegraphics[width=0.9\linewidth]{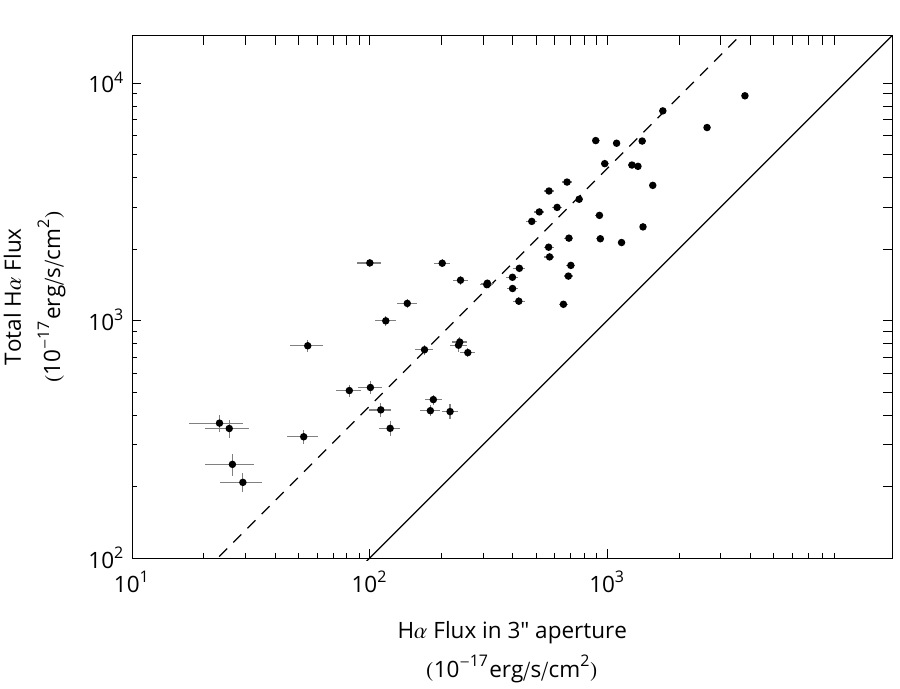}
  \caption{%
    \label{fig:ap-effects}%
    Ratio of fibre aperture flux to total flux (excluding masked
    regions) for the \Ha\ emission line. The solid line shows the 1--1
    relation, while the dashed line is offset by the median 0.63~dex
    observed across our sample. %
  }
\end{figure}

The effect on our sample selection of the 3-arcsec-diameter SDSS
optical fibres has been noted in \S~\ref{sec:target-selection}.  We
now quantify the amount by which the actual H$\alpha$ luminosities of
our galaxies exceed those predicted from SDSS spectra.  We use our
integral-field data cubes to compare the \Ha\ flux within a
3-arcsec-diameter aperture to the total emission-line flux (excluding masked regions; Section \ref{sec:emiss-line-fitt}). These fluxes are
shown in Figure~\ref{fig:ap-effects}. On average, the H$\alpha$ flux
of the whole galaxy is 0.63~dex larger than the region covered by the
SDSS fibre, with a root-mean-square scatter of 0.26~dex. This offset is
indicated by the dashed line in Figure~\ref{fig:ap-effects}. We use
the total \Ha\ flux in our subsequent analysis.

\subsection{Classification of galaxies by kinematic morphology}
\label{sec:nebular-morphology}

Integral-field spectroscopy of galaxies at high redshift reveals their
kinematic morphologies to be markedly different from those of galaxies
at low redshift.  Rotating galaxies at high redshift often have
turbulent, geometrically thick discs \citep{2006Natur.442..786G,NFS2006,
  2009ApJ...706.1364F, 2006ApJ...650..644E}. Additionally, there are
both many merging galaxies and many `dispersion-dominated' galaxies
(having velocity shears smaller than their velocity
dispersions). Conversely, most of the galaxies sampled by the GHASP
survey look more like quiescent, thin discs
\citep{2010MNRAS.401.2113E}. Results from the IMAGES survey suggest
that the fraction of rotating, disc-like galaxies changes with time
\citep{2008A&A...477..789Y}.  We define a classification scheme and
classify our galaxies for comparison with other galaxy samples.

Our sample of galaxies is classified into three different general
kinematic morphologies using a variation of the system described by
\cite{Flores2006} for the IMAGES survey. Although other approaches
have been suggested for classifying the kinematic morphologies of
high-redshift galaxies \citep[e.g.,][]{2008ApJ...682..231S, 2009ApJ...706.1364F},
the IMAGES system is based on simple visual criteria that can be applied to data with high noise or low resolution.  The IMAGES
classification defines three distinct categories: rotating discs (RD),
perturbed rotators (PR) and complex kinematics (CK). We
supplement the classification with a sub-categories for
compact galaxies that otherwise fit into the original three
categories. Our classification system is defined as follows:
\begin{description}
\item[Rotating Discs (RD, {\color{blue} \large $\bullet$})] %
  Rotating disc galaxies show typical kinematic features of a disc;
  the velocity field is qualitatively consistent with rotation, the
  velocity dispersion peak corresponds to the centre of
  rotation\footnote{This peak in velocity dispersion corresponds to
    where the local velocity gradient is steepest, and is caused by
    beam smearing as described in \S~\ref{sec:sigm-bs}}, and the axis
  of rotation aligns with the minor-axis seen in SDSS $gri$
  imaging. These are the same features displayed by the best-case
  rotation-dominated galaxies of
  \cite{2009ApJ...706.1364F}. 
\item[Perturbed Rotators (PR, {\color{green} $\blacksquare$})] %
  Galaxies with perturbed rotation meet all the criteria for the RD
  classification except those for the velocity-dispersion map. For PR
  classification, either the peak of the velocity dispersion is offset
  from the centre of rotation by more than $3\kpc$, or there is no
  distinct peak.
\item[Complex Kinematics (CK, {\color{red} $\blacktriangle$})] %
  Galaxies that do not meet the criteria for either the RD or PR
  categories are classified as complex kinematics. This category
  includes galaxies for which both velocity and velocity dispersion
  vary significantly from regular discs, or where the rotation axis is
  significantly misaligned with the optical axis. Objects with
  multiple approaching and receding regions (`multi-polar' velocity
  fields) are also assigned this classification.
\item[Compact (cRD, {\color{blue} \large $\circ$}; cPR, {\color{green}
    $\square$}; cCK, {\color{red} $\triangle$})] %
  A large fraction of our sample is compact (\n{compact} of
  \n{sample}), and these are given separate classifications. Compact
  galaxies have $r$-band exponential scale-lengths of less than
  $3\kpc$ as measured by SDSS\footnote{SDSS's {\tt expRad\_r} in the
    {\tt PhotoObj} view} (2\farcs3 at $z \simeq 0.07$, 1\farcs2 at $z
  \simeq 0.14$). They are otherwise classified exactly as above. The
  classification is less reliable for these compact galaxies because
  they are not as well resolved, which is why we found this a useful
  distinction.
\end{description}

The results of our classification are included in
Table~\ref{tab:kinematic}. Of the sample of 67, 25 galaxies are
classified as RD, and six are classified as PR. We are confident these
31 RD and PR galaxies are indeed rotating, disc-like galaxies, as the
classification is corroborated by results from the Tully-Fisher relation, which we
discuss below in \S~\ref{sec:tf-relation}. A further 20 galaxies are
cRD, and five are classified cPR. These galaxies, while meeting the
classification criteria for their RD and PR counterparts, have small
angular sizes, and so the classification may be affected by limited
resolution. Of the remaining galaxies, five are classified CK and six
as cCK. These galaxies with complex kinematics are most likely
mergers, although the classification only clearly defines them as not
disc-like. Of the whole sample, 46 per~cent show all the signatures of
rotating disc galaxies, while 16 per~cent show no indication of global
rotation. Most of the remaining 37 per~cent show some evidence of
rotating discs. For the 27 galaxies in our sample with star formation
rates greater than $10\Msunyr$, seven are classified RD and one is PR,
making 30 per~cent rotating. A further 9 are classified cCK or CK. The
remaining 10 galaxies show marginally resolved indications global
rotation (cRD and cPR).

\subsection{Disc fitting}
\label{sec:disc-fitting}

Model discs have been fit to the velocity maps for all of our sample
galaxies. A simple disc model has been adopted, in line with disc models
used to fit integral-field spectra of galaxies at high redshifts
(e.g., \citealt{2009ApJ...706.1364F}, see \citealt{2010MNRAS.401.2113E} for a review of
different models).

The model velocity field is created as follows. First, we create a
spatial map of intrinsic velocities projected along the line of
sight. The rotation curve is parametrized by
\begin{equation}
  {\bf V}({\bf r}) = \frac{2 V_{\rm asym}}{\pi} \arctan \left( \frac{|{\bf r}|}{r_t} \right) (\hat{r}\times \hat{z})
\end{equation}
where $V_{\rm asym}$ is the asymptotic circular velocity, and $r_t$ is
the kinematic scale radius \citep{1997AJ....114.2402C}.  The spatial
vector ${\bf r}$ points to the location within the plane of the
disc. The unit vector $\hat{r}$ is parallel to ${\bf r}$, and the unit
vector $\hat{z}$ points perpendicular to the plane of the disc. The
velocity ${\bf V}$ is projected along the line of sight to produce the
velocity map. Second, we use the velocity map to define an intensity
cube with the same spatial coordinates and a velocity coordinate. The
spatial intensities are for an exponential surface-brightness profile
(using scale lengths from SDSS $r$-band photometry).  The intensities
along the velocity dimension are a Gaussian profile that is centred
on the line-of-sight velocity from the map. The width of the profile
corresponds to a constant intrinsic velocity dispersion across the
model, $\sigma_{\rm model}$.  Also included as free parameters are the
kinematic centre of the disc, and a systemic velocity offset (which
allows for errors in redshift). Next, the intensity cube is convolved
with a three dimensional Gaussian kernel. The spatial full width at
half maximum (FWHM) of the kernel is set to the seeing FWHM measured
during observation of the corresponding galaxy, and the velocity FWHM
is set to the instrumental resolution measured from arc-lamp
lines. This convolution or `beam smearing' raises the central
velocity dispersion of the model. Finally, the intensity cube, which
is analogous to an observed data cube, is collapsed into maps of
velocity and velocity dispersion by computing moments of the intensity
distribution. The final velocity and velocity-dispersion maps can then
be compared with the observed data.


We define the quality of the fit as $\chi^2$. This standard quality is
defined as
\begin{align}
  &\begin{aligned}
    \chi^2 &= \sum_{pix}\frac{{\left( v_{\rm pix,model} - v_{\rm pix,obs} \right)}^2}{
      {E(v_{\rm pix,obs})}^2} \\
    &\qquad+ W \sum_{pix}\frac{{\left( \sigma_{\rm pix,model} - \sigma_{\rm pix,obs} \right)}^2}{
      {E(\sigma_{\rm pix,obs})}^2}
  \end{aligned}
\end{align}
Here, $v$ and $\sigma$ are the velocity and velocity dispersions,
respectively. $E(x)$ gives the error on the corresponding
observation. The weighting, $W$, sets the relative importance of the
two different physical phenomenon to quality of the fit. We weight the
velocity map higher to ensure a good fit to the observed velocity
field ($W = 0.2$). $\chi^2$ is minimized by varying the free
parameters using a Levenberg-Markwardt routine. Our model has six free
parameters ($V_{\rm asym}$, $\sigma_{\rm model}$, $r_t$, position
angle, horizontal and vertical spatial centring, and velocity
offset). Inclination and circular velocity are nearly degenerate
parameters \citep{1989A&A...223...47B, 2010MNRAS.401.2113E}, so we fix
inclination, the seventh parameter, to that measured by the SDSS
photometric pipeline for the $r$-band exponential disc
fit\footnote{This is parameter {\tt expAB\_r} in the {\tt PhotoObj}
  view.}. Because we have weighted the velocity map highly in the fit,
we find the model velocity dispersion parameter often is not
representative, and we do not present it here. The velocity map of the
best fitting model and its difference from the observed velocity map
for each galaxy are included in the online data for this paper. For galaxies
visually classified as CK or cCK (\S~\ref{sec:nebular-morphology}),
these fits are questionable, so they are not used in our analysis.

The rotation curve that fits best is used to compute a characteristic
circular velocity for Tully-Fisher analysis in
\S~\ref{sec:tf-relation}.  The $V_{\rm asym}$ of our fit is not
suitable as a characteristic velocity because it is often much higher
than the maximum velocity observed due to the shape of the
arctangent-like rotation curve of our model. The model velocity at a
radius of 2.2 $r$-band exponential disc scale lengths,\footnote{{\tt
    expRad\_r} in the {\tt PhotoObj} view of the SDSS database.}
$\Vcirc$, is more suitable \citep{2011ApJ...741..115M}. This is where
the rotation curve of an ideal, self-gravitating, exponential disc
would peak \citep{1970ApJ...160..811F}, and is more comparable to
other Tully-Fisher type analyses. Therefore we adopt this value as the
characteristic circular velocity for all galaxies in our sample except
those classified as CK and cCK. The values are listed in
Table~\ref{tab:kinematic}.

\subsection{Non-parametric velocity dispersion}
\label{sec:sigm-bs}

We characterise the velocity dispersion of the gas in our galaxies
using the $\sigmean$ measure of \cite{2009ApJ...697.2057L} on the \Ha\ emission,
but corrected for the effects of unresolved disc rotation. The
quantity $\sigmean$ is defined to be the intensity ($I$) weighted mean
of the velocity dispersion measured in individual spatial pixels
($\sigpix$):
\begin{equation}
  \sigmean = \frac{\sum \sigpix I_{\rm pix}}{\sum I_{\rm pix}}
\end{equation}
Within each pixel, The intensity, $I_{\rm pix}$, is the flux, and the
velocity dispersion, $\sigpix$, is the width, of the best-fitting
Gaussian profile of \Ha{} emission
(\S~\ref{sec:emiss-line-fitt}). Effectively, $\sigmean$ measures the
velocity dispersion after removing velocity shear larger than the
spatial resolution element. The statistical error on $\sigmean$ for
our sample is typically 1--$2\kms$, and we have verified this estimate
with the difference of $\sigmean$ computed on alternating
checker-board masks of the full data for each galaxy. Note that the
variation of $\sigpix$ across each galaxy is typically much larger
than the error, and reflects real variation in the turbulence across
the galaxy.

The flux-weighted mean velocity dispersion, $\sigmean$, has been
widely adopted for characterising the velocity dispersion of
star-forming gas in galaxies at high redshift \citep{2009ApJ...697.2057L, Law2007,
  Lemoine-Busserolle2010, 2011MNRAS.417.2601W, 2010MNRAS.404.1247J,
  Epinat2009}. Intensity weighting makes $\sigmean$ robust against
errors from lower signal-to-noise ratio regions within the
integral-field spectroscopy. The $\sigmean$ estimate of velocity
dispersion is independent of any assumptions inherent in a parametric
model of the galaxy kinematics, making it useful for galaxies that
are not rotating discs.

The quantity $\sigmean$ can be biased by velocity shear on scales
smaller than the resolution element of the observation, as discussed by
\cite{2011ApJ...741...69D}. This effect, well known in radio astronomy
as `beam smearing,' can artificially inflate $\sigmean$ over the
intrinsic velocity dispersion. In the extreme case of a galaxy with a
strong velocity shear due to rapid rotation and observed with poor
resolution, the effect renders $\sigmean$ useless as a measure of
velocity dispersion. Correcting for beam smearing requires knowledge
of the underlying velocity field at infinite resolution. For a well
understood velocity field, such as a rotating disc, the unresolved
velocity shear can be estimated using a model and removed from the
observed velocity dispersion within each pixel, $\sigpix$, before
computing $\sigmean$. For a poorly understood velocity field, such as
in a merger, it is not possible to remove the unresolved velocity
shear because we do not know the underlying velocity field and how it
affects $\sigmean$ through beam smearing. Alternately, fitting a disc
model that includes velocity dispersion, as suggested by
\citeauthor{2011ApJ...741...69D}, would also be invalid where the
velocity field is not representative of a disc or if the intrinsic velocity dispersion 
was not constant.

In \cite{2010Natur.467..684G}, we corrected for beam-smearing using
the {\it measured} velocity field. \cite{2011ApJ...741...69D} point
out this is not strictly valid as the measured velocity field is
already convolved with the point-spread function. Given the spatial resolution of
our data, we estimated this was a small effect except in the inner
regions of each galaxy. In this paper, we provide an improved
correction for beam smearing based on our best-fitting disc model. The
disc model already accounts for beam smearing, avoiding including the
effects of seeing twice. This correction is better in principle for
disc galaxies, but is not applicable to non-disc velocity fields and
so we do not apply it, as explained below. For such non-discs, we
still compute (uncorrected) $\sigmean$, as it is meaningful for those
classes of galaxies.

Our correction for beam smearing from disc rotation is applied to
galaxies classified above as rotating as follows. First a map of
unresolved velocity shear present in the best-fitting disc model is
created using the prescription described in
Appendix~\ref{sec:bs-corr}. For this map, we assume the \Ha{}
intensity distribution is that of the exponential surface-brightness
profile from SDSS. Then, this map of unresolved shear is subtracted in
quadrature from the observed map of velocity dispersion (the observed map
has already had instrumental broadening removed,
\S~\ref{sec:emiss-line-fitt}). In the rare cases where the subtraction
would produce an imaginary (non-physical) result, we have set the
velocity dispersion to zero at that point in the resultant
map. Finally, we compute $\sigmean$ from the resulting map.

The correction does not address beam smearing arising from sources
other than the disc-like rotation of a galaxy. Consequently, we do
not apply it to galaxies classified as CK or cCK. We have checked,
using model data, that where the assumptions of surface brightness and
rotation accurately represent the intrinsic velocity map of the
galaxy, the correction fully removes any signature of beam smearing
on the velocity dispersion of observed (beam smeared) data. For
galaxies where these assumptions are not correct, observations with
greater spatial resolution will be the best way to reduce the impact
of beam smearing.

\begin{figure}
  \centering
  \includegraphics[width=0.8\linewidth]%
  {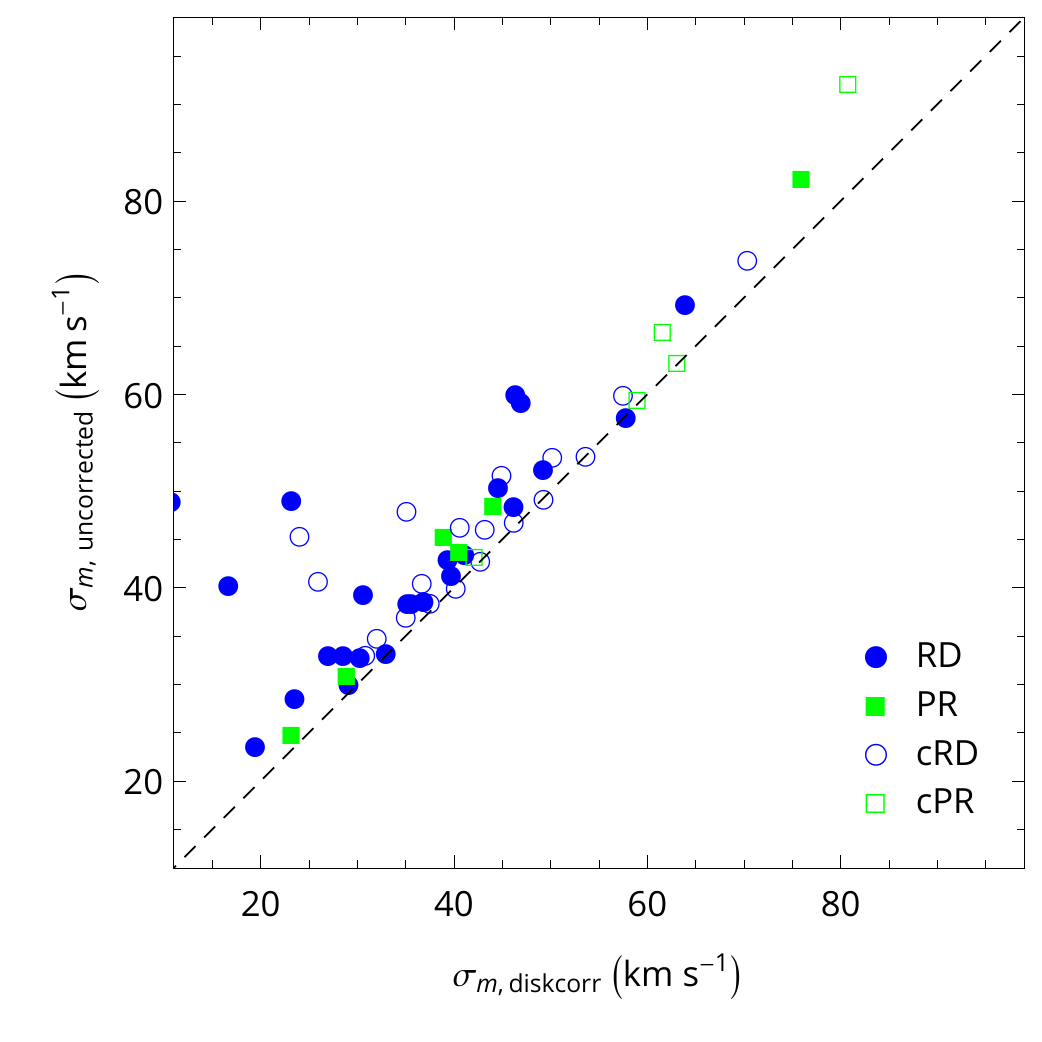}

  \caption{%
    \label{fig:sigma-corr}%
    The effect of our correction method on the value of
    $\sigmean$. The value after correcting for the velocity gradient
    based on our disc model and assuming an exponential surface
    brightness profile is shown on the horizontal axis, while the raw
    $\sigmean$ before correction is shown on the vertical axis. The
    correction can only reduce the velocity dispersion, so points are
    only scattered above the 1-1 relation shown by the dashed
    line. The CK and cCK objects are not corrected (and not shown)
    because they do not match the assumptions of the correction. %
  }
\end{figure}

For our sample, the correction for beam smearing due to disc rotation
does not qualitatively affect our results. The correction is typically
small, with a median correction of $3.6\kms$, although in some cases
(as expected for the most compact galaxies with the smallest
dispersions) it can be much larger. Figure~\ref{fig:sigma-corr} shows
the amount by which $\sigmean$ is corrected by beam smearing due to
disc rotation. The points are coded by their kinematic
classification. Galaxies showing the clearest signatures of rotation
(RD and cRD classifications, blue symbols) show the largest range in
correction. Galaxies that do not show signatures of rotation (CK and
cCK) have not been corrected and are not shown in
Figure~\ref{fig:sigma-corr}. The galaxies C~13-3, C~20-2, D~21-3,
B~14-1, A~4-3, and D~15-2 have the largest corrections, greater than
$20\kms$. These galaxies have very low intrinsic velocity
dispersions ($\sigma_{m,{\rm corr}} \lesssim 10 \kms$), unusually high
concentrations, large circular velocities along the line of sight, or more typically some combination of the three. In any of these scenarios, the central region, which is most affected by beam smearing, tends to dominate in the computation of $\sigmean$.

Independent of the correction, we have
obtained adaptive-optics-corrected integral-field spectroscopy at
$10\times$ higher spatial resolution for several of our galaxies as a
further test of the potential effects of beam smearing on
$\sigmean$. Those data show beam smearing has a negligible effect on
the value of $\sigmean$ for these galaxies\footnote{We will present
  these data in detail in a future paper.}. We conclude that while
beam smearing is present in our data, $\sigmean$, particularly with
this correction, remains a good relative indicator of velocity
dispersion in the ionized gas of galaxies in our sample, and the
sample of \cite{2010Natur.467..684G}.

\section{Results}
\label{sec:results}

\subsection{Qualitative results from integral-field spectroscopy}
\label{sec:ifu-results}

We first make a few qualitative observations about the galaxies in our
sample. The spatial maps of flux, relative velocity, and velocity
dispersion in gas emitting in the \Ha\ line created from the
integral-field spectroscopic data are included in the online
data. Also shown with these maps are relevant physical quantities
characterising each galaxy, which we will discuss in more detail
below. The velocity field of the best-fitting disc model is also
shown, and the differences between the model velocities and the
observed velocities.

A range of morphologies is seen in the maps of \Ha{} emission. Roughly
two-thirds of the galaxies in the sample show \Ha{} emission that is
centrally concentrated. This dominance of centrally concentrated star
formation is unsurprising given much of the sample is effectively
selected to have high \Ha{} surface brightness in the central 3
arcsec. Our sample is unlikely to include galaxies with strong star
formation only outside the central few arc-seconds of the galaxy, such
as in large rings. Despite this bias, our sample does include galaxies
with more complex \Ha{} morphologies. Doubles, as well as galaxies
with distributed, clumpy star formation, are present. The complexity
of the \Ha{} spatial distributions make clear that a one- or
two-component radial profile is inadequate to describe the \Ha{}
surface brightness distributions in many of these galaxies -- a common
assumption when considering the disc-like nature of galaxies at any
redshift. The distribution of \Ha{} emission, and consequently of star
formation, can be much more complex than the SDSS broad-band imaging
reveals.

About half of the galaxies in our sample have regular, rotation-like
symmetry. For comparison, in the largest sample observed with
integral-field spectroscopy at high redshift, the SINS survey,
approximately two-thirds of the galaxies are either compact,
dispersion-dominated systems or mergers
\citep{2009ApJ...706.1364F}. This high fraction of disturbed
kinematics has been seen in other high-redshift surveys as well
\citep{Flores2006,2012A&A...539A..92E}.  Interestingly, in our sample
even some obvious mergers, such as E~0--3 and G~21--2 still show
disc-like velocity fields within the individual merging
galaxies. Also, many objects with complex, clumpy distributions of
star formation do still show smooth rotation, such as D~14--1,
C~13--1, and C~0--1.  \cite{2011ApJ...733..101G} and
\cite{2012MNRAS.422.3339W} have made similar arguments for $z>1$
star-forming galaxies, where the star-forming clumps appear to be
embedded in a disc-like velocity field.

\subsection{Star-formation rates}
\label{sec:star-formation-rates}

Total star-formation rates of {\small DYNAMO} galaxies have been estimated from
their \Ha\ luminosities, after correcting for extinction due to
interstellar dust. The \Ha\ luminosities are measured by spatially
integrating the \Ha{} emission-line flux (regions masked during the
emission-line fitting are excluded, \S~\ref{sec:emiss-line-fitt}).

The extinction correction was derived from the Balmer decrement using
the method of \cite{1996ApJ...458..132C} and
\cite{1997AJ....113..162C}. The observed flux ratio of H$\alpha$ to
H$\beta$ was obtained from the MPA-JHU Value Added Catalogue fluxes
\citep{2004ApJ...613..898T}. The mean extinction at \Ha\ derived using
this method for the galaxies in our sample is 1.0 mag.

The Balmer-decrement method for estimating dust extinction used here
differs from that used in typical high-redshift analyses. For galaxies
at high redshift, stellar-population models are fit to the
spectral-energy distribution of the galaxy, as direct measurements of
both \Ha{} and H$\beta$ luminosity are difficult. Stellar population
fits provide estimates of dust attenuation, which can then be applied
to emission-line fluxes used to estimate star formation rates.
\cite{2009A&A...495..759A} explore whether these two methods provide
comparable estimates of star formation. They find the dust corrections
derived from spectral-energy-distribution (SED) fitting are not
applicable to measuring star-formation rates via emission-line fluxes.
Therefore, quantitative comparisons between star-formation rates in
galaxies presented here and those of high-redshift galaxies should be
treated cautiously.

The star-formation rate (SFR) was calculated from the
extinction-corrected \Ha\ luminosity using the conversion of
\cite{1998ARA&A..36..189K} modified for a \cite{2003PASP..115..763C}
IMF
\begin{equation}
  {\rm SFR} = 0.56
  L_{\text{\Ha, int}} \left( 7.9\times 10^{-42}\,\frac{\rm M_{\odot}/yr}{\rm
      erg/sec} \right).
\end{equation}

The star-formation rates for {\small DYNAMO} galaxies cover a broad range from
0.2 to 56.6$\,{\rm M}_{\odot}\,{\rm yr}^{-1}$, with a median of 9.1
and a mean of 12.7$\,{\rm M}_{\odot}\,{\rm yr}^{-1}$. For comparison,
SINS galaxies at $z \simeq 2$ range from a few to over $100\Msunyr$
(but with the caveat about the method of correcting for dust
extinction mentioned previously). Local galaxies in the GHASP sample
have star-formation rates ranging from less than $0.1\Msunyr$ to about
$10\Msunyr$. The star-formation rates of galaxies in our sample extend
across both nearby star-forming galaxies and the typical galaxies
studied at high redshifts (which have extreme star-formation rates compared to local galaxies).

\begin{figure}
  \centering
  \includegraphics[width=0.8\linewidth]{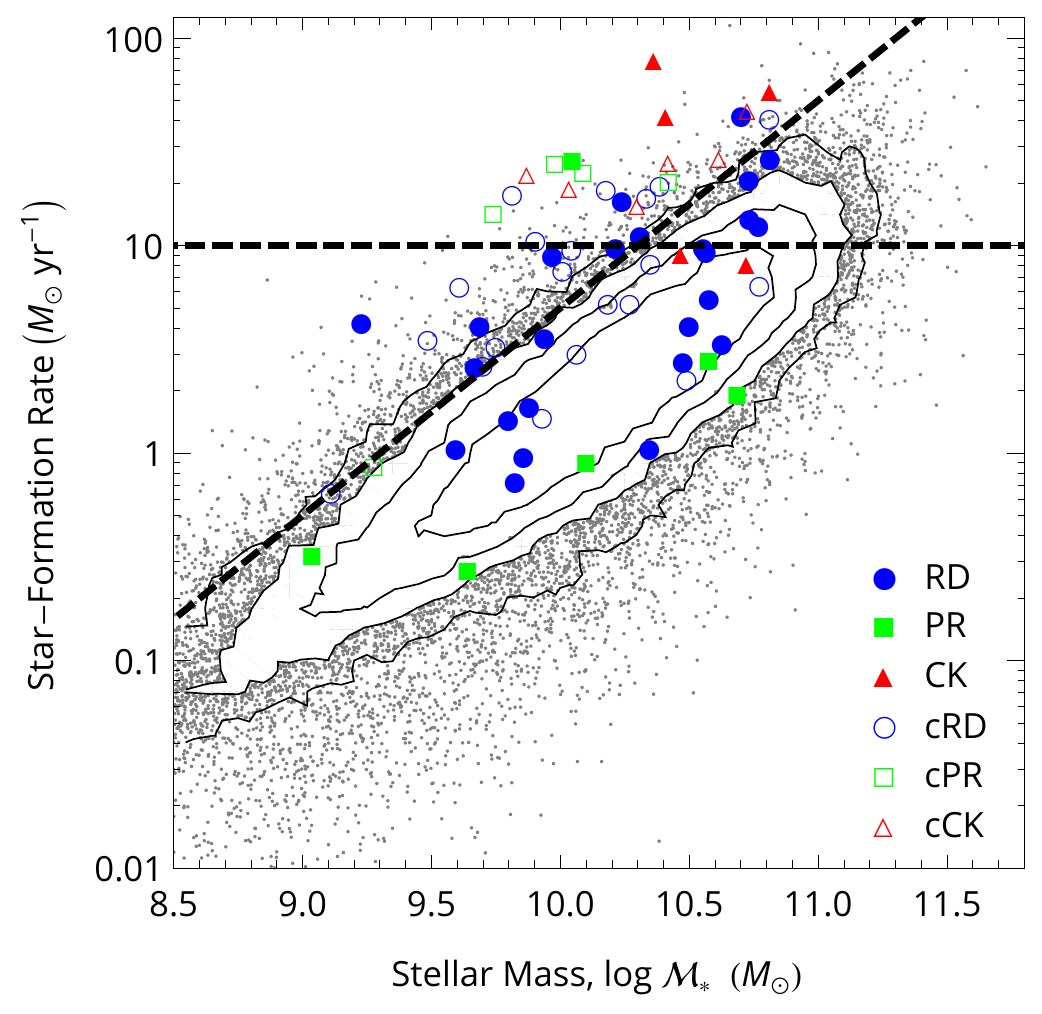}

  \caption{%
    \label{fig:sfr-mass}%
    The coloured symbols show the star-formation rates and stellar
    masses of the selected galaxies. The small grey points and
    logarithmic-density contours show the distribution of all
    star-forming galaxies from SDSS (classified as `{\bf SF}' by
    \citealt{2004MNRAS.351.1151B}). Stellar masses for both are
    derived by \cite{Kauffmann2003c}. Total (aperture-corrected) star
    formation rates for the broader SDSS sample are derived by
    \cite{2004MNRAS.351.1151B}, but for our sample the star-formation
    rates derived in \S~\ref{sec:star-formation-rates} are shown. Our
    estimates of the star-formation rate correlate well with those of
    \cite{2004MNRAS.351.1151B}. The dashed lines show specific
    star-formation rates of $5\times 10^{-10}\peryr$ and
    star-formation rates of $10\Msunyr$. These limits are used to
    divide the galaxy sample in subsequent sections.  %
  }
\end{figure}

Figure~\ref{fig:sfr-mass} shows the distribution of star-formation rates and stellar masses for our sample overlaid on the whole of the star-forming sample of SDSS. The star-formation rates shown for the SDSS galaxies are the aperture corrected values of \cite{2004MNRAS.351.1151B}. Unsurprisingly, the most disturbed galaxies, those classified as CK or cCK, tend to show the highest star formation rates because the disturbance has destabilised the gas reservoirs. Also shown in Figure~\ref{fig:sfr-mass} are limits in specific star-formation rate and star-formation rate. We will use these limits in later sections for internal comparisons of our sample.

\subsection{Total gas content}
\label{sec:gas-mass}

A large supply of gas is necessary to maintain high rates of star-formation in galaxies for any significant period. Unfortunately,
neutral and molecular gas reservoirs are difficult to detect with
current telescopes even at $z \lesssim 0.1$ because the gas is either
very diffuse or fairly cool, and consequently has low
surface-brightness. The relationship observed in local galaxies
between the surface density of the rate of star formation, $\Sigma_{\rm SFR}$,
and the total gas surface density, $\Sigma_{\rm gas}$, \citep[the
Kennicutt-Schmidt Law,][]{1998ApJ...498..541K}, has been used to infer
the quantity of gas in galaxies at high redshift. Many of those
galaxies have high gas fractions
\citep{2009ApJ...706.1364F,2006ApJ...646..107E,Lemoine-Busserolle2010}.  We use
this same empirical relationship to estimate the gas masses for the
galaxies in our sample.

The star-formation surface density is defined as the star-formation
rate per spatial pixel divided by the physical area covered by the
pixel at the galaxy redshift. We have included a correction for the
inclination of each galaxy as measured from the SDSS photometry. This
correction may not be the best approach for galaxies classified as CK
or those with a geometrically-thick disc, where a volumetric law may
be more appropriate \citep{2012ApJ...745...69K}.  Equation 4 of
\cite{1998ApJ...498..541K} is used to estimate the total gas surface
density (\ion{H}{1} and H$_2$) per pixel, which then gives the total
gas mass of the galaxy. Statistical error estimates include the
1-sigma confidence intervals of their equation 4. The conversions for
star-formation rate and the Kennicutt-Schmidt Law are both based on
the same IMF, making the gas masses computed independent of the
IMF. Note also that the Kennicutt-Schmidt Law is based on observed
rates of star formation up to $1000\Msunyr \, {\rm kpc}^{-2}$, and
therefore should remain valid even for the most extreme galaxies
observed here.  These masses are listed in Table
\ref{tab:basic-sfr}. They range from 0.82 to $50\times 10^{9}\Msun$,
which is similar to the SINS survey mass range of 1.4 to $40\times
10^{9}\Msun$ \citep{2009ApJ...706.1364F}.

\begin{figure*}
  \centering
  \includegraphics[width=0.95\linewidth]%
  {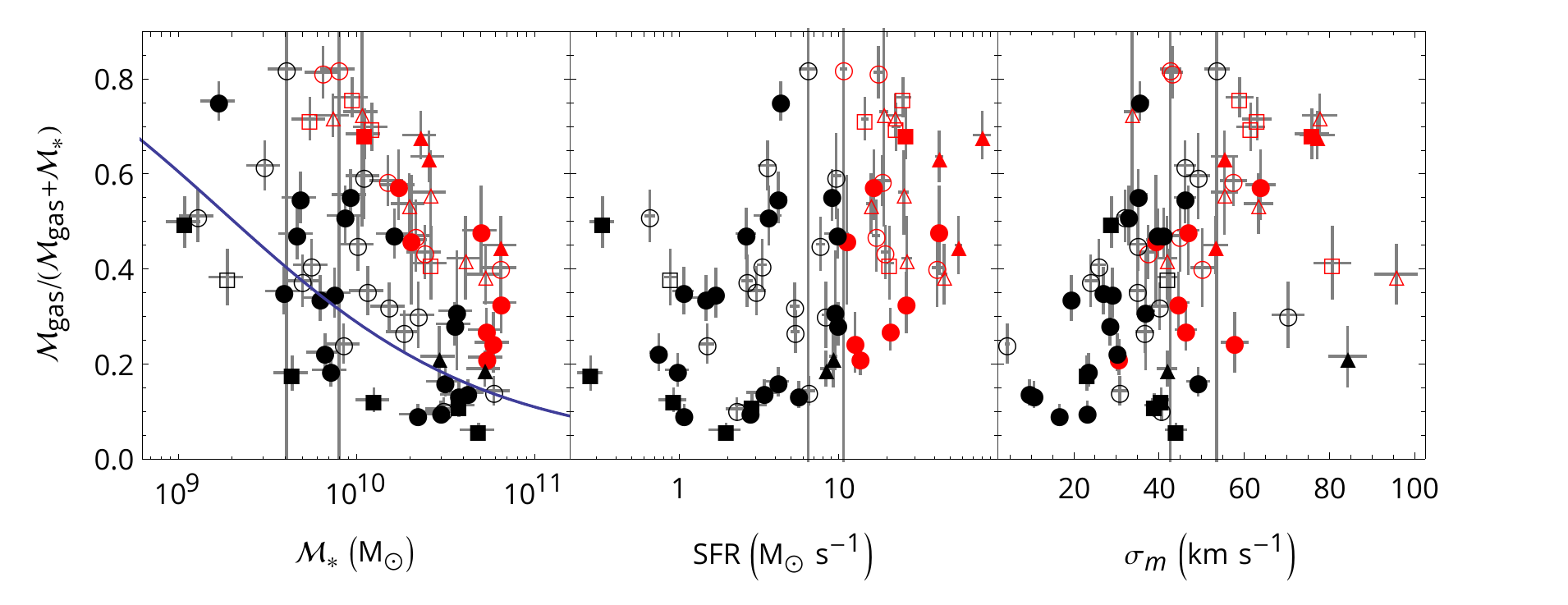}

  \caption%
  {\label{fig:gasfrac-sigma}%
    The estimated gas mass fraction for galaxies in our sample is
    shown as a function of stellar mass $(\Mstellar)$, star-formation
    rate (SFR) and mean local velocity dispersion ($\sigmean$). The
    symbol shapes are coded by the kinematic classification
    (\S~\ref{sec:nebular-morphology}). Red symbols correspond to
    galaxies with star-formation rates greater than $10\Msunyr$, and
    black symbols to those with lower star-formation rates. (A cut on
    specific-star-formation rate roughly corresponds to gas fractions
    of 0.4, a simple reflection of the method used to compute the
    fractions.)  Gas fractions have been estimated using the
    Kennicutt-Schmidt Law, and so are not independent of the
    star-formation rate. The smooth curve in the left panel is the
    relationship between gas fraction and stellar mass found in the
    COLD GASS Survey \citep{2011MNRAS.415...32S}. Errorbars show
    statistical, 1-sigma errors for all quantities. %
  }
\end{figure*}

The fraction of baryons in gas (by mass) is computed for each galaxy
in the sample:
\begin{equation}
f_{\rm gas} \equiv \frac{\Mgas}  {(\Mgas + \Mstellar)}
\end{equation}
We assume the total mass of baryons in each galaxy to be the sum of
the gas mass and the stellar mass. The stellar masses used are those
of \cite{Kauffmann2003a}. Dust is also present, but makes up a
negligible fraction of the mass of baryons. The estimated mass
fraction of gas varies from \n{mingasfrac} to \n{maxgasfrac} across
our sample, as shown in Figure~\ref{fig:gasfrac-sigma}. Higher gas
fractions are found mostly in galaxies with star-formation rates above
a few $\Msunyr$, although this may be a circular consequence of the
dependence of our method for estimating the gas mass on star-formation
rate. Across the whole sample, the mass fraction of gas does not correlate with the gas
velocity dispersion, $\sigmean$, or with the stellar mass of the
galaxy (Pearson's $R=-0.42$ and 0.44, respectively). However, when only the galaxies with star-formation rates below $10\Msunyr$ are considered, there is more correlation ($R=-0.68$),in agreement with \cite{2011MNRAS.415...32S}. At a given stellar mass, it is the galaxies forming stars at
the highest rates that have the highest estimated gas mass fractions.

\subsection{Tully-Fisher relation}
\label{sec:tf-relation}

We now plot the Tully-Fisher relation for our sample to confirm our kinematic
classification criteria and explore potential sources of evolution in
the relation. \cite{Tully1977} report a relationship between circular
velocity and luminosity (or mass) in disc galaxies. The TFR has
become a key kinematic relationship for disc galaxies
\citep[][etc.]{2001ApJ...563..694V, 2000ApJ...533L..99M,
  1997AJ....114.2402C, 1996ApJS..107...97M, 2007AJ....134..945P,
  2007MNRAS.377..806C}. Recent studies have argued for evolution in
the TFR at higher redshift \citep{2008A&A...484..173P,
  2009ApJ...697..115C, 2011A&A...528A..88G, 2012A&A...546A.118V}
although there is also evidence against this
\citep{2011ApJ...741..115M,2012ApJ...753...74M}. If part of our sample
is really representative of high-redshift galaxies, then an offset in
the TFR in that sub-sample may be apparent.

\subsubsection{Tully-Fisher relation as a test kinematic classification}

We compare galaxies in our sample to the TFR for disc galaxies to
help demonstrate that part of the sample is indeed disc-like. In
general, galaxies that are not rotating discs deviate from the
relationship, usually towards lower circular velocities as the action of
merging will convert systematic rotational motions to random ones and
reduce overall velocity gradients \citep{2010ApJ...710..279C}.
Observations with limited seeing of two galaxies merging could show
a velocity field similar to that of a rotating disc, but are less
likely to agree with the TFR. Consequently, we can use the
TFR as a discriminator between galaxies with disc-like kinematics
and those with velocity shear arising from other sources (such as
mergers), and check that our kinematic classifications as e.g., RD and
CK (\S~\ref{sec:nebular-morphology}), are valid.

\begin{figure}

  \center
  \includegraphics[width=0.9\linewidth]
  {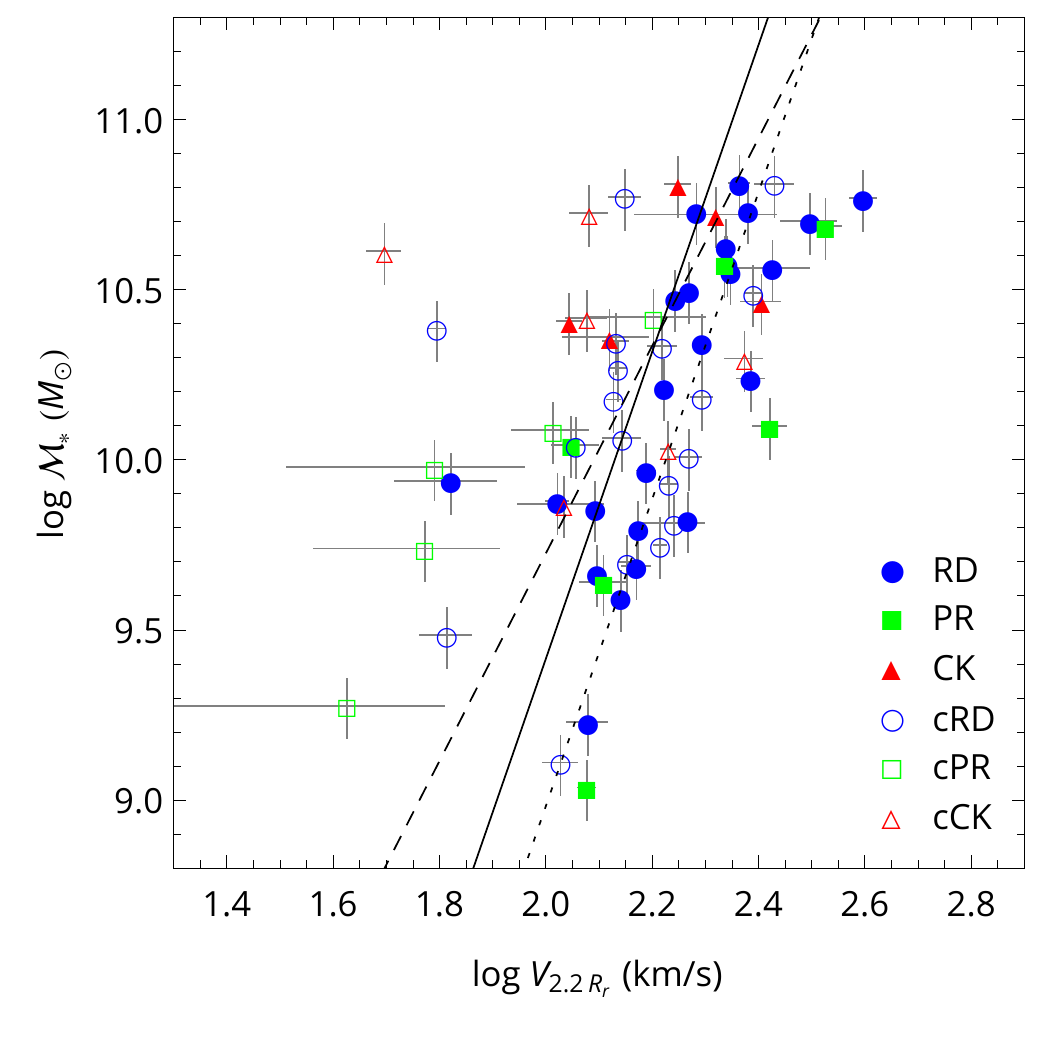}
  
  \caption{%
    \label{fig:tfr}%
    Stellar mass ($\mathcal{M}_*$, \citealt{Kauffmann2003a}) of
    galaxies in our sample plotted as a function of circular velocity
    ($\Vcirc$).  Points are coded by their kinematic classification as
    shown at the lower-right (\S~\ref{sec:nebular-morphology}). The
    circular velocity, $\Vcirc$ is inferred at 2.2 times the
    exponential disc scale radius in $r$-band, $2.2R_r$, using the
    best-fitting disc model (\S~\ref{sec:disc-fitting}). For galaxies
    with complex kinematics (CK and cCK), $v_{\rm shear}/2$ is shown
    in place of $\Vcirc$ (\S~\ref{sec:tf-relation}). These CK objects
    (red triangles) do not show a correlation, while more disc-like
    galaxies correlate as expected \citep{Tully1977}. The local
    Tully-Fisher relations of \citet[solid line]{2001ApJ...550..212B}
    and \citet[dashed line]{2005ApJ...633..844P} are shown for
    comparison. The dotted line shows the offset relation found by
    \cite{2009ApJ...697..115C} for $z \simeq 2$ galaxies. Statistical,
    $1\sigma$ error bars are included, but do not include any
    differences between the photometric and kinematic inclination. %
  }

\end{figure}

The TFR for galaxies in our sample is shown in
Figure~\ref{fig:tfr}. The coloured symbols show the positions of
galaxies in our sample, with the shape and colour of the symbol
corresponding to its kinematic morphology derived in
\S~\ref{sec:nebular-morphology}. Stellar masses for our galaxies have
been determined by \cite{Kauffmann2003a}, and are adjusted by a factor
of 0.88 to account for the difference in IMF. We adopt circular
velocities from disc fitting (\S~\ref{sec:disc-fitting}) or, where
those fits are not valid (e.g., CK and cCK classifications), from an
estimate, $\Vshear/2$, of the maximum velocity shear across the
galaxy.  The velocity $\Vshear$ is the difference between the 5th
percentile and the 95th percentile velocities observed in the velocity
map (similar to \cite{Law2007,2010ApJ...724.1373G,2011MNRAS.417.2601W}). 
This difference reflects the total velocity shear across the
velocity map, while avoiding outliers in the distribution of
velocities within the map. $\Vshear/2$ has been used in place of
$\Vcirc$ for CK- and cCK-classified galaxies in Figure~\ref{fig:tfr}
and is listed in parenthesis in the $\Vcirc$ column of
Table~\ref{tab:kinematic}.  Not shown are galaxies with inclinations
of less than $12\degr$ because the inclination correction introduces a
large error to such systems.

The correlation with the TFR (or lack thereof) matches
expectations from the kinematic morphology of galaxies within our
sample. Galaxies classified as having complex kinematics (those with
CK and cCK designations) are clearly offset to the left in
Figure~\ref{fig:tfr}, show almost no correlation (Pearson's $R=0.06$),
and do not follow the TFR. Some compact, disc-like galaxies (cRD
and cPR) are also scattered to the left in
Figure~\ref{fig:tfr}. However, this apparent disagreement could be due
to observational effects, as the atmospheric seeing can reduce
modelled circular velocity as the angular size of a galaxy approaches
the seeing limit \citep{2010MNRAS.401.2113E}. Also, seeing can affect
the estimate of inclination for a galaxy that is poorly
resolved. Therefore, the comparison with the Tully-Fisher relation
does not confirm or reject cRD or cPR galaxies as disc-like, although
they are better correlated ($R=0.55$) than galaxies with complex
kinematics. However, the remaining RD and PR classified galaxies do
have correlated circular velocities and masses (Pearson's correlation
coefficient of $R=0.73$). The TFR confirms that the \n{rd} RD and
\n{pr} PR classified galaxies are indeed discs.

The local TFRs of \cite{2001ApJ...550..212B} and
\cite{2005ApJ...633..844P} are also shown in Figure~\ref{fig:tfr} for
comparison (adjusted to a \citealt{2003PASP..115..763C}
IMF). There is still some disagreement between these two as to
the exact slope of the local TFR (and in general, see \citealt{2007ApJ...662..322H} and \citealt{2013arXiv1305.2469G} for
more discussion). Also
apparent is that many of the {\small DYNAMO} galaxies have lower stellar masses
than is typical for local galaxies of a fixed rotation velocity. We
have already noted (\S~\ref{sec:gas-mass}) that many of these galaxies
are extremely gas-rich, and high gas content can offset galaxies to higher circular velocities in
the stellar-mass TFR
\cite{2000ApJ...533L..99M,2005ApJ...632..859M}. Therefore, the
disagreement between rotating galaxies in our sample and the local
stellar mass TFR is not unexpected.

\begin{figure}

  \center
   \includegraphics[width=0.9\linewidth]
   {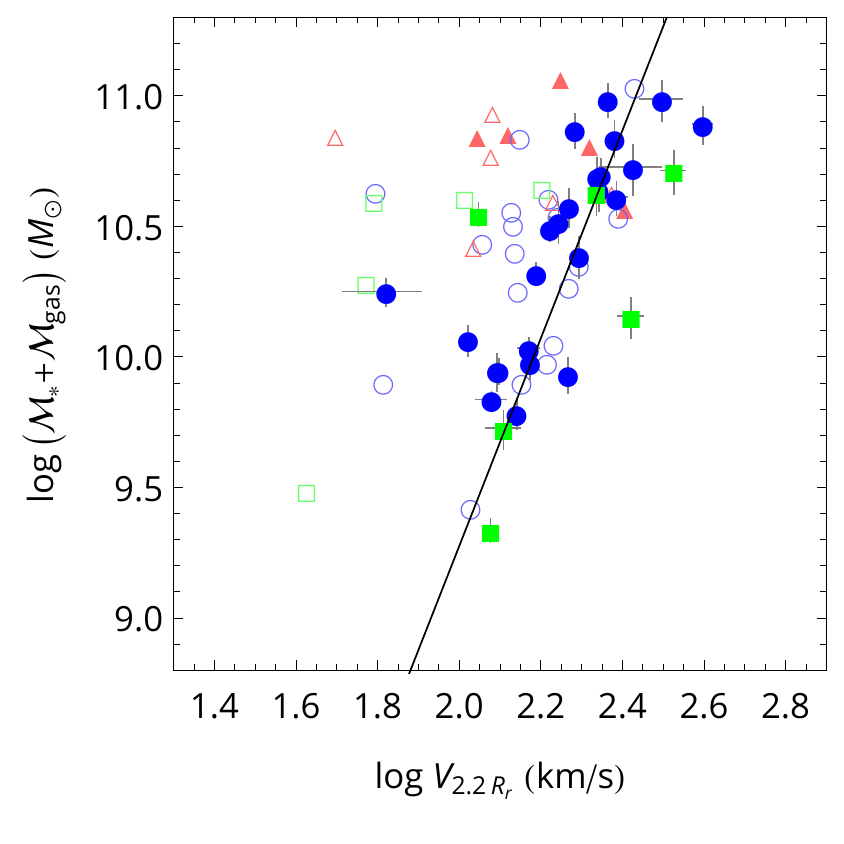}

  \caption{%
    \label{fig:tfr-offset}%
    The baryonic Tully-Fisher relation for the DYNAMO sample. Galaxies
    from the sample are plotted as in Figure~\ref{fig:tfr}, but
    galaxies classified as rotating discs (RD and PR) have been
    highlighted. Error bars are omitted from the other points for
    clarity. The local baryonic TFR of \cite{2000ApJ...533L..99M} is
    shown by the solid line, and agrees well for the rotating disc
    galaxies. Comparing with Figure~\ref{fig:tfr}, it is apparent the
    offset from the stellar mass TFR results from the high gas
    fractions inferred for many of the galaxies in our sample. The RD
    classified galaxy B~11-2 shows much lower circular velocity than
    expected; this galaxy is very round in appearance -- the
    photometric inclination may not accurately reflect the kinematic
    inclination. %
  }

\end{figure}

The RD and PR galaxies in our sample are compared with the baryonic
TFR of \cite{2000ApJ...533L..99M} for local galaxies in
Figure~\ref{fig:tfr-offset} (also adjusted to a
\citealt{2003PASP..115..763C} IMF). The positions of other
galaxies in our sample are also shown. The trimmed sample agrees very
well with the reference relation, further confirming that these
galaxies host regular rotating discs. Residual scatter is 0.5~dex RMS
in total mass for the galaxies classified as RD or PR. Including the
gas mass for these galaxies brings them into agreement with the local
TFR, further supporting their gas-rich nature.

\subsubsection{Evolution in the Tully-Fisher relation}
\label{sec:tf-offsets}

We compare rotating galaxies in our sample with two recent reports of
evolution in the TFR at higher redshifts. While we would not
expect to see any redshift evolution of the TFR in our sample, if
the offset at high redshifts is a result of changes in star formation
or gas velocity dispersion then it may also be apparent in our
sample. Figure~\ref{fig:tfr} shows the `evolved' relation of
\cite{2009ApJ...697..115C} for SINS survey galaxies at $z \simeq 2$ as
a dotted line in addition to the local reference relations introduced
above. \cite{2008A&A...484..173P} also see an offset of the same scale
as \citeauthor{2009ApJ...697..115C} for the IMAGES galaxies at $z
\simeq 0.6$. However, when the mass in gas is included for gas-rich
{\small DYNAMO} discs, they do not show a significant offset, as seen
in Figure~\ref{fig:tfr-offset}. This result is also found in IMAGES
\citep{2010A&A...510A..68P}, and in MASSIV
\citep{2012A&A...546A.118V}.

\begin{figure}
  \center
  \includegraphics[width=0.9\linewidth]
  {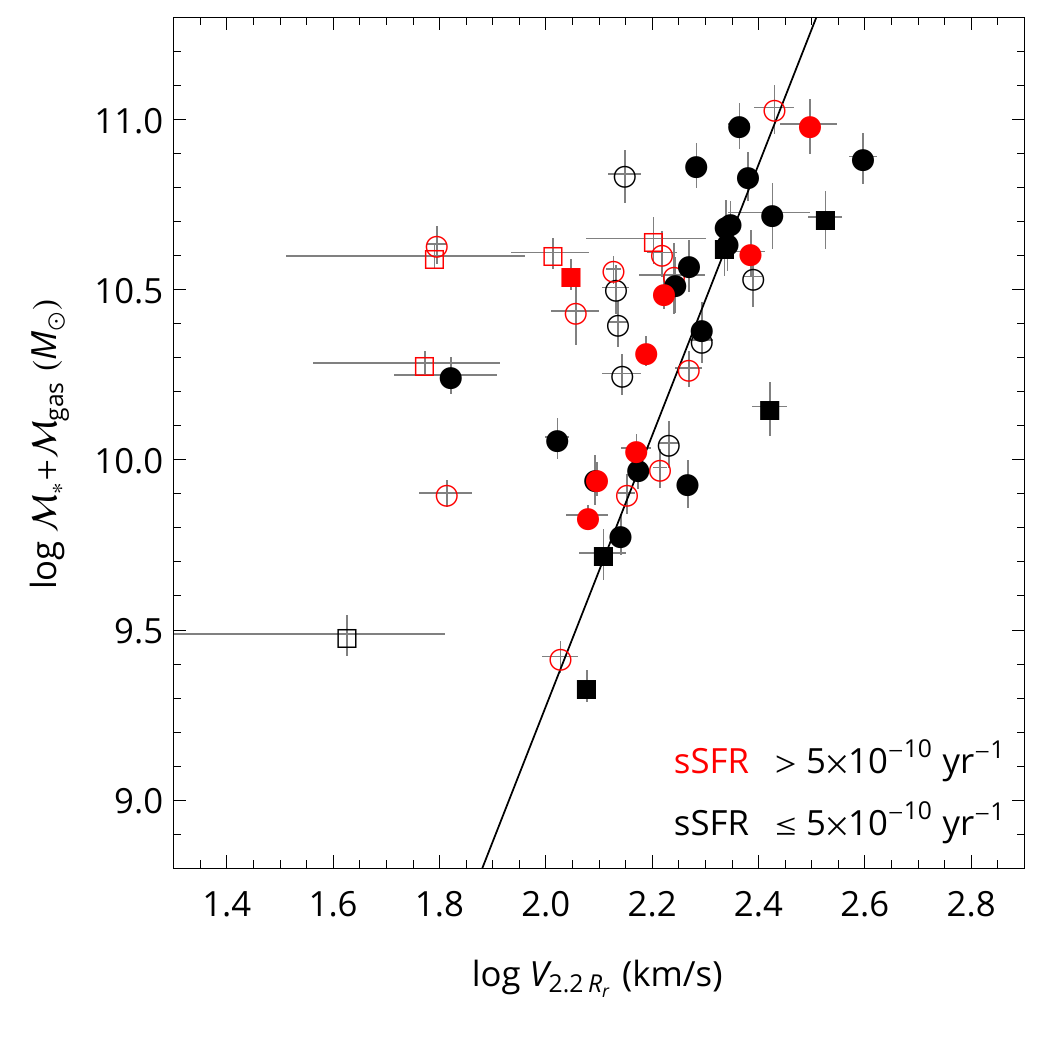}

  \caption{%
    \label{fig:tfr-evo}%
    The impact of specific-star-formation rate on the Tully-Fisher
    relation. Red symbols correspond to galaxies in our sample with
    specific-star-formation rates greater than $5\times
    10^{-10}\peryr$, black symbols correspond to those with lesser
    specific-star-formation rates. The symbol shape corresponds to
    kinematic classification as in Figure~\ref{fig:tfr}. The solid
    line shows the same local Tully-Fisher relation of
    \cite{2001ApJ...550..212B} shown in
    Figure~\ref{fig:tfr-offset}. The galaxies with high
    specific-star-formation rate that are expected to be most similar
    to galaxies at high redshift show no systematic offset from more
    typical low-redshift galaxies.  %
  }
\end{figure}

The rotating galaxies in our sample can be divided into two groups
based on their specific-star-formation rate to check for any change in
the TFR within our sample. Figure~\ref{fig:tfr-evo} shows this
division. The red symbols correspond to galaxies in our sample with
specific-star-formation rates greater than $5\times 10^{-10}\peryr$
(reflecting the position of the star-formation main sequence at high
redshift, \citealt{2007ApJ...660L..43N}), while the galaxies shown by
the black symbols have lower specific-star-formation rates (more
comparable to typical local disc galaxies). The red symbols show no
systematic offset to higher circular velocities than the rest of our
sample. Instead, many of these galaxies scatter to lower circular
velocities than local galaxies, although the compactness of some
galaxies complicates our estimate of their circular velocities. In our
sample, the TFR of rotating disc galaxies with high
specific-star-formation rates, which place them off the local
star-formation main sequence, is indistinguishable from the TFR
for rotating disc galaxies lying on the main sequence.

We have divided our sample using several other properties to check for
possible systematic differences. A division by star-formation rate of
$10 \Msunyr$, corresponding roughly to \Ha{} luminosities of
$10^{42}\ergsec$, reveals no offset. Similarly, dividing the sample by
gas velocity dispersion, $\sigmean$, around $40\kms$, also does not
reveal any clear offset or deviation from the local relation. While we
cannot rule out true redshift evolution in the TFR, there is no
evidence for evolution due to differences in total star-formation
rate, specific-star-formation rate, or velocity dispersion, several of
the primary differences observed between local and high-redshift
galaxies. The stellar-mass TFR is offset for galaxies that are
gas rich. However, the baryonic TFR remains constant. Our
conclusion is similar to that of \cite{2010A&A...510A..68P} and \citep{2012A&A...546A.118V}, namely
that the offset in the TFR claimed by e.g.,
\citeauthor{2009ApJ...697..115C} and \cite{2008A&A...484..173P}
reflects the gas richness of the samples concerned.

\subsection{Disc stability and velocity dispersions}

\begin{figure}
  \centering
  \includegraphics[width=0.9\linewidth]{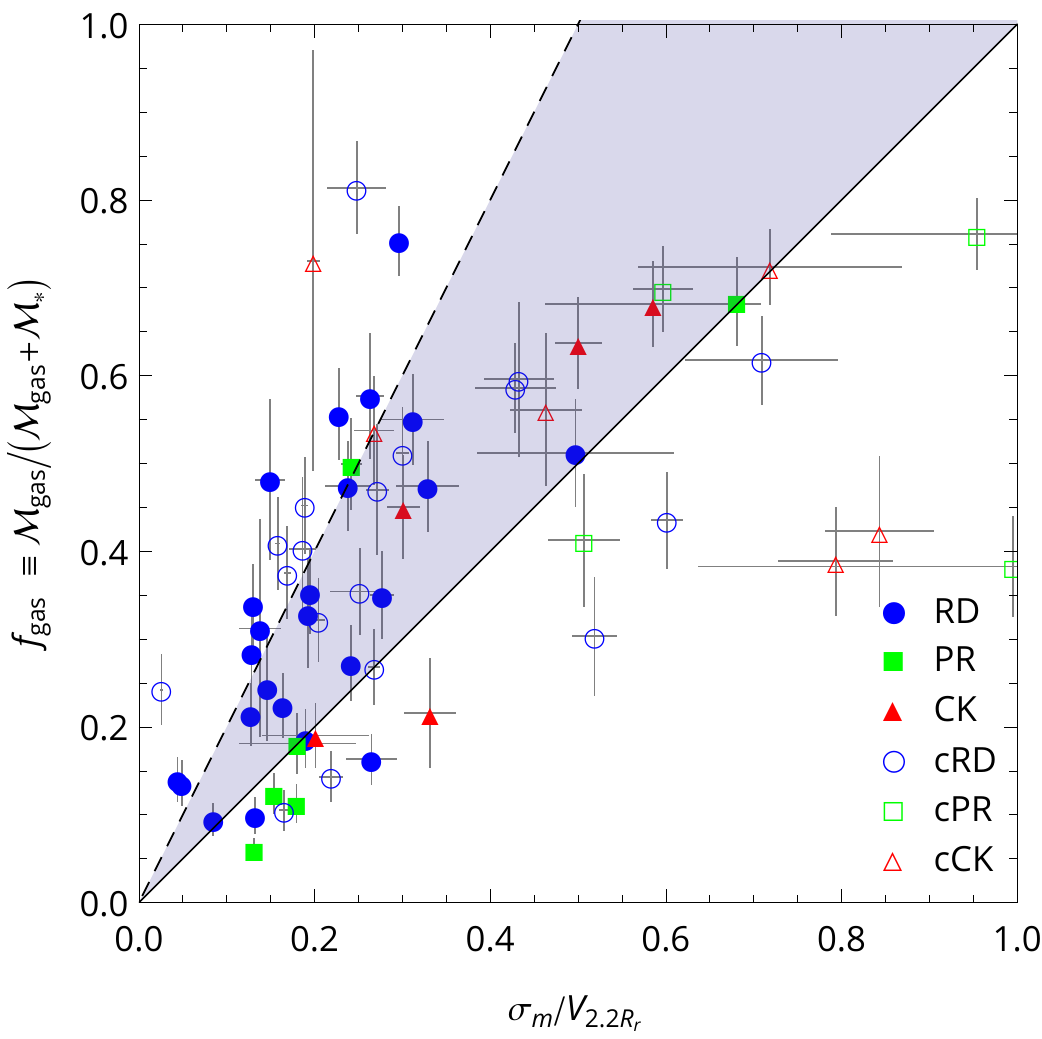}
  \caption{ %
    \label{fig:v-sigma-fgas}%
    Gas fraction ($f_{\rm gas}$) as a function of the ratio $\sigma/V$
    for DYNAMO galaxies. The velocities are those used for our Tully-Fisher relation
    analysis (\S~\ref{sec:tf-relation}). Lines show the stability
    limits for $a=1$ (solid) and $a=2$ (dashed). %
  }
\end{figure}

\cite{2011ApJ...733..101G} presents a simple argument relating $V/\sigma$ to gas mass fraction $f_{\rm gas}$. If one assumes that a gas disc is marginally stable (Toomre $Q\sim 1$) then one can derive:
\begin{equation}  \label{eq:fgas2}
\frac{V} {\sigma} \simeq  \frac{a}{f_{\rm gas}}
\end{equation}
where $a$ is a numerical factor of value $1<a<2$ that depends on the
shape of the disc's rotation curve. The
\citeauthor{2011ApJ...733..101G} argument implies that gas fraction
will correlate with the disc dynamical temperature (parametrized by
$\sigma$) for marginally stable discs at a fixed mass (circular
velocity, $V$, correlates with total mass, not gas
fraction). Figure~\ref{fig:v-sigma-fgas} shows the parameter space of
the stability criterion. A correlation, with Pearson's $R=0.60$,
between $\sigmean/\Vcirc$ and $f_{\rm gas}$ for disc-like galaxies
(i.e., RD, cRD, PR and cPR classifications) is apparent. This
correlation improves to $R=0.69$ if only RD and PR classified galaxies
are considered. Of these disc-like galaxies, 72 per~cent fall in the
region bounded by $a \geq 1$, the lower limit for disc stability, and
this fraction rises to 81 per~cent for the higher quality discs (RD
and cRD classifications). We conclude that most of these galaxies are
at least marginally stable discs.

What source drives the high velocity dispersions needed to keep these
gas-rich discs stable ($Q\gtrsim 1$) is less clear. Without driving,
turbulence decays rapidly on the time scale of a few times the
turbulent crossing time for the disc, $d/\sigma \sim 1\kpc / 50 \kms
\sim 20\,{\rm Myr}$, where $d$ is the size of the disc \citep[see][for
a review]{2004ARA&A..42..211E}. Unless the turbulence is sustained by
energy injection, $Q$ will drop below unity and the disc will undergo
run-away collapse. Gravitational energy from this collapse will be
converted to turbulence, but this normally provides only a fraction of
the energy required to maintain the observed turbulence for extended
periods \citep{2010ApJ...712..294E}. Once the gravitational energy has
been exhausted and star formation begins, stellar winds and supernova
may sustain the necessary turbulence \citep{2004RvMP...76..125M,
  2010ApJ...712..294E}.

\subsection{The relationship between star-formation rate and gas velocity dispersion}
\label{sec:lsig-theory}

\begin{figure*}
  \includegraphics[width=0.9\linewidth]{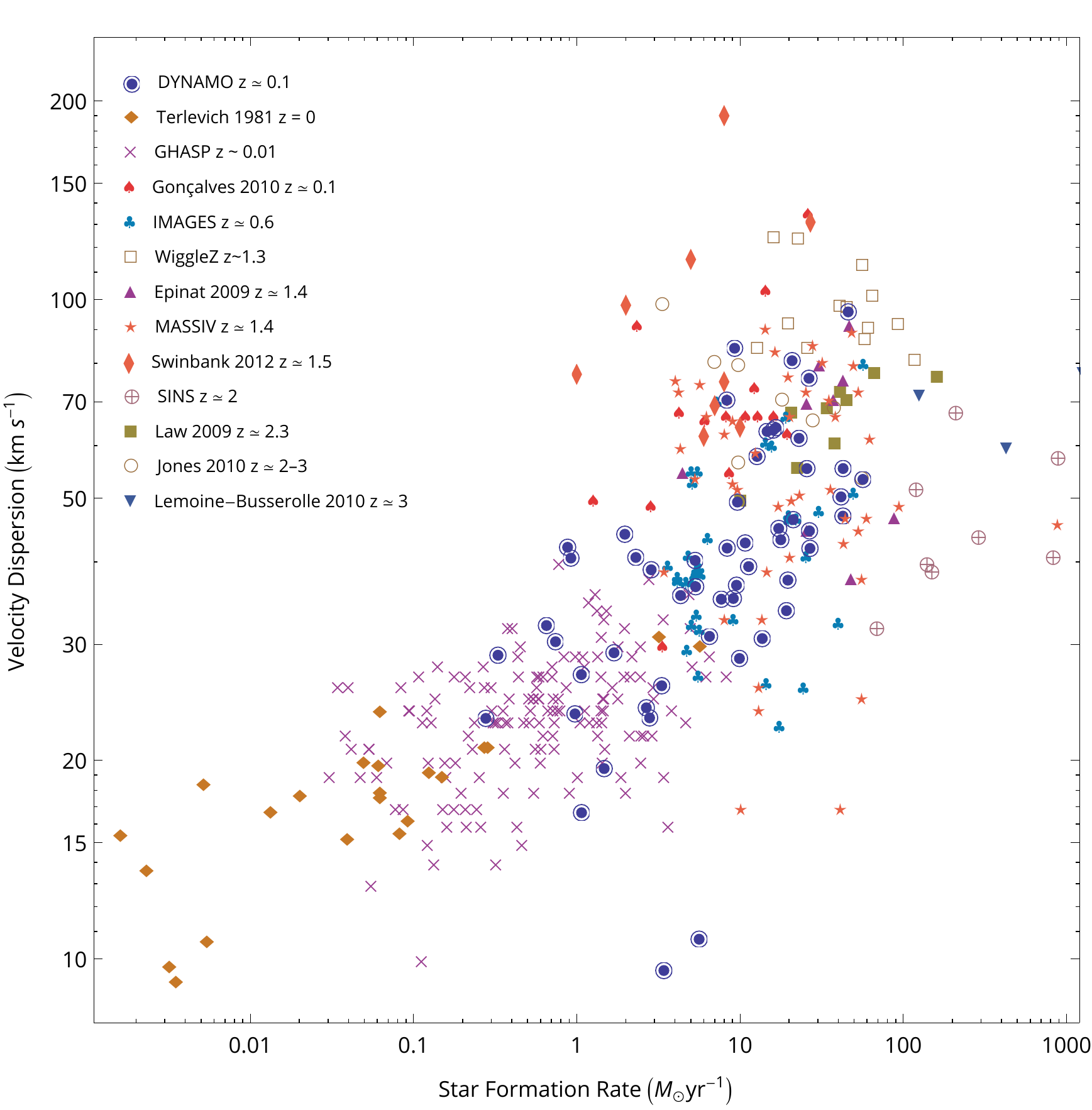}

  \caption{%
    \label{fig:L-sigma-label}%
    Star formation rate and turbulence are correlated in star-forming
    objects of all scales and redshifts. Data from a variety of low-
    and high-redshift galaxies are shown (described in detail in
    \S~\ref{sec:lsig-theory}). Also shown are (unresolved) velocity
    dispersion measurements of local \ion{H}{2} regions. The velocity
    dispersions do not include large scale velocity shear (e.g., from
    disc rotation). Methods for determining star-formation rates vary
    between samples, and systematic differences in method may account
    for much of the scatter. A correlation (Pearson's $R=0.72$) is
    seen in the data.  %
  }
\end{figure*}

We reproduce the result first presented in \cite{2010Natur.467..684G}, namely
the correlation between the \Ha\ luminosity and velocity dispersion in
star-forming galaxies. Figure~\ref{fig:L-sigma-label} shows that the
star-formation rate in galaxies with a range of distances (a few Mpc
to $z = 3.5$) and stellar masses ($10^{9}$ to $10^{11}\Msun$)
correlates with their gas velocity dispersions. Previously, a dichotomy
had been observed between typical gas velocity dispersions of nearby
galaxies and those at $z > 1$, but we show there is an empirical
relationship between these parameters which connects the two redshift
regimes.

Figure~\ref{fig:L-sigma-label} shows several data sets in addition to
those reported in \cite{2010Natur.467..684G}. Our earlier letter
included the samples of \cite{2009ApJ...697.2057L,Law2007}, GHASP
\citep{2008MNRAS.390..466E,2008MNRAS.388..500E,2002A&A...387..821G},
\cite{Epinat2009}, and \cite{2010MNRAS.401.1657L}.  We have added
several samples of galaxies at high redshift, including SINS
\citep{2009ApJ...697..115C}, MASSIV \citep{2012A&A...539A..91C},
WiggleZ \citep{2011MNRAS.417.2601W} and the samples of
\cite{2010MNRAS.401.1657L,2010MNRAS.404.1247J} and
\cite{2012ApJ...760..130S}.  Notable among these is the sample of
\cite{2010MNRAS.404.1247J}, which includes higher spatial resolution
data than is typically possible for these redshifts by leveraging
gravitational lenses. At intermediate redshift, we include the IMAGES
\citep{2008A&A...477..789Y} survey. In the local universe, we add the
Lyman-Break Analogs sample of \cite{2010ApJ...724.1373G}, and local giant
\ion{H}{2} regions and galaxies from \cite{Terlevich1981}.

In all cases, the star-formation rates have been corrected for dust
extinction and scaled to our adopted \cite{2003PASP..115..763C}
IMF. However, the methods used to determine the star-formation rates vary (\Ha-luminosity scaling, fits to spectral-energy distributions, ultra-violet-luminosity scaling) and therefore are not entirely consistent (see \S~\ref{sec:star-formation-rates}). The positions of the {\small DYNAMO}
galaxies are shown using the beam-smearing-corrected mean velocity
dispersion, $\sigma_{m,{\rm corr}}$. For most of the other samples,
galaxies are shown using the $\sigmean$ measure of velocity
dispersion. Exceptions are the GHASP sample, which uses an un-weighted
mean; the sample of \cite{Epinat2009}, who weight individual velocity
dispersions in the average by inverse-error; the IMAGES sample, which
uses an un-weighted mean with the central pixel removed; the SINS survey, which reports velocity dispersions derived from disc modelling; and the
sample of \cite{Terlevich1981}, who report an integrated velocity
dispersion for an object of comparable or smaller than a single
spatial pixel in the other samples. \cite{2011ApJ...741...69D} have
shown that the un-weighted mean is comparable to the intensity-weighted
mean.

The additional samples presented in Figure~\ref{fig:L-sigma-label} all
support the correlation (Pearson's $R=0.72$) between star-formation
rate and velocity dispersion in the ionised gas, as previously
reported. None of the samples included has been selected in way that
would affect the range of velocity dispersions measured. Consequently,
we infer galaxies with high dispersion and low star-formation rate or
low dispersion and high star-formation rate (corresponding to the
empty regions of Figure~\ref{fig:L-sigma-label}) must be rare. While
no selection bias affects the range in velocity dispersion observed,
the ways in which the samples were selected does affect the range of
star-formation rates observed. At higher redshifts, only the more
luminous highly star-forming galaxies are detected. Also, highly
star-forming galaxies are rarer in local samples because of the
declining density of global star formation as the Universe ages
\citep{2004ApJ...615..209H}. These effects that bias the range of
star-formation rates observed could combine with the
star-formation--velocity-dispersion relation to create an apparent
correlation of velocity dispersion with redshift. Samples that include
rarer, higher-luminosity nearby galaxies, such as our {\small DYNAMO}
sample, and those that probe lower luminosity galaxies at higher
redshift, such as that of IMAGES or the sub-$L_*$ galaxies of
\citeauthor{2010MNRAS.404.1247J}, help reject this possibility and
confirm the relationship between star-formation rate and velocity
dispersion.

The scatter in the correlation is not necessarily unexpected given the
varying methods for measuring star formation and velocity dispersion
across the various samples shown. Methods for measuring velocity
dispersion can be hampered by beam smearing as has already been
pointed out, particularly in data of high-redshift galaxies or poorly
resolved galaxies. Furthermore, star-formation rates are estimated by
a variety of indicators, and dust correction methods also vary. It is
especially for these reasons that it would be premature to argue for a
particular power-law fit until a better understanding of the
differences in analysis methods or a self-consistent analysis is
available.

\begin{figure}
  \centering
  \includegraphics[width=0.8\linewidth]
  {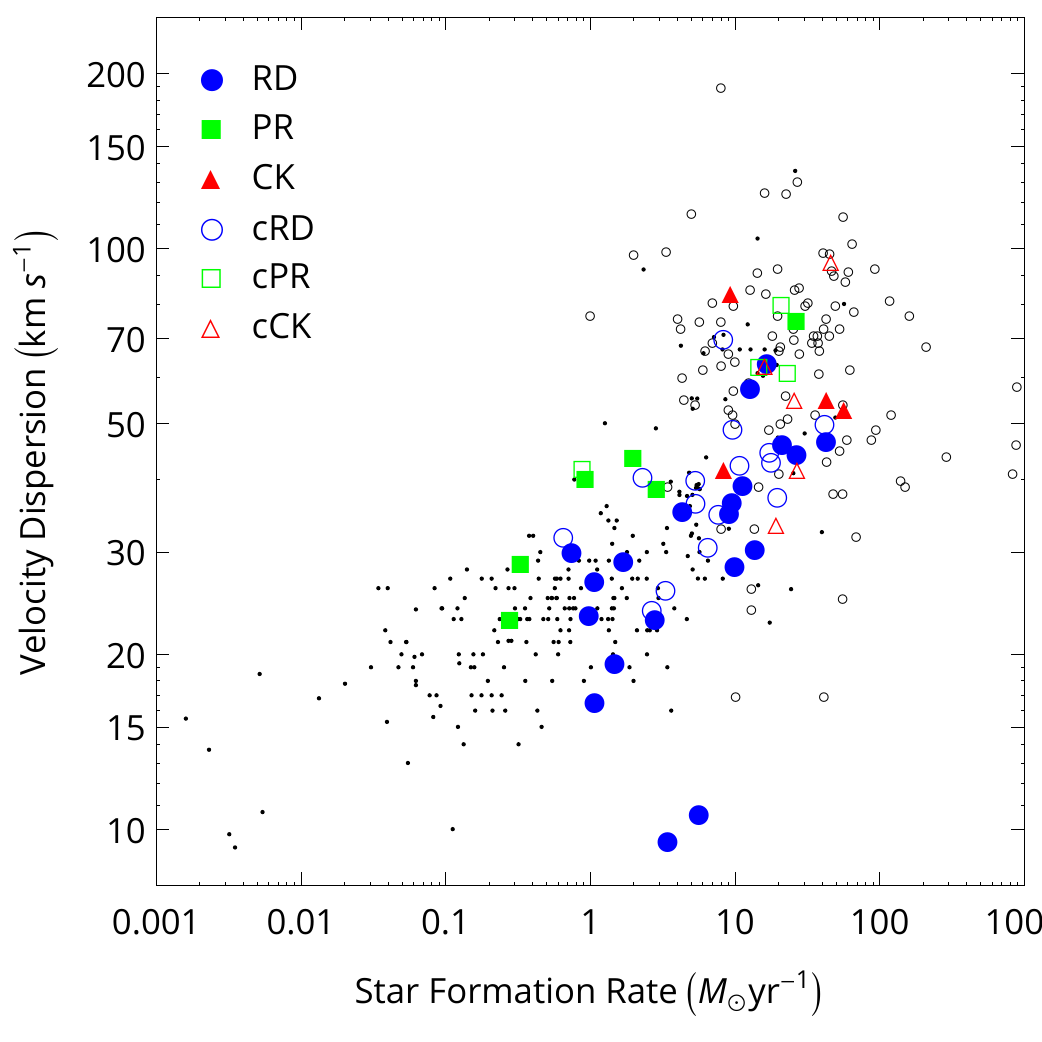}

  \caption{%
    \label{fig:L-sig-class}%
    Same as Figure~\ref{fig:L-sigma-label}, but with points for our
    sample of galaxies coded by their kinematic classification (see
    \S~\ref{sec:nebular-morphology}). Galaxies of other samples with
    $z < 1$ are shown with black points, and those with $z > 1$ are
    shown with black circles. The two outliers at the bottom of the
    figure correspond to galaxies A~4-3 and C~13-3. %
  }
\end{figure}

\cite{2010Natur.467..684G} also reported that disc-like galaxies with
star-formation rates and velocity dispersions comparable to galaxies
at $z > 1$ still exist at $z \simeq 0.1$.  To confirm that claim, we
reproduce Figure~\ref{fig:L-sigma-label}, but with {\small DYNAMO} galaxies coded
by their kinematic classification in Figure~\ref{fig:L-sig-class}. The
black points correspond to galaxies from other samples, with larger
points corresponding to $z > 1$ galaxies. Rotating and compact
rotating discs (RD, cRD) extend to both high dispersion (even after
correction for beam smearing, as measured by $\sigmeancorr$, see
Section~\ref{sec:sigm-bs}) and star-formation rate. Perturbed rotators
(PR, cPR) extend to the very highest velocity dispersions. Galaxies classified
as RD and PR also follow the TFR, reflecting their disc like
nature. The RD and PR classified galaxies show similar star-formation
rates and velocity dispersions to those of galaxies above $z > 1$,
confirming the claim of our earlier letter. Nineteen of the 31
galaxies with $\sigmean > 30\kms$ are classified RD or PR.  This
correlation supports the idea that turbulence can be sustained by
star-formation feedback which we discuss further in the next section.

\section{Discussion}
\label{sec:discussion}

\subsection{Connection between low and high redshift}

The sample presented here establishes that a rare but highly
star-forming population of galaxies at $z \simeq 0.1$ shows many
similar properties to galaxies at $1 < z < 4$. The high gas fractions,
gas velocity dispersions, and turbulent rotating discs seen in many
integral-field spectroscopic surveys of $z > 1$ galaxies can also be
found in our sample at $z \simeq 0.1$ and are not unique to that
earlier era of galaxy evolution. Identifying similar galaxies in these
two disparate eras provides a link that can be used to improve our
understanding of the physical processes of galaxy formation and
evolution.

The importance of gas velocity dispersion in star formation and galaxy
evolution makes it crucial to comparing galaxies from different
evolutionary eras. Most local galaxies have ionised gas velocity
dispersions of 20--$25\kms$ \citep{2010MNRAS.401.2113E,
  2006ApJS..166..505A}. These include centre-to-centre dispersions of
5--$7\kms$ between \ion{H}{2} regions (similar to giant molecular
clouds and \ion{H}{1}, \citealt{1989ApJ...339..763S}); the thermal
$10^4\kelvin$ dispersion of the hot \ion{H}{2} region gas; and the
feedback-driven turbulent motions within \ion{H}{2} regions
\citep{1967ApJ...147..490M, 1990ARA&A..28..525S}. As we have seen,
high-redshift galaxies have typical velocity dispersions of
30--$90\kms$. However, we can see in Figure~\ref{fig:L-sigma-label}
that our sample, and also the local sample of Lyman-Break Analogs of
\cite{2010ApJ...724.1373G} overlap with the high-redshift samples and link
low and high-redshift regimes.

The samples also overlap in $V/\sigma$ values. Most of {\small DYNAMO}
galaxies with $\sigmean> 40\kms$ have $1 < \Vcirc / \sigmean < 8$,
which compares with the large discs seen at
high-redshift. \citeauthor{2010ApJ...724.1373G} find $0.5<V/\sigma<2$
in their sample, which is more comparable to the
`dispersion-dominated' galaxies at high
redshift. \cite{2013A&A...557A..59B} suggest $V/\sigma$
anti-correlates with infrared luminosity in local galaxies, and may
reflect merger stage. In our sample, the discs are dynamically hot but
rotation is still dominant. The high inferred gas fractions, up to 70
per~cent, are also consistent with measurements at high-redshift
\citep{2010Natur.463..781T, 2013ApJ...768...74T, 2010ApJ...713..686D,
  2013arXiv1301.0371C} and the scaling of $V/\sigma$ with gas fraction
supports the idea of the existence of a gas rich disc.

The overlap of properties in galaxies from different eras is
suggestive of the idea that the same physics is regulating
star formation and velocity dispersion at low and high-redshift.
Physical processes that only operated at high-redshift are excluded
for our sample. However, these local galaxies having high
star-formation rates are rare so one can still allow processes that
were common at high redshift but are rare today. These could include
processes that are rare in time (for example mergers today) or rare in
space (for example regions of the universe with high cosmic accretion
could be rare today). Exactly how rare are {\small DYNAMO} galaxies?
For $L(\Ha)>10^{42}\ergsec$, the space density of {\small DYNAMO}
galaxies is $\sim 10^{-5}\,{\rm Mpc}^{-3}$. These represent 4 per~cent of
galaxies with masses $>10^{10}\Msun$ in our redshift range.

\begin{figure}
  \centering
  \includegraphics[width=0.9\linewidth]
  {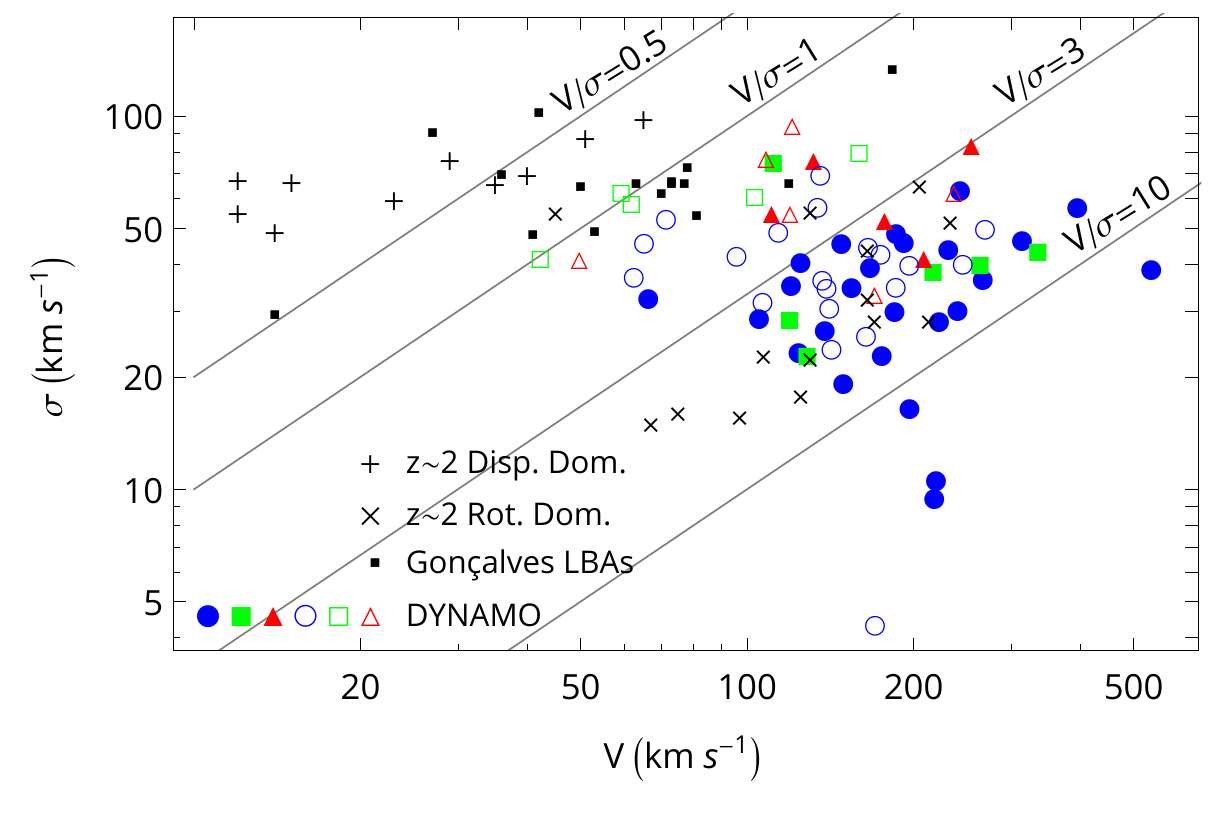}
  
\includegraphics[width=0.9\linewidth]
  {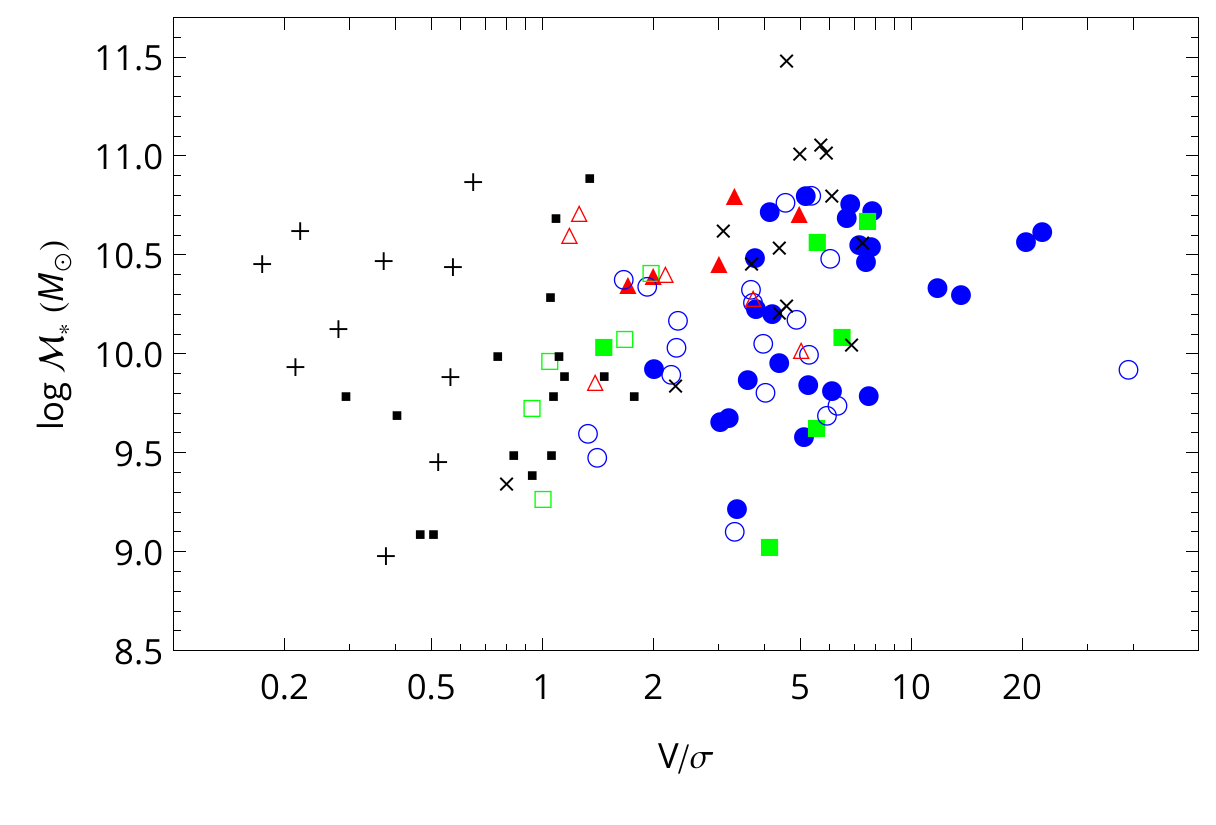}
  
  \caption{%
    \label{fig:LBA-comparison}%
    A comparison of our sample and the Lyman-Break Analogs sample
    \citep{2010ApJ...724.1373G} and our sample with dispersion-dominated and
    disc-like galaxies at high redshift. Symbols in both panel are
    coded the same, and those for DYNAMO galaxies match their
    kinematic classification as described in
    Figure~\ref{fig:L-sig-class}. The Lyman-Break Analogs sample is
    complementary to DYNAMO: our work probes galaxies with stronger
    rotation and slightly higher stellar masses -- more akin to the
    SINS disc sample of \cite{2009ApJ...697..115C} in the properties
    shown. Lyman-Break Analogs are more closely matched to the
    dispersion-dominated sample of \citeauthor{2009ApJ...697.2057L} at
    high redshift. %
  }
\end{figure}

Another local sample that is useful for comparison with high redshift
is the $z\sim 0.2$ Lyman-Break Analogs sample, which is similar to our own as a
sample of rare galaxies with high star-formation rates.  The Lyman-Break Analogs
were selected based on their near-ultraviolet luminosity by
\cite{Heckman2005} using the GALEX space telescope. Their
near-ultraviolet luminosity overlaps $2<z<3$ Lyman Break galaxies,
which are intensely star forming but somewhat lower in stellar mass
than near-infrared-selected galaxies at the same redshift
\citep{2009ApJ...706.1364F}. The most compact examples were followed
up using adaptive optics and integral-field spectroscopy in the
Paschen~$\alpha$ lines by \cite{2009ApJ...699L.118B} and
\cite{2010ApJ...724.1373G}. The sample observed with integral-field
spectroscopy is complementary to our own as shown in
Figure~\ref{fig:LBA-comparison}. Additionally, the Lyman-Break Analogs tend to
be more compact than DYNAMO galaxies because of their explicit compactness
selection criterion. The Lyman-Break Analogs are more similar in those
properties to dispersion-dominated galaxies at high redshift, while
our sample is more similar to rotation-dominated galaxies at high
redshift. Together, the Lyman-Break Analogs and {\small DYNAMO} samples provide
good comparisons with the range of star-forming galaxies seen at high
redshift.

\subsection{Relationship between turbulence and star formation}

The trend between velocity dispersion and star formation indicated by
Figure~\ref{fig:L-sigma-label} establishes a relationship for all
galaxies with $z < 4$. Instead of an evolution of galaxy velocity
dispersions with redshift, we should instead also consider an evolution 
of galaxies {\it along} this relation as high star-formation rate galaxies
become more common at high redshift.
Indications of this relationship have existed for some
time in the scaling relations for \ion{H}{2} regions \citep[e.g.,][to
name a few]{Terlevich1981, Hippelein1986, Roy1986,2009ApJ...699.1660L,2013arXiv1304.7734L,
  2012MNRAS.422.3339W}, but only by combining many samples has the
relationship, which spans five orders of magnitude in star-formation
rate, become clear for galaxies.

The relationship of Figure~\ref{fig:L-sigma-label} may reflect a
common mechanism for fuelling star formation in galaxies. In part to
explain the turbulent-disc galaxies in the high-redshift universe, a
mechanism whereby cool gas can flow directly onto the galaxy disc has
been proposed. In this model, gas is supplied via smooth accretion of
cold gas through an unstable hot-gas halo, \citep[`cold
flows,'][]{2005MNRAS.363....2K, 2006MNRAS.368....2D,
  2009Natur.457..451D}. Such cold flows are expected to shut down
below redshift $z \simeq 1$ \citep{2010MNRAS.407.1403C}, and are
therefore not expected to be relevant for galaxies in our
sample. \cite{2009ApJ...694..396B} give some evidence that these cold
flows may continue for lower mass galaxies beyond $z \simeq 1$, but
that cannot explain all of our sample.  Gas to fuel the high
star-formation rates observed must be delivered in another manner.  If
there were two different fuelling mechanisms, one for each redshift
range, the difference is not manifest in the smooth relationship
between star formation and velocity dispersion shown in
Figure~\ref{fig:L-sigma-label}. The existence of galaxies with similar
properties in both eras suggests that the mechanism may be the same
for both populations, or the relationship of Figure~\ref{fig:L-sigma-label} results from physics within the galaxy and is independent of the mechanism of gas supply. We caution though that recent simulations by
\cite{2013MNRAS.429.3353N} comparing multiple hydrodynamic codes
suggests that the delivery of large amounts of cold gas directly into
the centres of $z\sim 2$ galaxies may be a numerical artefact, and
cold mode accretion may be much less important in building galaxy mass.  The
distinction between `hot' and `'cold' modes may not be so clear cut
though accretion rates are still higher at high-redshift.
Observationally, \cite{2010ApJ...717..289S} see no evidence of
filamentary cold flows at $z \sim 2$, but considerable gas falling into
galaxy halos via other mechanisms.

Regardless of how the gas is delivered, once a turbulent, gas-rich
disc has formed, the processes governing the conversion of gas into
stars should be similar at high and low redshift (except for issues
associated with overall gas metallicity).  A turbulent disc must be
continuously sustained. Two key scenarios are outlined by
\cite{2008ApJ...687...59G}; gravitational energy from collapse, and
feedback from star formation. The first scenario includes energy
injection from cosmic accretion
\citep{2010ApJ...712..294E,2010ApJ...719.1230A} but also Toomre disc
instabilities, Jeans collapse \citep{2004ApJ...611...20I,
  2010MNRAS.409.1088B, 2010MNRAS.404.2151C, 2010ApJ...719.1230A} and
clump-clump gravitational stirring
\citep{2009ApJ...703..785D,2010MNRAS.404.2151C}.  In the latter
scenario energetic processes associated with star formation, such as
stellar winds and supernova, drive turbulence in the gas disc
\citep{2009ApJ...699.1660L,2011A&A...534L...4L,2013arXiv1304.7734L} as happens in the
Milky Way interstellar medium
\citep{2004RvMP...76..125M,2006ApJ...638..797D}.

Although both of these models probably contribute to the observed velocity
dispersion, the relative importance of each is still
unclear. \cite{2008ApJ...687...59G} and \cite{2011ApJ...733..101G}
argue for the importance of gravitational accretion, particularly cold
flows, in driving the high velocity dispersions in galaxies of their
sample. They quantify the energy in the in-falling gas and suggest a
time-scale for the dissipation of turbulent energy based on the
collision rate of individual large gas clouds. The gas accretion rate
evolves cosmologically as the reservoirs of free-floating gas are
depleted. Consequently, the energy available to drive turbulence from
gravitational accretion declines with time. The accretion rate is
further diminished as halos of hot gas form and block the subsequent
in-fall of cold gas, and as active galactic nuclei form. The decline
in accretion rate with time explains why galaxies with high
gas velocity dispersions are common early in the history of the
universe, but become rare in the present
era. \citeauthor{2008ApJ...687...59G} see very tentative evidence for
this decline. However, subsequent analysis by
\cite{2009ApJ...699.1660L,2013arXiv1304.7734L}, which is supported by the models of
\cite{2009ApJ...703..785D}, suggests that cold flows have insufficient
energy to account for the observed turbulence.

Alternatively, feedback processes associated with star formation could
drive the observed turbulence, regardless of gas fuelling
mechanism. The many processes associated with star formation have long
been understood to deliver considerable energy to the inter-stellar
medium, although there is still debate about the relative importance
of specific processes. \cite{2006ApJ...638..797D} show how supernova
can correlate with turbulence in cool gas using numerical
simulations. However, \cite{2009ApJ...703.1352K} argue that supernova
may not begin until after the typical star-forming \ion{H}{2} region
has dispersed. \cite{2004RvMP...76..125M} review the many sources of
mechanical energy from star formation, but are uncertain how
efficiently this energy is transferred to the interstellar
medium. \cite{2009ApJ...706.1364F} point out that regardless of the feeding
mechanism, high rates of star formation will invariably feed a lot of
energy back into the surrounding interstellar medium, leading to high
gas turbulence. Our relationship between rates of star formation and
gas turbulence, shown in Figure~\ref{fig:L-sigma-label} supports this
idea.

\cite{2009ApJ...699.1660L,2013arXiv1304.7734L,2011A&A...534L...4L} and
\cite{2012ApJ...760..130S} also find a relationship between
star-formation rate and dispersion, but express it as resolved
quantities (i.e., star-formation rate surface density and velocity
dispersion per spatial pixel). They argue that an energy input
proportional to star-formation surface-density predicts fairly well
the observed velocity dispersion. Numerical estimates of the
mechanical energy from star formation also match the energy
requirement to maintain the turbulence observed in the
\citeauthor{2009ApJ...699.1660L,2013arXiv1304.7734L} galaxy sample. Therefore, energy
feedback from star-formation processes alone could drive the observed
turbulence in galaxies at all redshifts.  However,
\citeauthor{2009ApJ...699.1660L,2013arXiv1304.7734L} in particular see a lot of
galaxy-to-galaxy variation in their sample. Their data were for
natural seeing with only 5--8 kpc resolution, making it difficult to
disentangle resolved quantities from global ones. In our own data, we
see only a weak correlation at best in spaxel-to-spaxel data, although
the much larger errors would weaken any
correlation. \cite{2011ApJ...733..101G} consider this issue with
adaptive-optics data on high-redshift galaxies and fail to see a
correlation on smaller (sub-galactic) scales.

Turbulence at high-redshift is most likely sustained by multiple
mechanisms. For example, \cite{2010ApJ...712..294E} explore a model where
gravitational accretion energy sets the initial structure in
primordial galaxies before star formation begins.  Then, once
widespread star formation starts, the associated feedback of energy
controls the later evolution of the galaxy. In this model, the initial
accretion of gas leads to very high velocity dispersions, such that
galaxy-size clouds of gas remain Toomre- and Jeans-stable against star
formation. As this initial energy dissipates, the cloud fragments into
large star-forming complexes, where energy from star formation then
controls the subsequent evolution of the turbulence. Such a
combination of drivers for turbulence is much more likely than any one
process remaining universally dominant, and this picture certainly
fits well with both existing observations and simulations, but is not
yet proven.

\section{Conclusions}
\label{sec:conclusions}

We have observed, using integral-field spectroscopy, a sample of
\samplesize{} galaxies covering a range in \Ha\ luminosities selected
from the Sloan Digital Sky Survey. The selection has been structured to ensure that a large fraction of the galaxies observed are rare, highly \Ha-luminous galaxies, which would be most easily detected with integral-field spectroscopy were they at $z \gtrsim 2$. Analysis of this galaxy sample has led to the following conclusions:
\begin{enumerate}
\item Narrow-band images of our galaxies centred on the \Ha\ emission
  line show a range in morphologies. Most of the sample are centrally
  concentrated, but some show two distinct concentrations or several
  smaller knots of decentralized star formation.
\item The majority of the sample show indications of disc-like
  rotation using common qualitative classification criteria. The confirmed
  disc galaxies show a reasonable agreement with the 
  Tully-Fisher relation.
\item Galaxies in our sample with star-formation rates greater than a
  few $\Msunyr$ often show high (greater than $40 \kms$) velocity
  dispersions in the ionised gas component. This subset also overlaps
  with typical star-formation rates and velocity dispersions of high
  redshift samples observed with integral-field spectroscopy.
\item We find $V/\sigma$ values similar to high-redshift discs, they
  are greater than `dispersion-dominated galaxies' (both at low and
  high redshift) but much less than modern large spiral galaxies.
\item The range in stellar masses for galaxies in our sample overlaps
  with the range covered by rotation-dominated galaxies at high
  redshift.
\item The same highly star-forming portion of our sample is also
  expected to be gas rich, with inferred gas fractions as high as
  70 per~cent. The median gas fraction of 41 per~cent is still much higher than
  typical local gas fractions in star-forming galaxies of $\sim 10$ per~cent
  \citep{2011MNRAS.415...32S}.  Gas fraction correlates only weakly,
  if at all, with stellar mass, star-formation rate, and velocity
  dispersion in our sample. However we do see a good correlation
  with $V/\sigma$ in the manner predicted for marginally stable
  gas-rich Toomre discs.
\item The gas velocity dispersion in galaxies correlates with their
  star-formation rate. This correlation holds across a broad range of
  samples and redshifts. Beam smearing due to disc rotation, which can
  affect the measure of the velocity dispersion, does not affect this
  result in our sample.
\item Some highly star-forming galaxies show inferred gas fractions,
  velocity dispersions, kinematic morphologies and star-formation
  rates similar to samples of galaxies at high redshift. Further study
  of the sample presented here can inform understanding of star
  formation in high-redshift galaxies.
\end{enumerate}

\subsection{Future work}
The sample of galaxies we have defined here, while diverse, overlaps
in many properties with typical galaxies at high redshift. If they are
true analogues then their proximity will allow a great range of
detailed study of astrophysical processes in such objects. For
example, they can be studied at greater spatial resolution (via
adaptive optics) and across multiple wavelengths not possible at high
redshift.

In this first paper of the series, we make a prediction which we will
test throughout the series.  We predict that there is no evolution
between $z \sim 0.1$ and $z \sim 2$ in the nature of star formation or
the fuelling mechanisms of star-forming galaxies. Instead, the
`evolution' in star formation and relative fractions of galaxy types
across that time we predict is a result of the reduction in the
available gas and buildup of cosmological structure. Here, and in
\cite{2010Natur.467..684G}, we already have tentative evidence that both the
kinematics and gas supplies remain similar for galaxies of similar
star-formation rates in these two eras.

In future papers in this series we will consider:

\begin{itemize}

\item High-resolution morphology both from ground-based imaging using
  adaptive optics and space-based Hubble imaging. Do we see kpc-sized
  clumps of star formation \citep{Elmegreen2005}, similar to
  high-redshift galaxies, as predicted by the Jeans instability
  scenario?

\item Can we confirm the high-gas fractions using molecular gas
  observations, for example using the CO line?

\item How does the Star Formation Law linking gas and star-formation
  rate surface densities compare with more normal galaxies and with
  high-redshift? What role does velocity dispersion play in
  self-regulation?

\item Can we detect thermal dust emission from far-infrared
  observations and relate the dust-mass and luminosity to those of
  classical LIRGs and ULIRGs?

\item Can we measure the {\it stellar kinematics} with deep resolved
  absorption-line spectra and confirm the ionised gas really is
  tracing a young stellar disc? Does the high dispersions correspond
  to young, {\it thick} stellar discs?

\item How do resolved quantities, such as metallicity, in local
  galaxies compare with high-redshift discs? For example, there has
  been some evidence for {\it positive} metallicity gradients in some
  samples \citep{2010Natur.467..811C, 2012A&A...539A..93Q,1964ApJ...139.1217T} interpreted
  as arising from cold flows.

\end{itemize}

We live in an exciting time in the study of galaxy assembly. The use of adaptive optics and integral-field spectroscopy have revolutionised our pictures of galaxy assembly. In the next few years, the advent of powerful new facilities such as ALMA and the next generation 
of large optical telescopes will allow cold gas and stellar motions at low and high redshift to be measured directly and
allow questions raised by the low- and high-redshift samples to be answered.

\hfill
\break

\section*{Acknowledgements}

In addition to the named authors who have contributed to this paper,
we thank additional DYNAMO team members Danail Obreschkow, Attila
Popping, Matthew Satterthwaite, Rob Bassett, Erin Mentuch-Cooper and
David Fisher who read and commented on this manuscript.

We thank the staff of the Australian Astronomical Observatory and the
Australian National University's 2.3 meter telescope for supporting
the observations presented here. Finally, we appreciate the referees
time, comments and suggestions, which improved this manuscript.

\bibliographystyle{agmnras}

\bibliography{dynamo-sdss_kinematics}

\appendix

\section{Removing the effects of beam smearing from dispersion maps of rotating disc galaxies}
\label{sec:bs-corr}

Here we present an approach to compute and remove the effects of beam
smearing from velocity dispersion maps obtained from integral field
spectroscopy. The method presented below fully accounts for any
additional dispersion arising from unresolved velocity gradient across
the sampling resolution.

Beam smearing describes the smearing of a velocity gradient
perpendicular to the line-of-sight into increased velocity dispersion
along the line-of-sight. Spatial pixel of an integral-field spectrograph covers a finite part
of the galaxy, and any velocity gradient across that area will be
detected only as increased velocity dispersion. Therefore, this effect
can bias spatially resolved velocity dispersion measurements,
particularly near the steep inner rotation curve found in most
rotating disc galaxies \citep{2011ApJ...741...69D}.

When the intrinsic velocity field is known, the velocity profile for
each spatial sample can be computed exactly, and then removed from
observations before computing the spatially resolved velocity
dispersion. The computation for a simple rotating disc with an
arctangent-like rotation curve is presented here.

We begin by defining a functional form to the iso-velocity contours in
the observed plane of the galaxy. In polar coordinates, we can write
the observed velocity at every point as a function of the rotation
curve, $V(r)$, inclination, $i$, and position angle on the sky, $\phi$,
\begin{equation}
  V_{\rm obs}(r, \phi) = V(r) \sin i \cos \phi
\end{equation}
For the velocity curve, we adopt that of our disc model,
\begin{equation}
  V(r) = V_{\rm circ} \frac{2}{\pi} \arctan (r/r_t)
\end{equation}
We can then compute the path of an iso-velocity contour by setting
this function equal to a particular velocity, $V_{\rm obs}=v$, and
solve to find
\begin{equation}
  r_{\rm iso}(v,\phi) =  r_t \tan{\left( \frac{\pi v \csc i \sec \phi}{2 V_{\rm
          circ}} \right)}
\end{equation}

Next, we assume that the projected intensity of the light associated
with this rotation is that of an exponential disc, $I(r)=I_0 \exp
r/h$. For our purposes, the overall normalisation $I_0$ is irrelevant,
so we set it to 1. The exponential scale radius of the galaxy is $h$.
We can then project this intrinsic intensity distribution to the observed plane, giving
\begin{equation}
  I(r,\phi) = \exp \left( \frac{r \sqrt{ (\cos \phi \sin i)^2 + (\sin \phi)^2 } }{h} \right)
\end{equation}

We can then recover the integrated velocity profile, $f(V)$, (i.e. the
integrated spectrum) by conducting a line integral of $I$ along each
iso-velocity contour:
\begin{align}
  f(V) &= \oint_V I\, ds
\end{align}
The length element is given by
\begin{equation}
  ds = \sqrt{ r_{\rm iso}^2 + {\left( \frac{d}{d \phi} r_{\rm
              iso} \right) }^2} \, d\phi
\end{equation}
which allows us to write the line integral as a normal integral,
\begin{equation}
  f(V) = \int_{\phi}I \sqrt{ r_{\rm iso}^2 + {\left( \frac{d}{d \phi} r_{\rm iso} \right) }^2}\, d\phi
\end{equation}
The limits of integration are given by the limits of the line, which
are the solutions to
\begin{equation}
  r_{\rm iso}(v,\phi) = \pm \pi/2
\end{equation}

With this formalism, we can generate the observed, integrated velocity
spectrum for the whole galaxy with $f(V)$. Furthermore, by introducing
a window function on the intensity, we can recover the velocity
spectrum for any sub region of the galaxy. For a square pixel, this window function is constructed of several Heaviside (unit) step functions in rectangular coordinates:
\begin{equation}
  W(x,y) = H(x + x_{\rm min}) H(y + y_{\rm min}) H(1-x + x_{\rm min}) H(1 - y + x_{\rm min})
\end{equation}
where we have set the size of the pixel to one, and the position of the lower left corner of the pixel is given by $(x_{\rm min}, y_{\rm min})$. This window can then be convolved with a Gaussian seeing function with full-width at half-maxiumum of $s$ to give:
\begin{equation}
W_{\rm seeing} = \frac{1}{4}\left[
  \erf {\left( \frac{x-x_{\rm min}-1}{\sqrt{2}s}\right)} -
  \erf {\left( \frac{x+x_{\rm min}}{\sqrt{2}s} \right)} \right]
\left[
  \erf {\left( \frac{y-y_{\rm min}-1}{\sqrt{2}s} \right) } -
  \erf {\left( \frac{y-y_{\rm min}}{\sqrt{2}s} \right)} \right]
\end{equation}

We can then write the velocity spectrum for an individual pixel,
$f_{\rm pix}(V)$ (including `beam smearing', i.e. seeing) as
\begin{equation}
  f_{\rm pix}(V) = \oint_V I W_{\rm seeing} ds
\end{equation}
This allows us to compute the spectrum due to beam smearing of the
velocity map within each spatial pixel. The beam smearing velocity
width can then be subtracted (in quadrature) from the observed
velocity dispersion in each pixel to recover a spatially resolved map
of velocity dispersion which is free of beam smearing arising from the
differential rotation of the disc. In the rare cases where the beam
smearing correction is larger than the observed velocity dispersion,
we set the velocity dispersion to zero. Thus these spaniels do not contribute
to the final $\sigmean$.


\clearpage
\LongTables
\begin{deluxetable*}{lr@{~}llcclc}

\tablecaption{\label{tab:observations}SDSS Target Information}
\tablewidth{0pt}
\tablecolumns{8}
\tablehead{
  \colhead{Name} &
  \multicolumn{2}{c}{Sel. ID} &
  \colhead{SDSS SpecObjID\tablenotemark{a}} &
  \colhead{Obs Date} &
  \colhead{Instrument} &
  \colhead{$T_{\rm exp}$\tablenotemark{b}} &
  \colhead{seeing\tablenotemark{c}} \\
  \colhead{~} &
  \colhead{~} &
  \colhead{~} &
  \colhead{~} &
  \colhead{~} &
  \colhead{~} &
  \colhead{seconds} &
  \colhead{arcsec}
}
\startdata
SDSS J040242.50$-$06:44:47.7 & A & 04-3 & 130827332481974272 & 16 Jan 2010 & AAT/SPIRAL  & 3600 & 2.1 \\    
SDSS J04023.187$-$05:16:29.2 & A & 04-4 & 130827334742704128 & 17 Jan 2010 & AAT/SPIRAL  & 7200 & 1.4 \\    
SDSS J083751.79+06:16:9.58   & A & 08-3 & 334337177892683776 & 16 Jan 2010 & AAT/SPIRAL  & 7200 & 1.3 \\    
SDSS J08488.126+04:15:33.5   & A & 08-4 & 335181439442092032 & 17 Jan 2010 & AAT/SPIRAL  & 7200 & 4.0 \\    
SDSS J104746.94$-$00:22:4.24 & A & 10-1 &  77628570993688576 & 16 Jan 2010 & AAT/SPIRAL  & 5400 & 1.0 \\    
SDSS J104638.40+01:48:31.6   & A & 10-2 & 142932668890218496 & 17 Jan 2010 & AAT/SPIRAL  & 7200 & 4.0 \\    
SDSS J135232.50+02:43:30.6   & A & 13-2 & 149405190099828736 & 5 Jun 2009  & AAT/SPIRAL  & 7200 & 1.1 \\    
SDSS J040828.64$-$04:52:19.8 & B & 04-3 & 131108818087051264 & 19 Jan 2010 & 2.3 m/WiFeS & 7200 & 1.2 \\    
SDSS J040228.88$-$05:20:24.5 & B & 04-4 & 131108817319493632 & 20 Jan 2010 & 2.3 m/WiFeS & 7200 & 1.2 \\    
SDSS J084735.35+09:31:2.56   & B & 08-3 & 495623963607564288 & 19 Jan 2010 & 2.3 m/WiFeS & 7200 & 1.2 \\    
SDSS J08543.867+07:34:52.2   & B & 08-4 & 365863508732018688 & 20 Jan 2010 & 2.3 m/WiFeS & 7200 & 1.2 \\    
SDSS J105918.14+09:48:37.1   & B & 10-1 & 343625916930326528 & 5 Jun 2009  & AAT/SPIRAL  & 3600 & 1.5 \\    
SDSS J111635.58+02:33:48.1   & B & 11-2 & 143777215144787968 & 20 Jan 2010 & 2.3 m/WiFeS & 7200 & 1.2 \\    
SDSS J144452.04+04:36:54.5   & B & 14-1 & 165449263151579136 & 4 Jun 2009  & AAT/SPIRAL  & 1200 & 1.5 \\    
                             &   &      &                    & 5 Jun 2009  & AAT/SPIRAL  & 2400 & 1.5 \\    
SDSS J154413.93+03:45:50.6   & B & 15-1 & 167419669911699456 & 5 Jun 2009  & AAT/SPIRAL  & 3600 & 1.0 \\    
SDSS J151129.57$-$02:05:21.1 & B & 15-2 & 260026477938999296 & 5 Jun 2009  & AAT/SPIRAL  & 3600 & 1.0 \\    
SDSS J205912.10$-$06:07:40.6 & B & 20-1 & 179242181897224192 & 5 Jun 2009  & AAT/SPIRAL  & 7200 & 1.1 \\    
SDSS J00288.639$-$00:39:10.2 & C & 00-1 & 110279118452424704 & 15 Jul 2008 & AAT/SPIRAL  & 3600 & 1.4 \\    
SDSS J04109.301$-$06:05:10.5 & C & 04-1 & 131108816509992960 & 21 Jan 2010 & 2.3 m/WiFeS & 3600 & 1.2 \\    
SDSS J040425.99$-$05:22:13.2 & C & 04-2 & 131108817516625920 & 21 Jan 2010 & 2.3 m/WiFeS & 3600 & 1.2 \\    
SDSS J132639.42+01:30:1.47   & C & 13-1 & 148562120957493248 & 15 Jul 2008 & AAT/SPIRAL  & 3600 & 1.3 \\    
SDSS J13430.490+03:53:19.3   & C & 13-3 & 240604571322286080 & 15 Jul 2008 & AAT/SPIRAL  & 3600 & 1.5 \\    
SDSS J143342.60$-$00:48:0.23 & C & 14-2 &  86353122502377472 & 15 Jul 2008 & AAT/SPIRAL  & 3100 & \nodata \\
SDSS J205259.06$-$06:04:28.7 & C & 20-2 & 179242181045780480 & 15 Jul 2008 & AAT/SPIRAL  & 3600 & 1.9 \\    
SDSS J213919.76$-$08:41:42.6 & C & 21-1 & 331522925825884160 & 15 Jul 2008 & AAT/SPIRAL  & 3600 & \nodata \\
SDSS J223949.34$-$08:04:18.0 & C & 22-2 & 203449235771752448 & 15 Jul 2008 & AAT/SPIRAL  & 3600 & 1.3 \\    
SDSS J004324.43+00:00:59.0   & D & 00-2 & 110842120222277632 & 14 Jul 2008 & AAT/SPIRAL  & 2400 & 1.3 \\    
SDSS J103235.52+07:06:6.84   & D & 10-4 & 281419573666775040 & 22 Jan 2010 & 2.3 m/WiFeS & 3600 & 1.4 \\    
SDSS J133047.05+01:42:31.5   & D & 13-1 & 148562120336736256 & 14 Jul 2008 & AAT/SPIRAL  & 3600 & 1.4 \\    
SDSS J13307.005+00:31:53.3   & D & 13-5 &  83821232935403520 & 13 Jul 2008 & AAT/SPIRAL  & 3600 & 1.4 \\    
SDSS J14468.569+00:51:51.4   & D & 14-1 & 151375517637935104 & 14 Jul 2008 & AAT/SPIRAL  & 3600 & 1.4 \\    
SDSS J15437.078$-$01:08:3.98 & D & 15-1 & 260870942786322432 & 13 Jul 2008 & AAT/SPIRAL  & 3600 & \nodata \\
SDSS J15435.269$-$01:46:42.4 & D & 15-2 & 260870941217652736 & 13 Jul 2008 & AAT/SPIRAL  & 2400 & 1.6 \\    
SDSS J153435.39$-$00:28:44.5 & D & 15-3 &  88886509838532608 & 14 Jul 2008 & AAT/SPIRAL  & 3600 & 1.2 \\    
SDSS J20529.089$-$00:30:39.3 & D & 20-1 & 276915143844036608 & 14 Jul 2008 & AAT/SPIRAL  & 3600 & 1.1 \\    
SDSS J211733.50$-$00:08:5.02 & D & 21-3 & 278041387637669888 & 14 Jul 2008 & AAT/SPIRAL  & 3600 & 1.3 \\    
SDSS J221614.50+12:23:43.8   & D & 22-1 & 207389871029878784 & 14 Jul 2008 & AAT/SPIRAL  & 3600 & 1.3 \\    
SDSS J223745.27$-$08:39:22.5 & D & 22-2 & 203449235545260032 & 14 Jul 2008 & AAT/SPIRAL  & 3600 & 1.3 \\    
SDSS J234957.74+00:05:22.9   & D & 23-1 & 108871770764738560 & 14 Jul 2008 & AAT/SPIRAL  & 3600 & \nodata \\
SDSS J00561.343+00:43:30.0   & E & 00-2 & 111124108015566848 & 16 Jul 2008 & AAT/SPIRAL  & 1800 & \nodata \\
SDSS J005453.56$-$00:59:8.52 & E & 00-3 & 111124105498984448 & 16 Jul 2008 & AAT/SPIRAL  & 3600 & 1.0 \\    
SDSS J040530.53$-$05:43:22.0 & E & 04-1 & 131108816992337920 & 19 Jan 2010 & AAT/SPIRAL  & 3600 & 0.9 \\    
SDSS J093412.85+10:40:35.4   & E & 09-1 & 490275791427338240 & 19 Jan 2010 & AAT/SPIRAL  & 3600 & 1.3 \\    
SDSS J101940.29$-$00:26:24.7 & E & 10-1 &  76502554946568192 & 19 Jan 2010 & AAT/SPIRAL  & 3600 & 1.6 \\    
SDSS J235833.89+14:54:45.3   & E & 23-1 & 211330581978415104 & 16 Jul 2008 & AAT/SPIRAL  & 1800 & 1.5 \\    
SDSS J08311.931+04:03:18.9   & F & 08-2 & 333773945712934912 & 21 Jan 2010 & 2.3 m/WiFeS & 3600 & 1.2 \\    
SDSS J091929.43+06:10:55.3   & F & 09-1 & 279168078305034240 & 21 Jan 2010 & 2.3 m/WiFeS & 3600 & 1.2 \\    
SDSS J102715.47+06:50:3.66   & F & 10-1 & 281419573087961088 & 21 Jan 2010 & 2.3 m/WiFeS & 3600 & 1.2 \\    
SDSS J122534.26$-$02:50:29.0 & F & 12-4 &  94235951454224384 & 21 Jan 2010 & 2.3 m/WiFeS & 3600 & 1.2 \\    
SDSS J034857.51$-$00:42:54.2 & G & 03-2 & 349819129932283904 & 18 Jan 2010 & AAT/SPIRAL  & 3600 & 1.7 \\    
SDSS J032718.58$-$00:28:30.5 & G & 03-4 & 116753553971740672 & 19 Jan 2010 & AAT/SPIRAL  & 3600 & 1.0 \\    
SDSS J041219.71$-$05:54:48.6 & G & 04-1 & 131108815939567616 & 18 Jan 2010 & AAT/SPIRAL  & 3600 & 1.8 \\    
SDSS J08456.474+02:46:15.4   & G & 08-1 & 158976187991851008 & 18 Jan 2010 & AAT/SPIRAL  & 3600 & 2.0 \\    
SDSS J085347.72+06:51:6.87   & G & 08-2 & 334899956504592384 & 18 Jan 2010 & AAT/SPIRAL  & 3600 & 2.0 \\    
SDSS J080412.52+07:09:58.5   & G & 08-3 & 494498037406629888 & 18 Jan 2010 & AAT/SPIRAL  & 3600 & 1.9 \\    
SDSS J08276.707+04:19:23.0   & G & 08-4 & 333773945297698816 & 19 Jan 2010 & AAT/SPIRAL  & 3600 & 1.1 \\    
SDSS J085418.73+06:46:20.6   & G & 08-5 & 334899956882079744 & 19 Jan 2010 & AAT/SPIRAL  & 3600 & 1.3 \\    
SDSS J091210.96+10:04:12.0   & G & 09-1 & 489712833139834880 & 19 Jan 2010 & AAT/SPIRAL  & 3600 & 1.3 \\    
SDSS J102142.47+12:45:18.7   & G & 10-1 & 491683209700966400 & 18 Jan 2010 & AAT/SPIRAL  & 3600 & 2.1 \\    
SDSS J113520.16+11:12:42.1   & G & 11-1 & 452558278207995904 & 18 Jan 2010 & AAT/SPIRAL  & 1800 & 1.8 \\    
SDSS J135022.74$-$02:51:57.5 & G & 13-1 & 257211852017106944 & 16 Jul 2008 & AAT/SPIRAL  & 3600 & 1.5 \\    
SDSS J145428.33+00:44:34.3   & G & 14-1 &  87199082266755072 & 16 Jul 2008 & AAT/SPIRAL  & 3600 & 1.1 \\    
SDSS J145435.35$-$02:00:49.7 & G & 14-3 & 259463424658898944 & 16 Jul 2008 & AAT/SPIRAL  & 3600 & 1.8 \\    
SDSS J203724.58$-$06:22:0.33 & G & 20-1 & 178679178873274368 & 16 Jul 2008 & AAT/SPIRAL  & 3600 & 1.1 \\    
SDSS J20442.915$-$06:46:57.9 & G & 20-2 & 178960572606316544 & 16 Jul 2008 & AAT/SPIRAL  & 3600 & 0.9 \\    
SDSS J211911.80+01:08:31.9   & G & 21-2 & 278041388019351552 & 16 Jul 2008 & AAT/SPIRAL  & 1800 & 1.1 \\    
SDSS J104431.76+12:09:25.2   & H & 10-2 & 450869566705238016 & 22 Jan 2010 & 2.3 m/WiFeS & 3600 & 1.4 \\    

\enddata


\tablenotetext{a}{The SDSS spectroscopic id number from data release
  four \citep{2006ApJS..162...38A}.}

\tablenotetext{b}{The total exposure time for the night. See Section
  \ref{sec:observations} for the lengths of individual exposures.}

\tablenotetext{c}{The full-width-at-half-maximum seeing observed on a
  nearby star as part of the observation sequence.}

\end{deluxetable*}

\clearpage
\LongTables
\begin{deluxetable*}{lllrrrrrrr}

  \tablecaption{\label{tab:basic-sfr}Basic and Star Formation Properties
    of the Sample}
  \tablewidth{0pt}
  \tablecolumns{10}
  \tablehead{
    \colhead{Sel. ID} &
    \colhead{$z$} &
    \colhead{$M_r$\tablenotemark{a}} &
    \colhead{\Mstellar\tablenotemark{b}} &
    \colhead{$L_{\rm fibre}(\Ha)$\tablenotemark{c}} &
    \colhead{$L_{\rm IFU}(\Ha)$\tablenotemark{d}} &
    \colhead{${\rm SFR}_{\rm B04}$\tablenotemark{e}} &
    \colhead{${\rm SFR}_{\Ha}$\tablenotemark{f}} &
    \colhead{Ext.\tablenotemark{g}} &
    \colhead{$\Mgas$\tablenotemark{h}} \\
    \colhead{~} &
    \colhead{~} &
    \colhead{~} &
    \colhead{$\left( 10^{9}\Msun \right)$} &
    \colhead{$\left( \log \ergsec \right)$} &
    \colhead{$\left( \log \ergsec \right)$} &
    \colhead{$\left( \Msunyr \right)$} &
    \colhead{$\left( \Msunyr \right)$} &
    \colhead{(mag)} &
    \colhead{$\left( 10^9\,\Msun \right)$}
  }
  \startdata
  A 04-3 & 0.06907 & -21.2 & 42.35 & 40.18 & ${41.36}_{-0.03}^{+0.03}$ &  1.78 & ${3.42}_{-0.73}^{+0.73}$  & ${1.32}_{-0.25}^{+0.20}$  & ${6.89}\pm{0.35}$ \\     
A 04-4 & 0.06617 & -19.4 &  1.09 & 39.93 & ${40.88}_{-0.04}^{+0.04}$ &  0.29 & ${0.33}_{-0.05}^{+0.05}$  & ${-0.02}_{-0.14}^{+0.13}$ & ${1.08}\pm{0.06}$ \\     
A 08-3 & 0.06408 & -20.3 &  6.28 & 40.02 & ${41.16}_{-0.02}^{+0.02}$ &  0.81 & ${1.47}_{-0.29}^{+0.29}$  & ${0.92}_{-0.24}^{+0.19}$  & ${3.22}\pm{0.13}$ \\     
A 08-4 & 0.05921 & -20.7 & 48.53 & 40.04 & ${41.18}_{-0.03}^{+0.03}$ &  3.66 & ${1.97}_{-0.41}^{+0.41}$  & ${1.18}_{-0.24}^{+0.20}$  & ${3.15}\pm{0.13}$ \\     
A 10-1 & 0.08138 & -20.5 & 37.66 & 40.29 & ${41.02}_{-0.03}^{+0.03}$ &  1.48 & ${2.86}_{-0.59}^{+0.59}$  & ${1.99}_{-0.24}^{+0.20}$  & ${4.80}\pm{0.35}$ \\     
A 10-2 & 0.06183 & -19.8 & 12.56 & 40.10 & ${40.97}_{-0.03}^{+0.03}$ &  0.59 & ${0.92}_{-0.16}^{+0.16}$  & ${0.87}_{-0.19}^{+0.16}$  & ${1.78}\pm{0.09}$ \\     
A 13-2 & 0.07690 & -20.0 &  4.36 & 40.12 & ${40.88}_{-0.02}^{+0.02}$ &  0.44 & ${0.28}_{-0.04}^{+0.04}$  & ${-0.20}_{-0.18}^{+0.15}$ & ${0.96}\pm{0.11}$ \\     
B 04-3 & 0.06611 & -20.6 & 31.53 & 40.47 & ${41.40}_{-0.05}^{+0.04}$ &  3.85 & ${4.18}_{-0.59}^{+0.59}$  & ${1.46}_{-0.12}^{+0.10}$  & ${6.14}\pm{0.28}$ \\     
B 04-4 & 0.07555 & -20.5 & 11.58 & 40.63 & ${41.74}_{-0.05}^{+0.04}$ &  2.00 & ${3.05}_{-0.38}^{+0.38}$  & ${0.26}_{-0.08}^{+0.08}$  & ${6.35}\pm{0.43}$ \\     
B 08-3 & 0.06313 & -20.8 & 16.36 & 40.38 & ${41.85}_{-0.05}^{+0.04}$ &  0.31 & ${9.87}_{-1.54}^{+1.54}$  & ${1.26}_{-0.14}^{+0.12}$  & ${14.77}\pm{0.42}$ \\    
B 08-4 & 0.05597 & -20.1 &  4.65 & 40.26 & ${41.53}_{-0.05}^{+0.04}$ &  0.68 & ${2.64}_{-0.33}^{+0.33}$  & ${0.62}_{-0.09}^{+0.08}$  & ${4.19}\pm{0.18}$ \\     
B 10-1 & 0.06661 & -19.9 &  3.93 & 40.26 & ${41.16}_{-0.01}^{+0.02}$ &  0.86 & ${1.07}_{-0.14}^{+0.14}$  & ${0.57}_{-0.15}^{+0.13}$  & ${2.15}\pm{0.10}$ \\     
B 11-2 & 0.07533 & -20.3 &  8.67 & 40.32 & ${41.67}_{-0.05}^{+0.04}$ &  3.05 & ${3.64}_{-0.66}^{+0.66}$  & ${0.62}_{-0.18}^{+0.15}$  & ${9.10}\pm{1.23}$ \\     
B 14-1 & 0.06068 & -20.3 & 22.15 & 40.31 & ${40.86}_{-0.02}^{+0.02}$ &  2.92 & ${1.07}_{-0.10}^{+0.10}$  & ${1.33}_{-0.09}^{+0.08}$  & ${2.31}\pm{0.16}$ \\     
B 15-1 & 0.06575 & -20.4 &  6.67 & 40.22 & ${41.12}_{-0.01}^{+0.01}$ &  1.11 & ${0.74}_{-0.08}^{+0.08}$  & ${0.26}_{-0.12}^{+0.11}$  & ${1.93}\pm{0.10}$ \\     
B 15-2 & 0.07661 & -20.0 &  7.19 & 40.33 & ${41.07}_{-0.02}^{+0.02}$ &  0.72 & ${0.98}_{-0.14}^{+0.14}$  & ${0.69}_{-0.16}^{+0.14}$  & ${1.66}\pm{0.12}$ \\     
B 20-1 & 0.08017 & -20.8 &  7.56 & 40.51 & ${41.40}_{-0.01}^{+0.01}$ &  1.01 & ${1.69}_{-0.17}^{+0.17}$  & ${0.48}_{-0.10}^{+0.10}$  & ${4.07}\pm{0.33}$ \\     
C 00-1 & 0.06083 & -19.5 &  1.29 & 40.77 & ${41.29}_{-0.01}^{+0.01}$ &  0.39 & ${0.65}_{-0.04}^{+0.04}$  & ${-0.29}_{-0.06}^{+0.05}$ & ${1.35}\pm{0.08}$ \\     
C 04-1 & 0.06657 & -20.0 &  4.86 & 40.74 & ${41.65}_{-0.05}^{+0.04}$ &  1.18 & ${4.18}_{-0.48}^{+0.48}$  & ${0.82}_{-0.06}^{+0.06}$  & ${5.94}\pm{0.32}$ \\     
C 04-2 & 0.07067 & -19.9 &  3.06 & 40.71 & ${41.64}_{-0.05}^{+0.04}$ &  1.31 & ${3.59}_{-0.41}^{+0.41}$  & ${0.69}_{-0.06}^{+0.06}$  & ${4.94}\pm{0.31}$ \\     
C 13-1 & 0.07876 & -21.7 & 35.78 & 40.72 & ${41.94}_{-0.01}^{+0.01}$ &  4.03 & ${9.93}_{-0.93}^{+0.93}$  & ${1.05}_{-0.10}^{+0.09}$  & ${14.22}\pm{1.02}$ \\    
C 13-3 & 0.07110 & -20.8 & 37.71 & 40.78 & ${41.41}_{-0.01}^{+0.01}$ &  3.97 & ${5.62}_{-0.50}^{+0.50}$  & ${1.74}_{-0.10}^{+0.09}$  & ${5.94}\pm{0.48}$ \\     
C 14-2 & 0.05620 & -20.2 &  5.61 & 40.78 & ${41.49}_{-0.01}^{+0.01}$ &  0.63 & ${3.32}_{-0.17}^{+0.17}$  & ${0.98}_{-0.05}^{+0.05}$  & ${3.88}\pm{0.29}$ \\     
C 20-2 & 0.07722 & -20.2 &  8.50 & 40.67 & ${41.29}_{-0.02}^{+0.02}$ &  0.75 & ${1.50}_{-0.13}^{+0.13}$  & ${0.61}_{-0.08}^{+0.08}$  & ${2.72}\pm{0.21}$ \\     
C 21-1 & 0.07850 & -21.1 & 30.92 & 40.72 & ${41.17}_{-0.03}^{+0.03}$ &  3.97 & ${2.30}_{-0.32}^{+0.32}$  & ${1.37}_{-0.14}^{+0.12}$  & ${3.63}\pm{0.45}$ \\     
C 22-2 & 0.07116 & -21.0 & 15.28 & 41.05 & ${41.68}_{-0.01}^{+0.01}$ &  4.89 & ${5.32}_{-0.28}^{+0.28}$  & ${1.00}_{-0.05}^{+0.05}$  & ${7.24}\pm{0.55}$ \\     
D 00-2 & 0.08130 & -21.6 & 24.30 & 41.59 & ${42.21}_{-0.01}^{+0.01}$ &  9.86 & ${19.65}_{-0.72}^{+0.72}$ & ${1.11}_{-0.04}^{+0.04}$  & ${18.74}\pm{1.74}$ \\    
D 10-4 & 0.06738 & -20.3 &  5.49 & 41.53 & ${42.24}_{-0.01}^{+0.01}$ &  5.43 & ${14.66}_{-0.55}^{+0.55}$ & ${0.70}_{-0.03}^{+0.03}$  & ${13.80}\pm{0.89}$ \\    
D 13-1 & 0.05881 & -20.0 &  1.69 & 41.08 & ${41.85}_{-0.01}^{+0.01}$ &  1.29 & ${4.33}_{-0.17}^{+0.17}$  & ${0.36}_{-0.04}^{+0.04}$  & ${5.17}\pm{0.33}$ \\     
D 13-5 & 0.07535 & -21.7 & 53.84 & 41.52 & ${42.10}_{-0.01}^{+0.01}$ & 15.14 & ${21.20}_{-0.86}^{+0.86}$ & ${1.46}_{-0.04}^{+0.04}$  & ${20.17}\pm{1.48}$ \\    
D 14-1 & 0.07362 & -21.0 & 20.37 & 41.24 & ${41.92}_{-0.01}^{+0.01}$ &  6.51 & ${11.30}_{-0.54}^{+0.54}$ & ${1.23}_{-0.05}^{+0.05}$  & ${17.46}\pm{8.76}$ \\    
D 15-1 & 0.05929 & -19.8 &  4.99 & 40.96 & ${41.45}_{-0.02}^{+0.02}$ &  0.96 & ${2.67}_{-0.17}^{+0.17}$  & ${0.83}_{-0.05}^{+0.05}$  & ${3.00}\pm{0.29}$ \\     
D 15-2 & 0.05619 & -19.1 &  1.89 & 40.97 & ${41.22}_{-0.02}^{+0.02}$ &  0.76 & ${0.88}_{-0.06}^{+0.06}$  & ${0.21}_{-0.04}^{+0.04}$  & ${1.17}\pm{0.15}$ \\     
D 15-3 & 0.06712 & -21.4 & 54.15 & 41.03 & ${41.79}_{-0.01}^{+0.01}$ &  7.69 & ${13.70}_{-1.00}^{+1.00}$ & ${1.77}_{-0.08}^{+0.07}$  & ${14.74}\pm{0.88}$ \\    
D 20-1 & 0.07049 & -20.9 & 18.56 & 41.22 & ${41.74}_{-0.01}^{+0.01}$ &  5.63 & ${5.35}_{-0.25}^{+0.25}$  & ${0.88}_{-0.05}^{+0.05}$  & ${6.80}\pm{0.55}$ \\     
D 21-3 & 0.05746 & -21.2 & 29.87 & 41.05 & ${41.40}_{-0.01}^{+0.01}$ &  6.10 & ${2.80}_{-0.18}^{+0.18}$  & ${1.02}_{-0.07}^{+0.06}$  & ${3.29}\pm{0.32}$ \\     
D 22-1 & 0.06793 & -21.9 & 59.24 & 41.06 & ${41.78}_{-0.01}^{+0.01}$ &  6.74 & ${6.51}_{-0.45}^{+0.45}$  & ${0.98}_{-0.07}^{+0.07}$  & ${9.90}\pm{1.13}$ \\     
D 22-2 & 0.08121 & -20.8 &  9.29 & 41.27 & ${41.91}_{-0.01}^{+0.01}$ &  5.10 & ${9.11}_{-0.37}^{+0.37}$  & ${1.02}_{-0.04}^{+0.04}$  & ${11.64}\pm{0.71}$ \\    
D 23-1 & 0.08089 & -21.1 & 10.20 & 41.34 & ${41.83}_{-0.01}^{+0.01}$ &  3.83 & ${7.69}_{-0.35}^{+0.35}$  & ${1.02}_{-0.05}^{+0.05}$  & ${8.42}\pm{0.74}$ \\     
E 00-2 & 0.14577 & -21.7 & 29.19 & 41.66 & ${41.97}_{-0.01}^{+0.01}$ & 10.61 & ${9.27}_{-0.32}^{+0.32}$  & ${0.90}_{-0.03}^{+0.02}$  & ${8.03}\pm{2.44}$ \\     
E 00-3 & 0.14662 & -22.2 & 52.58 & 41.28 & ${41.84}_{-0.01}^{+0.01}$ &  6.84 & ${8.35}_{-0.54}^{+0.54}$  & ${1.09}_{-0.07}^{+0.06}$  & ${12.35}\pm{1.46}$ \\    
E 04-1 & 0.14997 & -21.7 & 58.67 & 41.49 & ${41.93}_{-0.02}^{+0.02}$ & 10.25 & ${12.71}_{-0.98}^{+0.98}$ & ${1.33}_{-0.07}^{+0.07}$  & ${19.10}\pm{4.95}$ \\    
E 09-1 & 0.13892 & -21.7 & 36.67 & 41.20 & ${41.90}_{-0.02}^{+0.02}$ &  3.08 & ${9.53}_{-0.98}^{+0.98}$  & ${1.10}_{-0.11}^{+0.10}$  & ${16.65}\pm{8.94}$ \\    
E 10-1 & 0.13897 & -22.0 & 65.00 & 41.52 & ${42.23}_{-0.01}^{+0.01}$ & 28.08 & ${26.66}_{-1.73}^{+1.73}$ & ${1.39}_{-0.06}^{+0.06}$  & ${31.96}\pm{6.23}$ \\    
E 23-1 & 0.14485 & -21.5 & 19.82 & 41.69 & ${42.20}_{-0.01}^{+0.01}$ &  4.92 & ${15.99}_{-0.76}^{+0.76}$ & ${0.91}_{-0.04}^{+0.04}$  & ${23.04}\pm{3.27}$ \\    
F 08-2 & 0.06475 & -20.7 &  7.40 & 41.80 & ${42.47}_{-0.05}^{+0.04}$ &  8.61 & ${22.66}_{-2.36}^{+2.36}$ & ${0.60}_{-0.03}^{+0.03}$  & ${19.35}\pm{1.31}$ \\    
F 09-1 & 0.08279 & -21.0 & 22.95 & 41.77 & ${42.63}_{-0.05}^{+0.04}$ & 25.05 & ${79.79}_{-8.40}^{+8.40}$ & ${1.58}_{-0.04}^{+0.04}$  & ${49.12}\pm{4.53}$ \\    
F 10-1 & 0.08351 & -20.4 &  4.06 & 41.58 & ${42.06}_{-0.05}^{+0.04}$ &  2.00 & ${6.44}_{-0.69}^{+0.69}$  & ${0.26}_{-0.04}^{+0.04}$  & ${18.60}\pm{\infty\tablenotemark{i}}$ \\
F 12-4 & 0.06728 & -21.6 & 15.03 & 41.54 & ${42.42}_{-0.05}^{+0.04}$ &  8.90 & ${18.91}_{-1.99}^{+1.99}$ & ${0.53}_{-0.04}^{+0.04}$  & ${21.29}\pm{1.12}$ \\    
G 03-2 & 0.12946 & -21.3 &  6.51 & 41.81 & ${42.37}_{-0.01}^{+0.01}$ &  7.91 & ${17.79}_{-0.89}^{+0.89}$ & ${0.59}_{-0.05}^{+0.05}$  & ${28.43}\pm{7.93}$ \\    
G 03-4 & 0.13373 & -21.4 & 64.61 & 41.84 & ${42.35}_{-0.01}^{+0.01}$ & 41.03 & ${56.62}_{-2.08}^{+2.08}$ & ${1.90}_{-0.03}^{+0.03}$  & ${52.90}\pm{6.61}$ \\    
G 04-1 & 0.12981 & -22.0 & 64.74 & 41.75 & ${42.36}_{-0.01}^{+0.01}$ & 28.04 & ${41.61}_{-2.17}^{+2.17}$ & ${1.55}_{-0.05}^{+0.05}$  & ${43.68}\pm{11.65}$ \\   
G 08-1 & 0.13492 & -21.2 &  8.01 & 41.77 & ${42.26}_{-0.01}^{+0.01}$ &  5.98 & ${10.77}_{-0.38}^{+0.38}$ & ${0.31}_{-0.03}^{+0.03}$  & ${36.76}\pm{\infty\tablenotemark{i}}$ \\
G 08-2 & 0.13194 & -21.2 & 41.09 & 41.67 & ${42.13}_{-0.01}^{+0.01}$ & 19.24 & ${26.85}_{-1.67}^{+1.67}$ & ${1.64}_{-0.06}^{+0.06}$  & ${30.12}\pm{8.59}$ \\    
G 08-3 & 0.14270 & -21.6 & 25.54 & 41.91 & ${42.57}_{-0.01}^{+0.01}$ & 10.56 & ${42.84}_{-1.80}^{+1.80}$ & ${1.04}_{-0.04}^{+0.04}$  & ${44.83}\pm{4.25}$ \\    
G 08-4 & 0.13964 & -21.5 & 11.05 & 41.71 & ${42.12}_{-0.01}^{+0.01}$ &  5.65 & ${9.65}_{-0.54}^{+0.54}$  & ${0.55}_{-0.05}^{+0.05}$  & ${16.29}\pm{4.94}$ \\    
G 08-5 & 0.13217 & -21.2 & 17.31 & 41.69 & ${42.05}_{-0.01}^{+0.01}$ &  9.77 & ${16.62}_{-1.02}^{+1.02}$ & ${1.31}_{-0.06}^{+0.05}$  & ${23.60}\pm{4.96}$ \\    
G 09-1 & 0.13996 & -21.5 & 26.13 & 41.77 & ${42.18}_{-0.01}^{+0.01}$ & 11.96 & ${25.70}_{-1.47}^{+1.47}$ & ${1.46}_{-0.05}^{+0.05}$  & ${33.42}\pm{9.57}$ \\    
G 10-1 & 0.14372 & -21.7 & 12.24 & 41.92 & ${42.37}_{-0.01}^{+0.01}$ &  9.64 & ${23.00}_{-1.00}^{+1.00}$ & ${0.87}_{-0.04}^{+0.04}$  & ${28.39}\pm{3.21}$ \\    
G 11-1 & 0.13705 & -22.5 & 50.27 & 41.78 & ${42.46}_{-0.01}^{+0.01}$ & 20.23 & ${42.74}_{-2.05}^{+2.05}$ & ${1.33}_{-0.05}^{+0.05}$  & ${46.67}\pm{14.12}$ \\   
G 13-1 & 0.13876 & -21.7 & 11.11 & 42.21 & ${42.58}_{-0.01}^{+0.01}$ & 24.54 & ${26.47}_{-0.78}^{+0.78}$ & ${0.50}_{-0.03}^{+0.03}$  & ${24.09}\pm{2.78}$ \\    
G 14-1 & 0.13233 & -21.7 & 22.33 & 41.69 & ${41.93}_{-0.01}^{+0.01}$ & 10.11 & ${8.30}_{-0.33}^{+0.33}$  & ${0.86}_{-0.04}^{+0.04}$  & ${9.72}\pm{2.38}$ \\     
G 14-3 & 0.14464 & -21.4 & 26.27 & 41.94 & ${42.16}_{-0.01}^{+0.01}$ & 19.98 & ${20.87}_{-0.95}^{+0.95}$ & ${1.30}_{-0.05}^{+0.04}$  & ${18.44}\pm{4.33}$ \\    
G 20-1 & 0.13277 & -22.0 & 53.06 & 42.30 & ${42.63}_{-0.00}^{+0.00}$ & 26.65 & ${46.03}_{-1.23}^{+1.23}$ & ${0.98}_{-0.03}^{+0.03}$  & ${33.72}\pm{5.43}$ \\    
G 20-2 & 0.14113 & -22.0 & 21.56 & 41.91 & ${42.26}_{-0.01}^{+0.01}$ & 19.16 & ${17.27}_{-0.66}^{+0.66}$ & ${0.83}_{-0.04}^{+0.04}$  & ${19.12}\pm{4.16}$ \\    
G 21-2 & 0.13499 & -21.1 & 10.78 & 41.76 & ${42.21}_{-0.01}^{+0.01}$ &  9.83 & ${19.26}_{-0.95}^{+0.95}$ & ${1.07}_{-0.05}^{+0.05}$  & ${29.25}\pm{35.03}$ \\   
H 10-2 & 0.14907 & -22.1 &  9.50 & 42.02 & ${42.68}_{-0.05}^{+0.04}$ & 12.48 & ${25.35}_{-2.68}^{+2.68}$ & ${0.21}_{-0.04}^{+0.04}$  & ${30.28}\pm{2.73}$ \\    

  \enddata

  \tablenotetext{a}{The $r$-band absolute magnitude.}

  \tablenotetext{b}{The stellar mass of the object reproduced from
    \cite{Kauffmann2003a}. Their masses have been scaled by 0.88 to
    convert from their \cite{2001MNRAS.322..231K} initial-mass function.}

  \tablenotetext{c}{The \Ha\ luminosity as measured by the Sloan
    spectro-photometry from a 3 arcsec diameter fibre aperture.}

  \tablenotetext{d}{The \Ha\ luminosity as measured by our IFU
    observations in the masked region (\S~\ref{sec:emiss-line-fitt}).}

  \tablenotetext{e}{The star formation rate reported by
    \cite{2004MNRAS.351.1151B}. These have been scaled by 0.88 to
    convert from their \cite{2001MNRAS.322..231K} initial-mass function.}

  \tablenotetext{f}{The star formation rate measured from our IFU
    observations, including a dust correction (Section
    \ref{sec:star-formation-rates}).}

  \tablenotetext{g}{The extinction of \Ha\ due to dust in magnitudes.}

  \tablenotetext{h}{The gas mass (Section \ref{sec:gas-mass}).}

  \tablenotetext{i}{These values are effectively unconstrained because
    of the inclination correction.}

\end{deluxetable*}

\clearpage
\LongTables
\begin{deluxetable*}{lrrrrrrrr}

  \tablecaption{\label{tab:kinematic}Kinematic Properties of the Sample}
  \tablewidth{0pt}
  \tablecolumns{7}
  \tablehead{
    \colhead{Sel. ID} &
    \colhead{$r_{{\rm petro},r}$} &
    \colhead{$r_{{\rm exp},r}$} &
    \colhead{$r_{t}$} &
    \colhead{$i$} &
    \colhead{$\Vcirc$\tablenotemark{a}} &
    \colhead{$\sigmean$\tablenotemark{b}} &
    \colhead{Kinematic} \\
    \colhead{~} &
    \colhead{kpc} &
    \colhead{kpc} &
    \colhead{kpc} &
    \colhead{~} &
    \colhead{km/s} &
    \colhead{km/s} &
    \colhead{Class}
  }
  \startdata
  A 04-3 &  9.8 & 4.9 & $0.7\pm0.2$  & 58 & $218\pm 2$  & $10$ & RD \\ 
A 04-4 & 11.4 & 6.2 & $2.8\pm0.1$  & 54 & $119\pm 4$  & $29$ & PR \\ 
A 08-3 & 10.7 & 5.9 & $2.6\pm0.0$  & 59 & $149\pm 3$  & $19$ & RD \\ 
A 08-4 & 13.0 & 7.5 & $17.0\pm0.1$ & 80 & $336\pm 23$ & $44$ & PR \\ 
A 10-1 & 10.7 & 7.2 & $2.5\pm0.1$  & 76 & $217\pm 3$  & $39$ & PR \\ 
A 10-2 & 10.4 & 7.4 & $17.7\pm0.1$ & 81 & $264\pm 19$ & $41$ & PR \\ 
A 13-2 & 13.4 & 5.6 & $3.7\pm0.2$  & 35 & $128\pm 12$ & $23$ & PR \\ 
B 04-3 &  8.8 & 5.2 & $4.0\pm0.2$  & 69 & $186\pm 4$  & $49$ & RD \\ 
B 04-4 &  6.8 & 2.9 & $1.8\pm0.2$  & 28 & $139\pm 11$ & $35$ & cRD \\
B 08-3 & 10.2 & 6.5 & $3.6\pm0.0$  & 63 & $167\pm 2$  & $40$ & RD \\ 
B 08-4 &  6.0 & 3.8 & $1.7\pm0.1$  & 56 & $125\pm 3$  & $41$ & RD \\ 
B 10-1 &  7.1 & 4.7 & $2.7\pm0.1$  & 66 & $138\pm 4$  & $27$ & RD \\ 
B 11-2 &  8.5 & 3.3 & $3.9\pm0.0$  & 20 & $66\pm 14$  & $33$ & RD \\ 
B 14-1 &  8.3 & 5.9 & $0.6\pm0.1$  & 78 & $197\pm 2$  & $17$ & RD \\ 
B 15-1 &  9.2 & 5.6 & $2.5\pm0.1$  & 44 & $185\pm 6$  & $30$ & RD \\ 
B 15-2 & 11.0 & 5.4 & $1.4\pm0.1$  & 55 & $124\pm 3$  & $24$ & RD \\ 
B 20-1 & 10.5 & 5.2 & $2.3\pm0.1$  & 35 & $105\pm 5$  & $29$ & RD \\ 
C 00-1 &  6.3 & 3.0 & $2.9\pm0.1$  & 56 & $107\pm 8$  & $32$ & cRD \\
C 04-1 &  6.4 & 3.5 & $3.9\pm0.3$  & 49 & $148\pm 9$  & $46$ & RD \\ 
C 04-2 &  5.0 & 2.7 & $4.2\pm0.6$  & 59 & $65\pm 7$   & $46$ & cRD \\
C 13-1 & 10.8 & 5.3 & $2.3\pm0.1$  & 31 & $223\pm 8$  & $29$ & RD \\ 
C 13-3 &  8.1 & 5.1 & $0.7\pm0.2$  & 72 & $220\pm 2$  & $11$ & RD \\ 
C 14-2 &  4.5 & 2.7 & $0.8\pm0.0$  & 58 & $164\pm 4$  & $26$ & cRD \\
C 20-2 &  5.1 & 2.6 & $0.9\pm0.1$  & 45 & $170\pm 6$  & $4$  & cRD \\
C 21-1 &  4.5 & 1.9 & $0.7\pm0.2$  & 58 & $246\pm 9$  & $41$ & cRD \\
C 22-2 &  6.8 & 2.8 & $1.5\pm0.1$  & 35 & $197\pm 9$  & $40$ & cRD \\
D 00-2 &  6.3 & 2.8 & $1.1\pm0.1$  & 35 & $62\pm 2$   & $38$ & cRD \\
D 10-4 &  5.4 & 2.8 & $19.1\pm0.2$ & 68 & $59\pm 23$  & $63$ & cPR \\
D 13-1 &  6.2 & 3.6 & $4.8\pm0.1$  & 76 & $120\pm 10$ & $36$ & RD \\ 
D 13-5 &  9.0 & 4.4 & $0.7\pm0.2$  & 49 & $192\pm 2$  & $46$ & RD \\ 
D 14-1 & 10.3 & 4.3 & \nodata      & 12 & \nodata     & $39$ & RD \\ 
D 15-1 &  4.0 & 2.0 & $0.6\pm0.1$  & 55 & $142\pm 5$  & $24$ & cRD \\
D 15-2 &  2.4 & 0.6 & $0.5\pm0.1$  & 38 & $42\pm 22$  & $42$ & cPR \\
D 15-3 &  8.9 & 4.7 & $0.8\pm0.0$  & 47 & $240\pm 3$  & $31$ & RD \\ 
D 20-1 &  6.1 & 2.9 & $1.3\pm0.0$  & 39 & $137\pm 5$  & $37$ & cRD \\
D 21-3 & 10.9 & 4.2 & $0.5\pm0.1$  & 57 & $175\pm 2$  & $23$ & RD \\ 
D 22-1 &  5.2 & 2.0 & $0.6\pm0.2$  & 20 & $141\pm 10$ & $31$ & cRD \\
D 22-2 &  9.1 & 5.0 & $5.5\pm0.1$  & 54 & $155\pm 6$  & $35$ & RD \\ 
D 23-1 &  5.7 & 3.0 & $2.8\pm0.1$  & 54 & $186\pm 10$ & $35$ & cRD \\
E 00-2 &  7.9 & 3.4 & \nodata      & 36 & (127)       & $84$ & CK \\ 
E 00-3 & 19.6 & 7.4 & \nodata      & 64 & (104)       & $42$ & CK \\ 
E 04-1 &  8.2 & 3.5 & $1.3\pm0.3$  & 35 & $395\pm 22$ & $58$ & RD \\ 
E 09-1 &  8.8 & 4.2 & $1.6\pm0.2$  & 19 & $267\pm 46$ & $37$ & RD \\ 
E 10-1 &  9.6 & 3.9 & $1.2\pm0.3$  & 31 & $231\pm 10$ & $45$ & RD \\ 
E 23-1 &  7.4 & 2.6 & \nodata      & 44 & (118)       & $63$ & cCK \\
F 08-2 &  3.7 & 1.8 & \nodata      & 65 & (54)        & $78$ & cCK \\
F 09-1 &  7.5 & 3.7 & \nodata      & 62 & (66)        & $77$ & CK \\ 
F 10-1 &  3.3 & 1.1 & \nodata      &  1 & \nodata     & $54$ & cRD \\
F 12-4 &  4.3 & 1.9 & $0.6\pm0.2$  & 35 & $134\pm 4$  & $57$ & cRD \\
G 03-2 &  7.0 & 2.8 & $1.1\pm0.3$  & 24 & $174\pm 24$ & $43$ & cRD \\
G 03-4 &  9.7 & 4.5 & \nodata      & 58 & (89)        & $53$ & CK \\ 
G 04-1 &  8.2 & 3.1 & $1.2\pm0.1$  & 24 & $269\pm 22$ & $50$ & cRD \\
G 08-1 &  6.3 & 1.6 & \nodata      &  1 & \nodata     & $43$ & cRD \\
G 08-2 &  5.8 & 1.9 & \nodata      & 29 & (25)        & $42$ & cCK \\
G 08-3 &  8.6 & 4.2 & \nodata      & 52 & (55)        & $55$ & CK \\ 
G 08-4 &  8.2 & 2.6 & $1.6\pm0.2$  & 30 & $114\pm 11$ & $49$ & cRD \\
G 08-5 &  8.3 & 3.1 & $1.2\pm0.1$  & 36 & $243\pm 15$ & $64$ & RD \\ 
G 09-1 &  6.8 & 2.7 & \nodata      & 29 & (60)        & $55$ & cCK \\
G 10-1 &  6.2 & 2.5 & $6.1\pm0.7$  & 49 & $103\pm 17$ & $62$ & cPR \\
G 11-1 &  9.7 & 4.3 & $2.8\pm0.3$  & 23 & $314\pm 37$ & $47$ & RD \\ 
G 13-1 &  7.4 & 3.2 & $1.2\pm0.3$  & 69 & $112\pm 2$  & $76$ & PR \\ 
G 14-1 &  7.1 & 2.5 & $1.2\pm0.3$  & 35 & $136\pm 8$  & $70$ & cRD \\
G 14-3 &  5.5 & 1.6 & $3.9\pm0.5$  & 41 & $159\pm 40$ & $81$ & cPR \\
G 20-1 &  5.8 & 2.4 & \nodata      & 43 & (60)        & $96$ & cCK \\
G 20-2 &  5.7 & 2.0 & $1.2\pm0.3$  & 34 & $166\pm 10$ & $45$ & cRD \\
G 21-2 &  6.5 & 2.5 & \nodata      & 15 & (85)        & $34$ & cCK \\
H 10-2 &  7.4 & 2.7 & $21.8\pm3.3$ & 49 & $62\pm 29$  & $59$ & cPR \\

  \enddata

  \tablenotetext{a}{For objects classified as CK and cCK, the value of
    $\Vshear/2$ is shown in parenthesis instead of $\Vcirc$.}

  \tablenotetext{b}{The statistical error on $\sigmean$ is typically
    1--$2\kms$ (see \S~\ref{sec:sigm-bs}.}

\end{deluxetable*}



\end{document}